\documentclass[prd,showpacs,letterpaper,10pt,aps,notitlepage,nofootinbib,superscriptaddress]{revtex4-2}
\usepackage[utf8]{inputenc}
\usepackage{color}

\usepackage{physics}
\usepackage{slashed}
\usepackage{amsmath}
\usepackage{mathtools}
\usepackage{placeins}
\usepackage{amsfonts}
\usepackage{amssymb}

\usepackage{hyperref}
\hypersetup{colorlinks, linkcolor = [rgb]{0,0.0,0.75}, citecolor = [rgb]{0,0.0,0.75}, urlcolor = [rgb]{0,0.0,0.75}}

\begin{document}

\title{Methods for high-precision determinations of radiative-leptonic decay form factors using lattice QCD}

\author{Davide Giusti}
\affiliation{Fakultät für Physik, Universität Regensburg, 93040, Regensburg, Germany}

\author{Christopher F. Kane}
\affiliation{Department of Physics, University of Arizona, Tucson, AZ 85721, USA}

\author{Christoph Lehner}
\affiliation{Fakultät für Physik, Universität Regensburg, 93040, Regensburg, Germany}

\author{Stefan Meinel}
\affiliation{Department of Physics, University of Arizona, Tucson, AZ 85721, USA}

\author{Amarjit Soni}
\affiliation{Brookhaven National Laboratory, Upton, NY 11973, USA}

\date{April 19, 2023}

\begin{abstract}
    We present a study of lattice-QCD methods to determine the relevant hadronic form factors for radiative leptonic decays of pseudoscalar mesons. We provide numerical results for $D_s^+ \to \ell^+ \nu \gamma$. Our calculation is performed using a domain-wall action for all quark flavors and on a single RBC/UKQCD lattice gauge-field ensemble. The first part of the study is how to best control two sources of systematic error inherent in the calculation, specifically the unwanted excited states created by the meson interpolating field, and unwanted exponentials in the sum over intermediate states. Using a 3d sequential propagator allows for better control over unwanted exponentials from intermediate states, while using a 4d sequential propagator allows for better control over excited states. We perform individual analyses of the 3d and 4d methods as well as a combined analysis using both methods, and find that the 3d sequential propagator offers good control over both sources of systematic uncertainties for the smallest number of propagator solves. From there, we further improve the use of a 3d sequential propagator by employing an infinite-volume approximation method, which allows us to calculate the relevant form factors over the entire allowed range of photon energies. We then study improvements gained by performing the calculation using a different three-point function, using ratios of three-point functions, averaging over positive and negative photon momentum, and using an improved method for extracting the structure-dependent part of the axial form factor. The optimal combination of methods yields results for the $D_s^+ \to \ell^+ \nu \gamma$ structure-dependent vector and axial form factors in the entire kinematic range with statistical plus fitting uncertainties of order 5\%, using 25 gauge configurations with 64 samples per configuration.
\end{abstract}

\maketitle

\tableofcontents

\section{Introduction}

In this paper, we develop and test lattice-QCD methods for computing the hadronic matrix elements describing \emph{radiative leptonic decays} of pseudoscalar mesons, i.e., $H\to \ell^-\bar{\nu}\gamma$ or $H\to \ell^+\ell^-\gamma$. Such transitions are of interest both for soft photons and for hard photons, as discussed in the following.

Knowledge of the radiative leptonic decay rate in the region of small (soft) photon energies is required to include ${\mathcal O}(\alpha_{em})$ electromagnetic corrections to purely leptonic decays, needed for sub-percent precision determinations of Cabibbo-Kobayashi-Maskawa (CKM) matrix elements. According to the well-known Bloch-Nordsieck mechanism \cite{Bloch:1937pw}, the integral of the radiative decay rate in the phase space region corresponding to soft photons must be added to the decay rate with no real photons in the final states (the so-called virtual electromagnetic contribution to the decay rate, which has recently been computed on the lattice \cite{Giusti:2017dwk,DiCarlo:2019thl,Boyle:2022lsi}) in order to cancel infrared divergent contributions appearing in unphysical quantities at intermediate stages of the calculations. While for $\pi^- \to \mu^- \bar \nu_\mu (\gamma)$ and $K^- \to \mu^- \bar \nu_\mu (\gamma)$, at the current level of precision it is sufficient to evaluate the real soft-photon contributions in an effective theory in which the meson is treated as a point-like particle, structure-dependent contributions to the real photon emission are significant for $\pi^- \to e^- \bar \nu_e (\gamma)$ and $K^- \to e^- \bar \nu_e (\gamma)$ \cite{ParticleDataGroup:2022pth}.

In the region of hard (experimentally detectable) photon energies, radiative leptonic decays represent important probes of the internal structure of the mesons, and also provide sensitive probes of physics beyond the Standard Model inducing non-standard currents and/or non-universal corrections to the lepton couplings. For example, the rare decays $B_s^0\to\ell^+\ell^-\gamma$ and $B^0\to\ell^+\ell^-\gamma$ are sensitive to all operators in the $b\to s \ell^+\ell^-$ and $b\to d \ell^+\ell^-$ effective Hamiltonians, respectively, unlike their purely leptonic counterparts \cite{Kruger:2002gf,Melikhov:2004mk,Dettori:2016zff,Albrecht:2019zul,Beneke:2020fot,Chen:2020szf,Carvunis:2021jga}. There are presently hints of lepton-flavor-universal new physics contributing to the Wilson coefficient $C_9^{bs\ell\ell}$ \cite{Greljo:2022jac}, to which $B_s\to\mu^+\mu^-$ is insensitive, but which can be probed in a novel way with $B_s\to\mu^+\mu^-\gamma$. In addition, because the hard photon in the final state removes the helicity suppression, $B_{(s)}^0\to\ell^+\ell^-\gamma$ decays can also be used to test electron-vs-muon lepton universality \cite{Guadagnoli:2017quo}, which would not be possible with purely leptonic $B_{(s)}^0\to\ell^+\ell^-$ decays. Radiative leptonic $B^- \to \ell^- \bar{\nu} \gamma$ decays at high photon energy can provide novel determinations of $|V_{ub}|$ using light leptons, and are also well suited to constrain the first inverse moment of the $B$-meson light-cone distribution amplitude, an important input in QCD-factorization predictions for non-leptonic $B$ decays that is presently poorly determined \cite{Korchemsky:1999qb,Beneke:1999br,Descotes-Genon:2002crx,Lunghi:2002ju,Braun:2012kp,Wang:2016qii,Beneke:2018wjp,Wang:2018wfj,Shen:2018abs,Shen:2020hsp}.

The experimental status for radiative leptonic decays (with detected photons of energy above some specified lower limit) can be summarized as follows. For the kaon and pion decays $K^- \to e^- \bar \nu \gamma$, $K^- \to \mu^- \bar \nu \gamma$, $\pi^- \to e^- \bar \nu \gamma$, and $\pi^- \to \mu^- \bar \nu \gamma$, there are already precise measurements of the differential branching fractions \cite{E787:2000ehx,Bychkov:2008ws,KLOE:2009urs,ISTRA:2010smy,OKA:2019gav,Kobayashi:2022hwh,ParticleDataGroup:2022pth}. For the charmed-meson radiative leptonic decays $D^+ \to e^+ \nu \gamma$ and $D_s^+ \to e^+ \nu \gamma$, the BESIII collaboration has reported upper limits on the branching fractions with $E_\gamma^{(0)} > 10$ MeV of $3.0 \times 10^{-5}$ and $1.3\times 10^{-4}$, respectively \cite{BESIII:2017whk,BESIII:2019pjk}. In the bottom sector, the Belle collaboration reported an upper limit $\mathcal{B} (B^- \to \ell^- \bar \nu \gamma, E_\gamma^{(0)} > 1 ~ {\rm GeV}) < 3.0 \times 10^{-6}$, close to the Standard-Model expectation \cite{Belle:2018jqd}. It is expected that Belle II will eventually measure the $B^- \to \ell^- \bar \nu \gamma$ branching fractions with 3.6\% statistical uncertainty \cite{Belle-II:2018jsg}. For the flavor-changing neutral current decays $B^0\to\ell^+\ell^-\gamma$, BaBar reported upper limits for the branching fractions of order $10^{-7}$ in Ref.~\cite{BaBar:2007lky}. More recently, LHCb obtained the result $\mathcal{B} (B_s^0\to\mu^+\mu^-\gamma)<2.0\times10^{-9}$ for $m_{\mu\mu}>4.9$ GeV \cite{LHCb:2021awg}.

In the Standard Model, the hadronic contributions to the $H\to \ell^-\bar{\nu}\gamma$ decay rate at leading order in $\alpha_{\rm em}$ are the decay constant, $f_H$, and two form factors $F_V$ and $F_{A,SD}$, which are functions of the photon energy in the meson rest frame and are the focus of this work. The form factors parametrize, in momentum space, a meson-to-vacuum QCD matrix element of two currents at different spacetime points: the flavor-changing quark weak current and the quark electromagnetic current.

For low photon energies, the form factors can be studied using Chiral Perturbation Theory (ChPT), which has been done for light-meson radiative leptonic decays in Refs.~\cite{Bijnens:1996wm,Geng:2003mt,Mateu:2007tr,Unterdorfer:2008zz,Cirigliano:2011ny}. Although these ChPT calculations represent a systematic  effective-field-theory approach to the problem, the low-energy constants entering in the final results at ${\mathcal O}(p^6)$ have been estimated in phenomenological analyses relying in part on model-dependent assumptions.
Heavy-meson radiative leptonic decays have been studied theoretically using quark models \cite{Atwood:1994za,Colangelo:1996ct,Chang:1997re,Geng:2000fs,Chelkov:2001qx,Hwang:2005uk,Barik:2008zza,Shen:2013oua,Kozachuk:2017mdk,Dubnicka:2018gqg}, QCD factorization/soft-collinear effective theory/perturbative QCD \cite{Korchemsky:1999qb,Beneke:1999br,Descotes-Genon:2002crx,Lunghi:2002ju,Braun:2012kp,Wang:2016qii,Beneke:2018wjp,Wang:2018wfj,Shen:2018abs,Beneke:2020fot,Shen:2020hsp}, light-cone sum rules \cite{Khodjamirian:1995uc,Ali:1995uy,Eilam:1995zv,Aliev:1996ud,Ball:2003fq,Janowski:2021yvz}, heavy-hadron ChPT \cite{Burdman:1994ip}, and dispersion relations \cite{SalehKhan:2004kj,Kurten:2022zuy}. These approaches again have various limitations, being either model-dependent, making truncations in the $1/m_Q$ and $\alpha_s$ expansions, or requiring a large number of external inputs.

All of these limitations can be overcome, at least in principle, using  lattice gauge theory, a nonperturbative formulation of QCD that does not introduce new parameters beyond those of QCD itself and whose precision is limited only by the available computing resources. Numerical lattice-QCD calculations based on the path-integral formulation are performed in Euclidean spacetime, which may pose challenges for time-dependent matrix elements. As we showed in Ref.~\cite{Kane:2019jtj} and discuss again here (and as was also shown independently in Ref.~\cite{Desiderio:2020oej}), for on-shell photons, the hadronic tensor describing radiative leptonic decays can be obtained directly from a large-Euclidean-time limit of a Euclidean three-point function. Nevertheless, in practice, it is necessary to account for the sub-leading time dependence when analyzing the simulation results \cite{Kane:2021zee}.

While the present work was in progress, an independent lattice study of radiative leptonic decays was published in Ref.~\cite{Desiderio:2020oej}. That work used the twisted-mass formulation of lattice fermions and considered decays of charged pions, kaons, $D$ and $D_s$ mesons. For the charmed mesons, the energy of the final-state emitted photon $E_\gamma^{(0)}$ was less than about 400 MeV in the rest frame of the decaying hadron. For the pion and kaon radiative leptonic decays, the results of Ref.~\cite{Desiderio:2020oej} cover the full kinematic range and were compared to experimental data in Ref.~\cite{Frezzotti:2020bfa}. Significant deviations between theory and experiment were found, in particular for $K\to \mu\nu\gamma$ at large photon energy.

Here we extend our preliminary work \cite{Kane:2019jtj,Kane:2021zee} and present a detailed study of nonperturbative lattice methods to extract the structure-dependent form factors contributing to the amplitudes of radiative three-body decays $H \to \ell^- \bar{\nu} \gamma$.
For that purpose, the relevant non-local matrix elements are calculated using two different methods, which we call the ``3d method'' and the ``4d method'', in order to control the two major sources of systematic errors related to unwanted exponentials in the sum over intermediate states and to unwanted excited states created by the meson interpolating field (Ref.~\cite{Desiderio:2020oej} used only a 4d method and use constant fits to the data where it had plateaued.)
To explore a wider range of photon energies, we perform new calculations using an infinite-volume approximation technique. We also implement more sophisticated fits to remove unwanted exponentials. In this study we make use of one of the ``24I'' RBC/UKQCD lattice gauge-field ensembles with 2+1 flavors of domain-wall fermions and the Iwasaki gauge action \cite{RBC:2010qam}, with inverse lattice spacing $a^{-1} = 1.785 (5)$ GeV and pion mass $m_\pi = 340 (1)$ MeV \cite{RBC:2014ntl}. We consider the process $D_s^- \to \ell^-\bar{\nu}\gamma$, for which we provide, for the first time, model-independent determinations of the form factors in the full kinematically allowed photon-energy range. This paper focuses on a detailed investigation of lattice data-generation and data-analysis methods. Computations at the physical pion mass and for mesons other than the $D_s$, extrapolations to the continuum limit, and phenomenological studies of the decay observables are left for future work.

The structure of the remainder of this paper is as follows. In Sec.~\ref{ssec:Tmunu_minkowski} we review the Minkowski-space hadronic tensor and in Sec.~\ref{ssec:Euclidean} demonstrate how it is related to a Euclidean time three-point correlation function. We describe the 3d and 4d methods in Sec.~\ref{section:sequential_propagators}. The details of the lattice gauge-field ensemble and the lattice actions and parameters are given in Sec.~\ref{sec:lattice_params}. Section~\ref{sec:noise_vs_point} compares the statistical precision of the vector form factor using noise and point sources. The fit methods used to remove unwanted exponentials from intermediate and excited states are described in Sec.~\ref{section:fit_method}. We compare form factors calculated from individual analyses of the 3d and 4d methods, as well as a combined analysis using both methods, in Sec.~\ref{sec:compare_3d_4d}. The infinite-volume approximation technique is reviewed in Sec.~\ref{ssec:infinite_volume_approximation}. In Sec.~\ref{ssec:em_curr_at_origin} we show how the Minkowski hadronic tensor can be calculated using a different three-point function with the electromagnetic current instead of the weak current at the coordinate origin. We explain a number of improvements for determining the relevant form factors and demonstrate the level of improvement of each method in Sec.~\ref{sec:improved_FF}. The final improved analysis procedure, as well as the final form factors results, are presented in Sec.~\ref{sec:final_procedure}, and we conclude in Sec.~\ref{sec:conclusions}. Appendix \ref{app:WI} contains a discussion of discretization effects using the lattice vector Ward-Takahashi identity.

\section{Theoretical setup}
\subsection{Decay amplitude and correlation functions in Minkowski spacetime}
\label{ssec:Tmunu_minkowski}
In this work, we focus on charged-current decays $H \to \gamma \ell \bar{\nu}$ mediated by the $V-A$ weak current in the Standard Model, but most of our methods are also applicable to other types of currents. Here, $H$ is a pseudoscalar meson composed of quarks $q_1$ and $\bar{q}_2$. Using the weak effective Hamiltonian, and assuming that $H$ is negatively charged for concreteness, the amplitude for this process can be written as \cite{Beneke:2011nf,Beneke:2018wjp}
\begin{equation}
    \mathcal{A}(H^- \to \gamma \ell^- \bar{\nu}) = \frac{G_F V_{q_1 q_2}}{\sqrt{2}} \bra{\ell^- \bar{\nu} \gamma} \bar{\ell} \gamma^\mu (1-\gamma_5) \nu \cdot \bar{q}_1 \gamma_\mu (1-\gamma_5) q_2 \ket{H}
\end{equation}
(the decay process for the positively charged pseudoscalar meson is given by replacing $\ell \to \bar{\ell}$ and $\bar{\nu} \to \nu$). Note the appearance of the CKM matrix element $V_{q_1 q_2}$. The electromagnetic component of the amplitude is computed to first order in perturbation theory, resulting in 
\begin{equation}
    \mathcal{A}(H^- \to \gamma \ell \bar{\nu}) = \frac{G_F V_{q_1 q_2}}{\sqrt{2}} \Big[ e (\epsilon^*)^\mu \bar{\ell} \gamma^\nu (1-\gamma_5) \nu \cdot T_{\mu \nu}(p_H, p_\gamma) - i e Q_\ell f_H \cdot \bar{\ell} \slashed{\epsilon}^* (1-\gamma_5) \nu \Big],
\end{equation}
where $e$ is the elementary electric charge, $\epsilon_\mu$ is the photon polarization vector, $Q_\ell$ is the charge of the lepton in units of $e$, and $f_H$ is the $H$ meson decay constant. The remaining hadronic piece is contained in the hadronic tensor
\begin{equation}
    T_{\mu \nu}(p_H, p_\gamma) = -i \int \dd t_\text{em} \int \dd^3 x \ e^{i p_\gamma \vdot x} \bra{0} \textbf{T} \big( J^\text{em}_\mu(t_\text{em},\vec{x}) J^{\text{weak}}_\nu(0) \big) \ket{H(\vec{p}_H)},
\end{equation}
where the electromagnetic current (EM) is given by $J^\text{em}_\mu = \sum_q Q_q \bar{q} \gamma_\mu q$, and the weak current is given by $J^{\text{weak}}_\nu = \bar{q}_1\gamma_\nu(1-\gamma_5) q_2 $. The hadronic tensor can be written as the sum $T_{\mu \nu}=T_{\mu \nu}^<+T_{\mu \nu}^>$ of the contributions from the two different time orderings of the currents, corresponding to the integrals over  $t_\text{em}$ from $-\infty$ to 0 and from 0 to $+\infty$, respectively. The form factor decomposition for real photons, \textit{i.e.} $p_\gamma^2=0$, is given by \cite{Beneke:2018wjp}
\begin{equation}
\begin{split}
    T_{\mu \nu} = \epsilon_{\mu \nu \tau \rho} p^{\tau}_\gamma v^{\rho} F_V + i\big[ -g_{\mu \nu}(v \vdot p_\gamma) + v_\mu(p_\gamma)_\nu \big] F_A &+ i Q_\ell \frac{v_\mu v_\nu}{(v \vdot p_\gamma)} m_H f_H 
    \\
    &+ (p_\gamma)_\mu (p_\gamma)_\nu F_1 + (p_\gamma)_\mu v_\nu F_2,
    \label{eq:Tmunu_FF_decomp}
\end{split}
\end{equation}
where $p_H^\mu = m_H v^\mu$. To calculate the decay rate, $T_{\mu \nu}$ is contracted with the photon polarization vector $\epsilon_\mu$. Because $\epsilon_\mu \cdot p_\gamma^\mu=0$, the form factors $F_1$ and $F_2$ do not contribute to the decay rate. For a given meson $H$, the axial form factor $F_A$ and vector form factor $F_V$ are functions of $v \vdot p_\gamma$, which is the photon energy seen in the rest frame of the pseudoscalar meson, denoted by $E_\gamma^{(0)}$. We define a convenient dimensionless variable $x_\gamma \equiv 2E_\gamma^{(0)}/m_H$, which takes values $0 \leq x_\gamma \leq 1-m_\ell^2/m_H^2$ for physically allowed values of $E_\gamma^{(0)}$. 

Unlike the vector form factor, the axial form factor is composed of two pieces, namely a structure-dependent contribution and a point-like contribution. The point-like contribution describes the part of the decay amplitude when the photon does not probe the internal structure of $H$ and is given by $(-Q_\ell f_H/E_\gamma^{(0)})$. Note that this piece is divergent as $E_\gamma^{(0)}$ goes to zero. The structure-dependent part of the axial form factor is finite and can be calculated by subtracting the point-like contribution, $F_{A,SD} = F_A - (-Q_\ell f_H/E_\gamma^{(0)})$. Note that in Ref~\cite{Desiderio:2020oej}, $F_{A,SD}$ is denoted as $F_A$. Additionally, the sign convention in Ref~\cite{Desiderio:2020oej} for $F_{A,SD}$ is flipped relative to the convention used in this work. 

In Sec.~\ref{ssec:Euclidean}, we demonstrate how to relate the hadronic tensor to a Euclidean three-point function. This is done by comparing the spectral decompositions of $T_{\mu \nu}^<$ and $T_{\mu \nu}^>$ to the spectral decompositions of the corresponding time orderings of the Euclidean three-point function. Here, we first consider the spectral decomposition of the hadronic tensor in Minkowski spacetime. By inserting a complete set of energy-momentum eigenstates and performing the integrals over time, we find
\begin{equation}
	\begin{split}
		T^<_{\mu \nu} &= -i \int_{-\infty(1-i\epsilon)}^0 dt_{\text{em}} \int d^3x \ e^{-i p_\gamma \vdot x} \bra{0} J_\nu^{\text{weak}}(0) J^\text{em}_\mu(t_{\text{em}}, \vec{x}) \ket{H(\vec{p}_H)}
		\\
		&= -\sum_{n} \frac{\bra{0} J_\nu^{\text{weak}}(0) \ket{n(\vec{p}_H-\vec{p}_\gamma)} \bra{n(\vec{p}_H-\vec{p}_\gamma)} J^\text{em}_\mu(0) \ket{H(\vec{p}_H)}}{2 E_{n,\vec{p}_H-\vec{p}_\gamma}(E_\gamma + E_{n,\vec{p}_H-\vec{p}_\gamma} - E_{H,\vec{p}_H} - i \epsilon)}
	\end{split}
	\label{eq:hadronic_tensor_spectral_lessthan}
\end{equation}
and
\begin{equation}
	\begin{split}
		T^>_{\mu \nu} &= -i \int_{0}^{\infty(1-i\epsilon)} dt_{\text{em}} \int d^3x \ e^{-i p_\gamma \vdot x} \bra{0} J^\text{em}_\mu(t_{\text{em}}, \vec{x}) J_\nu^{\text{weak}}(0) \ket{H(\vec{p}_H)}
		\\
		&= -\sum_{m} \frac{\bra{0} J^\text{em}_\mu(0) \ket{m(\vec{p}_\gamma)} \bra{m(\vec{p}_\gamma)} J_\nu^\text{weak}(0) \ket{H(\vec{p}_H)}}{2 E_{m,\vec{p}_\gamma}(E_\gamma - E_{m,\vec{p}_\gamma} - i \epsilon)}.
	\end{split}
	\label{eq:hadronic_tensor_spectral_greaterthan}
\end{equation}
Here we use notation appropriate for the case of a finite spatial volume in which the spectrum is discrete. In infinite volume, the sums $\sum_n$ and $\sum_m$ would also contain integrals over the continuous spectrum of multi-particle states.

\subsection{Correlation functions in Euclidean spacetime}
\label{ssec:Euclidean}

In this section, we show how to extract $T_{\mu \nu}$ from the Euclidean-time three-point correlation function
\begin{equation}
	C_{3, \mu \nu}(t_{\text{em}}, t_H) = \int d^3x \int d^3y \ e^{-i \vec{p}_\gamma \vdot \vec{x}} e^{i \vec{p}_H \vdot \vec{y}}  \langle J_{\mu}^\text{em}(t_{\text{em}}, \vec{x}) J_\nu^\text{weak}(0) \phi^\dagger_H(t_H, \vec{y}) \rangle,
	\label{eq:three_point}
\end{equation}
where the meson interpolating field is given by $\phi_H^\dagger = -\bar{q}_2 \gamma_5 q_1$ (the momentum arguments of $C_{3,\mu \nu}(t_{\text{em}}, t_H)$ are omitted for brevity). For a finite integration range $T>0$, we define the time-integrated correlation functions, for both time orderings, as
\begin{equation}
    I^<_{\mu \nu}(t_H, T) = \int^0_{-T} dt_{\text{em}}\: e^{E_\gamma t_{\text{em}}} C_{3, \mu \nu}(t_{\text{em}}, t_H), \hspace{0.1in} I^>_{\mu \nu}(t_H, T) = \int^T_{0} dt_{\text{em}}\: e^{E_\gamma t_{\text{em}}} C_{3, \mu \nu}(t_{\text{em}}, t_H).
\end{equation}
Inserting two complete sets of energy-momentum eigenstates and performing the integrals over Euclidean time, we find
\begin{align}
    \begin{split}\label{eq:Imunu_spectral_lessthan}
        I^<_{\mu \nu}(t_H, T) &= \sum_{l,n} \frac{\bra{0} J^\text{weak}_\nu(0) \ket{n(\vec{p}_H - \vec{p}_\gamma)} \bra{n(\vec{p}_H - \vec{p}_\gamma)} J^\text{em}_\mu(0) \ket{l(\vec{p}_H)} \bra{l(\vec{p}_H)} \phi^\dagger_H(0) \ket{0}}{2E_{n, \vec{p}_H - \vec{p}_\gamma} 2E_{l, \vec{p}_H}(E_\gamma + E_{n,\vec{p}_H - \vec{p}_\gamma} - E_{l,\vec{p}_H})}
       \\
       &\hspace{0.25in} \times e^{E_{l, \vec{p}_H} t_H}\Big[1 - e^{-(E_\gamma - E_{l, \vec{p}_H} + E_{n, \vec{p}_H-\vec{p}_\gamma})T}\Big],
    \end{split}
    \\
    \begin{split}\label{eq:Imunu_spectral_greaterthan}
        I^>_{\mu \nu}(t_H, T) &= \sum_{l,m} \frac{\bra{0} J^\text{em}_\mu(0) \ket{m(\vec{p}_\gamma)} \bra{m(\vec{p}_\gamma)}  J^\text{weak}_\nu(0)  \ket{l(\vec{p}_H)} \bra{l(\vec{p}_H)}  \phi^\dagger_H(0) \ket{0}}{2E_{m, \vec{p}_\gamma}2E_{l, \vec{p}_H}(E_\gamma - E_{m,\vec{p}_\gamma})} 
        \\
        &\hspace{0.25in} \times e^{E_{l,\vec{p}_H} t_H} \big[ e^{(E_\gamma - E_{m, \vec{p}_\gamma})T} - 1 \big].
    \end{split}
\end{align}
We can achieve saturation by the ground state for the initial-state pseudoscalar meson $H$ by taking the limit $t_H \to -\infty$. For large $|t_H|$ with $t_H<0$, we find
\begin{align}
    \begin{split}\label{eq:Imunu_spectral_infinitetH_lessthan}
        I^<_{\mu \nu}(t_H, T) &\to  \frac{\bra{H(\vec{p}_H)} \phi^\dagger_H(0) \ket{0} e^{E_{H, \vec{p}_H} t_H}}{2E_{H, \vec{p}_H}} \\ &\times \sum_{n} \Big[1 - e^{-(E_\gamma - E_{H, \vec{p}_H} + E_{n, \vec{p}_H-\vec{p}_\gamma})T}\Big]\frac{\bra{0} J^\text{weak}_\nu(0) \ket{n(\vec{p}_H - \vec{p}_\gamma)} \bra{n(\vec{p}_H - \vec{p}_\gamma)} J^\text{em}_\mu(0) \ket{H(\vec{p}_H)}}{2E_{n, \vec{p}_H - \vec{p}_\gamma} (E_\gamma + E_{n,\vec{p}_H - \vec{p}_\gamma} - E_{H, \vec{p}_H})},
    \end{split}
    \\
    \begin{split}\label{eq:Imunu_spectral_infinitetH_greaterthan}
        I^>_{\mu \nu}(t_H, T) &\to \frac{\bra{H(\vec{p}_H)} \phi^\dagger_H(0) \ket{0} e^{E_{H, \vec{p}_H} t_H}}{2E_{H, \vec{p}_H}} \\ &\times
\sum_{m} \big[ e^{(E_\gamma - E_{m, \vec{p}_\gamma})T} - 1 \big] \frac{\bra{0} J^\text{em}_\mu(0) \ket{m(\vec{p}_\gamma)} \bra{m(\vec{p}_\gamma)}  J^\text{weak}_\nu(0)  \ket{H(\vec{p}_H)}}{2E_{m, \vec{p}_\gamma}(E_\gamma - E_{m,\vec{p}_\gamma})} .
    \end{split}
\end{align}
Each term in the sum over intermediate states in Eq.~(\ref{eq:Imunu_spectral_infinitetH_lessthan}) differs from the desired Minkowski-space result (\ref{eq:hadronic_tensor_spectral_lessthan}) by a factor of $\Big[1 - e^{-(E_\gamma - E_{H, \vec{p}_H} + E_{n, \vec{p}_H-\vec{p}_\gamma})T}\Big]$, and each term in the sum over intermediate states in Eq.~(\ref{eq:Imunu_spectral_infinitetH_greaterthan}) differs from the desired Minkowski-space result (\ref{eq:hadronic_tensor_spectral_greaterthan}) by a factor of $\big[ e^{(E_\gamma - E_{m, \vec{p}_\gamma})T} - 1 \big]$. We now argue that these factors become equal to 1 (i.e., the exponentials vanish) for large $T$.

Starting with the $t_{\text{em}}<0$ time ordering, we notice that, because the electromagnetic current operator cannot change the flavor quantum numbers of a state, the lowest-energy state appearing in the sum over $n$ is the pseudoscalar meson $H$. The unwanted exponential will vanish if $|\vec{p}_\gamma| + \sqrt{m_H^2+(\vec{p}_H-\vec{p}_\gamma)^2} > \sqrt{m_H^2 + \vec{p}_H^2}$, which is always true for $|\vec{p}_\gamma|>0$. Looking now at the $t_{\text{em}}>0$ time ordering, because the states in the sum over $m$ have mass, $|\vec{p}_\gamma|-\sqrt{m_m^2+\vec{p}_\gamma^2} < 0$ is also satisfied for $|\vec{p}_\gamma|>0$. The hadronic tensor can therefore be extracted by
\begin{equation}
	T_{\mu \nu} = - \lim_{T \to \infty} \lim_{t_H \to -\infty} \frac{2 E_{H, \vec{p}_H} e^{-E_{H, \vec{p}_H} t_H}}{\bra{H(\vec{p}_H)} \phi^\dagger_H(0) \ket{0}} I_{\mu \nu}(t_H, T),
	\label{eq:Tmunu_from_correlation}
\end{equation}
where $I_{\mu \nu}(t_H, T) = I^<_{\mu \nu}(t_H, T)+I^>_{\mu \nu}(t_H, T)$. We denote linear combinations of $I^<_{\mu \nu}(t_H, T)$ and $I^>_{\mu \nu}(t_H, T)$ that are used to extract the form factor $F=F_V, F_A, F_{A,SD}, f_H$ as $F^<(t_H, T)$ and $F^>(t_H, T)$, respectively, such that $F(t_H, T) = F^<(t_H, T) + F^>(t_H, T)$. For example, in the rest frame of the meson with photon momentum $\vec{p}_\gamma=(0,0,p_{\gamma,z})$, for the $t_\text{em}<0$ time ordering we have $F_V^<(t_H, T) = (I_{21}^<(t_H, T)-I_{12}^<(t_H, T))/(2 p_{\gamma, z})$.

\begin{figure}
    \centering
    \includegraphics[width=0.5\textwidth]{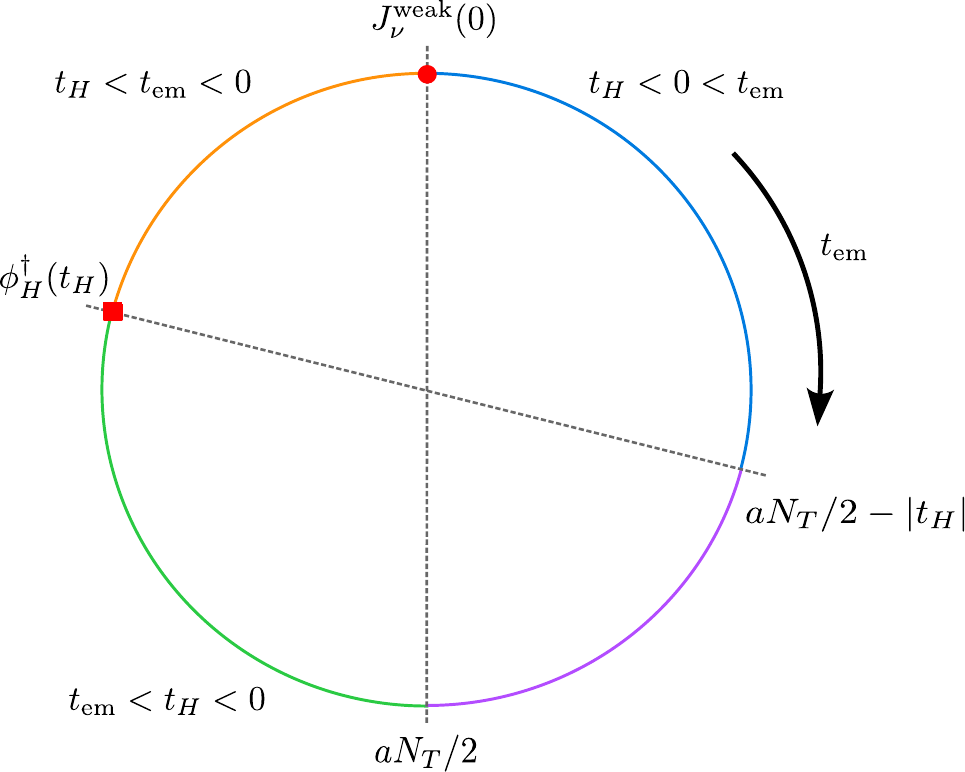}
    \caption{Schematic visualization of the different time orderings for the Euclidean-time three-point function in Eq.~\eqref{eq:three_point}. The $t_\text{em}$ coordinate describes the location of the electromagnetic current. The coordinate $t_\text{em}$ increases in the clockwise direction, and forms a circle due to periodic boundary conditions. The weak current is at time $t_\text{em}=0$ and the interpolating field is at time $t_\text{em}=t_H$. The orange and blue segments of the circle correspond to the time orderings $t_\text{em}<0$ and $t_\text{em}>0$, respectively. The green segment corresponds to the unphysical situation where the electromagnetic current is at an earlier time than the interpolating field. The purple segment is also unphysical. For the $t_\text{em}<0$ time ordering, one must use values of the integration range such that $T < |t_H|$. For the $t_\text{em}>0$ time ordering on the other hand, one must use values of the integration range such that $T < aN_T/2-|t_H|$, where $N_T$ is the number of lattice sites in the Euclidean time direction.}
    \label{fig:time_orders}
\end{figure}

Before proceeding, it is worth noting that, on a periodic lattice, one must be careful however when taking the $T \to \infty$ limits. Figure~\ref{fig:time_orders} depicts the different time orderings for the three-point correlation function in Eq.~\eqref{eq:three_point} on a periodic lattice. For the $t_{\text{em}}>0$ time ordering, the largest possible value of $T$ is $aN_T/2+t_H$, where $N_T$ is the number of lattice sites in the Euclidean time direction. Integrating past this time will incur systematic errors from wrap-around effects. For the $t_{\text{em}}<0$ time ordering, the largest possible value of $T$ is $-t_H$. Additionally, as one integrates closer to the interpolating field, excited state effects become larger. We will discuss these effect further in Sec.~\ref{section:fit_method}.

\section{Sequential propagators} 
\label{section:sequential_propagators}

\begin{figure}[h]
	\centering
	\begin{minipage}{0.45\textwidth}
	\centering
		\includegraphics[width=0.8\textwidth]{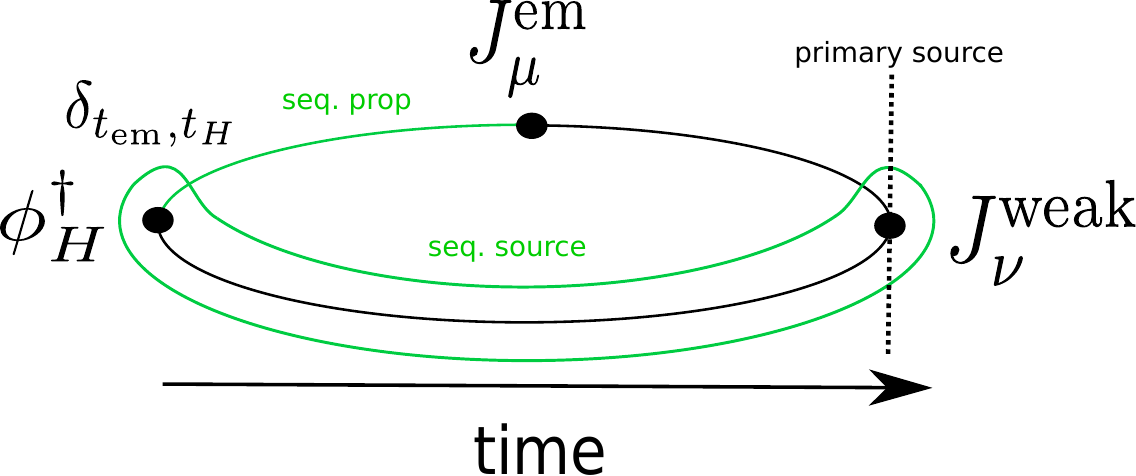}	
	\end{minipage}
	\begin{minipage}{0.45\textwidth}
	\centering
		\includegraphics[width=0.8\textwidth]{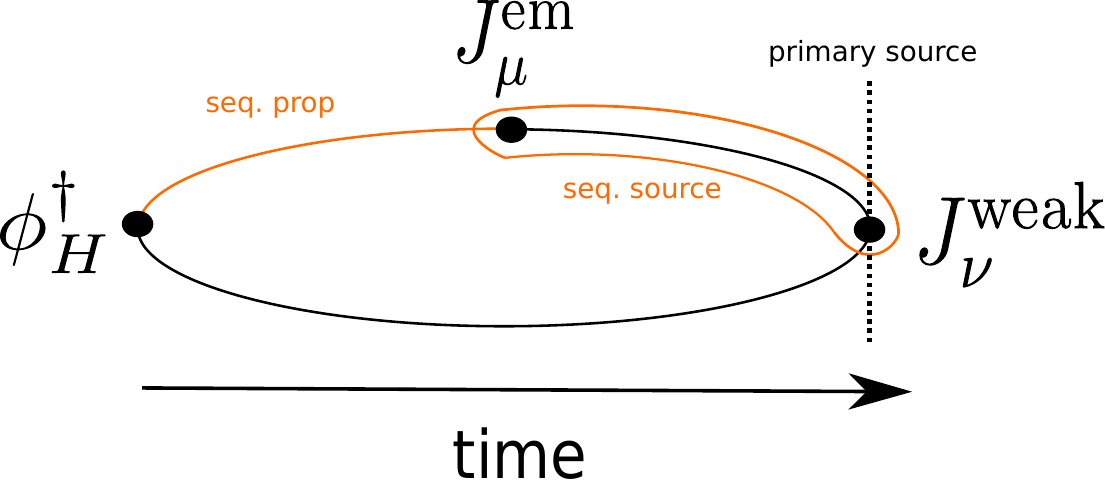}	
	\end{minipage}
 
	\caption{The left (right) figure is a schematic drawing of the 3d (4d) methods. For both methods, the initial noise source is located at the weak-current time. The sequential propagator for the 3d (4d) method is shown in green (orange) and the sequential source is circled in green (orange).}
	\label{fig:seq_prop_quark_lines}	
\end{figure}

In this section, we describe two different methods of calculating the time-integrated correlation function $I_{\mu \nu}(t_H, T)$ on the lattice, which are illustrated in Fig.~\ref{fig:seq_prop_quark_lines}. One method, which we refer to as the 3d method, uses a three-dimensional (timeslice source) sequential propagator through the interpolating field $\phi_H^\dagger$. In this way, for a fixed value of the source-sink separation $t_H$, one calculates the three-point function in Eq.~\eqref{eq:three_point} for all values of $t_{\text{em}}$. The second method, which we refer to as the 4d method, uses a four-dimensional sequential propagator through the electromagnetic current. The four-dimensional sequential source is non-zero on the range $-T \leq t_\text{em} \leq T$, where $T$ is the desired integration range, and must be multiplied by the factor $e^{E_\gamma t_\text{em}}$ (the details of the four-dimensional sequential source depend on the specific method used to calculate the time integrals $I^>_{\mu \nu}(t_H, T)$ and $I^<_{\mu \nu}(t_H, T)$). Using the 4d method, for a fixed value of the integration range $T$, one calculates the time-integrated correlation function $I_{\mu \nu}(t_H, T)$, directly on the lattice, for all values of the source-sink separation $t_H$. From this, we see that the 3d method is better suited to control unwanted exponentials from finite integration range $T$, while the 4d method is better suited to control unwanted exponentials from excited states created by the interpolating field $\phi_H^\dagger$. The results in Ref.~\cite{Desiderio:2020oej} were calculated using the 4d method, integrating over the full time extent of the lattice, \textit{i.e.} $T=N_T/2$.

One limitation of the 4d method is that, because the integral over $t_{\text{em}}$ is performed directly on the lattice, the two different time orderings of $I_{\mu \nu}(t_H, T)$ cannot be resolved. Because the intermediate states of the two time orderings are not the same in general, at finite $T$, one must use a fit form with multiple exponentials to remove the unwanted exponentials that come with the intermediate states. It is possible, however, to modify the 4d method such that one calculates the two time orderings separately. To do so, one performs two sequential solves through the electromagnetic current, but limits the extent of the sources in the time direction to only be non-zero for the desired time ordering. We will refer to this method as the 4d$^{>,<}$ method.

In this work, in order to control systematic errors from the unwanted exponentials, we perform the calculation for multiple values of $t_H$ when using the 3d method, and multiple values of $T$ when using the 4d or 4d$^{>,<}$ methods. To properly compare the methods, it is important to consider the number of propagator solves required for each. Table~\ref{tab:num_solves} shows the number of propagator solves required in terms of the number of meson momenta $N_{p_H}$, photon momenta $N_{p_\gamma}$, source-sink separations $N_{t_H}$ (for the 3d method), and integration ranges $N_T$ (for the 4d and 4d$^{>,<}$ methods). Note that these numbers are for a single source on a single configuration. The factor of 2 in front of every entry accounts for the two components of the electromagnetic current. Using point sources allows one to get all values of $p_\gamma$ for free if using the 3d method, and all values of $p_H$ for free if one uses the 4d or 4d$^{>,<}$ methods. In the 4d method, one must perform a sequential solve for each $\gamma_\mu$ matrix, which is the source of the factor of 4 in front of $N_{p_\gamma}$. The same is true for the 4d$^{>,<}$ method, except one solve must now be done for each time ordering, resulting in the factor of 8. The 3d method on the other hand only requires a single sequential solve for a given $p_{H}$.

\begin{table}[]
    \centering
    \begin{tabular}{ccccccc}
        \hline \hline
        Source & & 3d & & 4d & & 4d$^{>,<}$ \\
        \hline
        point & & $2(1+N_{t_H}N_{p_H})$ & & $2(1+4N_T N_{p_\gamma})$ & & $2(1+8N_T N_{p_\gamma})$ \\ 
        $\mathbb{Z}_2$ wall & & $2(1+N_{t_H} N_{p_H}+N_{p_H}N_{p_\gamma})$ & & $2(1+4 N_T N_{p_\gamma}+N_{p_\gamma} N_{p_H})$ & & $2(1+8 N_T N_{p_\gamma}+N_{p_\gamma} N_{p_H})$ \\
        \hline\hline
    \end{tabular}
    \caption{Number of propagator solves required for a single configuration for a single source in terms of the desired number of meson momenta $N_{p_H}$, photon momenta $N_{p_\gamma}$, number of source-sink separations $N_{t_H}$ for the 3d method, and number of integration ranges $N_T$ for the 4d methods. Results for the 3d, 4d, and 4d$^{>,<}$ methods are shown.}
    \label{tab:num_solves}
\end{table}

\section{Lattice parameters}
\label{sec:lattice_params}
In this section, we describe the properties of the lattice we perform calculations on as well as the details of our numerical setup. As previously described in the introduction, we have performed two sets of calculations. We start with the common parameters between them, and then discuss the differences.

Both calculations were performed on a single RBC/UKQCD ensemble \cite{RBC:2010qam} (one of the ``24I'' ensembles) which was generated using the Iwasaki gauge action and 2+1 flavors of domain-wall fermions using $N_5=16$ lattice sites in the fifth dimension. The sea-quark masses and gauge coupling are $am_{u,d}=0.005, am_s^\text{sea}=0.04, \beta=2.13,$ and the ensemble has an inverse lattice spacing of $a^{-1}=1.785(5) \text{ GeV}$ \cite{RBC:2014ntl}. For the valence strange quarks, we use the same domain-wall action as used for the sea quarks \cite{RBC:2010qam}, except that we use the physical mass $am_s^\text{val}=0.0323$. The charm valence quark is implemented using a M\"{o}bius domain-wall action with $L_5/a=12$, $aM_5=1.0$, $am_f=0.6$, and stout-smeared gauge links using three iterations with $\rho=0.1$ \cite{Boyle:2018knm}. The charm-quark mass obtained from these parameters is close to physical. All calculations use all-mode averaging \cite{Shintani:2014vja}. We currently neglect the disconnected diagrams that correspond to self-contracting the quark and antiquark in the electromagnetic current. These contributions are expected to be small due to combined $1/N_c$ and flavor-$SU(3)$ suppression (the sum of the up, down, and strange disconnected contributions would vanish for equal quark masses because the electric charges sum to zero).

We use local currents in our calculation. The matching factors of the individual quark components of the electromagnetic current were computed nonperturbatively using charge conservation. We employ ``mostly non-perturbative'' renormalization of the weak axial-vector and vector currents \cite{Hashimoto:1999yp,El-Khadra:2001wco} and use the tree-level values for the residual matching factors. For the strange-quark nonperturbative matching factor, we use the value calculated by the RBC-UKQCD collaborations \cite{RBC:2010qam} of $Z_V^{(ss)}=0.71651(46)$. We calculated the charm-quark matching factor to be $Z_V^{(cc)}=1.0205(57)$. Notice that the errors of both $Z_V^{(ss)}$ and $Z_V^{(cc)}$ are at the sub-percent level and therefore have a negligible effect on the final values of the form factors presented in this work.

The results in Secs.~\ref{sec:noise_vs_point}, \ref{section:fit_method}, and \ref{sec:compare_3d_4d} were calculated using either $\mathbb{Z}_2$ random-wall sources or point sources on $N_{\rm cfg}=25$ configurations, both using one exact and 16 sloppy samples per configuration. We use gauge-covariant Gaussian smearing for the strange quark field using a width of $\sigma=4.35$ and $n_S=30$ smearing iterations. For the strange quark, we combined conjugate gradient (CG) with low-mode deflation where we calculated the lowest 400 eigenvectors of the domain-wall-fermion operator. The strange-quark sloppy solves were performed using 110 CG iterations. For the charm quark, we always performed exact solves and did not implement low-mode deflation. For all 3d-method data in these sections, we performed calculations using three values of the source-sink separation $-t_H/a\in \{6,9,12\}$. For all 4d-method and 4d$^{>,<}$-method data in these sections, we performed calculations using three values of the integration range $T/a\in \{6,9,12\}$. Further details of the calculations performed in these sections are shown in Table \ref{tab:parameter_details}.

\begin{table}[h]
\centering
\begin{tabular}{ccccccc}
	\hline
	\hline
	Method && Source && Meson Momentum && Photon Momentum \\
	\hline
	3d && $\mathbb{Z}_2$-wall && $\vec{p}_{D_s}=(0,0,0)$ && $|\vec{p}_\gamma|^2\in(2\pi/L)^2\, \{1,2,3,4\}$ 
	\\
	3d && point && $p_{D_s,z}\in 2\pi/L\, \{0,1,2\}$ && all
	\\
	4d && $\mathbb{Z}_2$-wall && $p_{D_s,z}\in2\pi/L\,\{-1,0,1,2\}$ && $p_{\gamma,z}=2\pi/L$
	\\
	4d$^{>,<}$ && $\mathbb{Z}_2$-wall && $p_{D_s,z}\in 2\pi/L\, \{-1,0,1,2\}$ && $p_{\gamma,z}=2\pi/L$
	\\
    \hline\hline
\end{tabular}
\caption{The methods, sources, and momenta for which we performed calculations in Secs.~\ref{sec:noise_vs_point}, \ref{section:fit_method}, and \ref{sec:compare_3d_4d}. When only the $z$-component of the momentum is listed, the other momentum components are zero. For 3d point sources, ``all'' indicates these momenta can be calculated for free for a given value of $\vec{p}_{D_s}$. We did not perform calculations using point sources for the 4d or 4d$^{>,<}$ methods.}
\label{tab:parameter_details}
\end{table}
\FloatBarrier

The calculations in Secs. \ref{sec:improved_C3}, \ref{sec:improved_FF}, and \ref{sec:final_procedure} were performed using only the 3d method for two values of source-sink separation $-t_H/a \in \{9,12\}$. We use a combination of point sources and $\mathbb{Z}_2$ random-wall sources and perform calculations on $N_{\rm cfg}=25$ configurations with four and two exact samples per configuration, respectively. Sixty-four sloppy samples per configuration were used for both noise and point sources. As will be described in Sec.~\ref{ssec:infinite_volume_approximation}, using point sources, for a given value of $\vec{p}_H$, we are able to extract all values of $\vec{p}_\gamma$, even non-integer multiples of $2\pi/L$. We performed calculations in the meson rest frame for photon momenta in the $\hat{z}$ direction $p_{\gamma,z} \in 2\pi/L\{0.1, 0.2, 0.4, 0.6, 0.8, 1.0, 1.4, 1.8, 2.2, 2.4\}$. Using $\mathbb{Z}_2$ random-wall sources we performed calculations in the rest frame of the meson for two values of photon momenta $p_{\gamma,z}\in 2\pi/L\{0, 1\}$. As explained in Sec.~\ref{ssec:ratio_method}, the $\mathbb{Z}_2$ random-wall source data is used to reduce statistical noise of the point source data.

Another set of questions are the particular details of how the time integrals $I_{\mu \nu}^>(t_H, T)$ and $I_{\mu \nu}^<(t_H, T)$ are calculated. In particular, how the $t_{\text{em}}=0$ contribution is distributed between the two time orderings, and how the time integrals are approximated. For the 3d method, these details can be decided during the analysis stage. For the 4d$^{>,<}$ method however, these details must be decided while calculating the propagators. Note that, because the 4d method does not resolve the two time orderings, the question of how to distribute the $t_\text{em}=0$ contribution is irrelevant. The results shown in Secs.~\ref{sec:noise_vs_point}, \ref{section:fit_method}, and \ref{sec:compare_3d_4d} assign the entire $t_\text{em}=0$ contribution to $I_{\mu \nu}^>(t_H, T)$, approximate $I^>_{\mu \nu}(t_H, T)$ by summing from $t_\text{em}=0$ to $t_\text{em}=T$ with equal weights, and approximate $I^<_{\mu \nu}(t_H, T)$ by summing from $t_\text{em}=-a$ to $t_\text{em}=-T$ with equal weights. On the other hand, the results shown in Secs. \ref{sec:improved_C3}, \ref{sec:improved_FF}, and \ref{sec:final_procedure} assign half of the $t_\text{em}=0$ contribution to each time ordering, and approximate $I^>_{\mu \nu}(t_H, T)$ and $I^<_{\mu \nu}(t_H, T)$ using the trapezoid rule. These differences lead to discrepancies between some of the results shown in the different sections. In particular, discrepancies could appear for intermediate form factor data as a function of $T$, and as well as form factor results for individual time orderings. We found that changing how the time integral is approximated had no statistically significant effect on the final values of the form factors.

\section{Comparing statistical precision of noise and point sources}
\label{sec:noise_vs_point}
In this section, we compare the statistical precision of the vector form factor calculated using both noise and point sources\footnote{Note that we did not perform the necessary calculations to extract $F_{A,SD}$ using the improved method presented in Sec.~\ref{ssec:sub_pgamma_zero_FA_SD} and therefore do not consider it here.}. For both noise and point sources, calculations were done on the same $N_{\text{cfg}}=25$ configurations, both using one exact and 16 sloppy solves per configuration. Before proceeding, we point out that for $N_\text{cfg}=25$, the error on the error is $\sim 15\%$. Additionally, while we did not perform calculations using point sources for the 4d and 4d$^{>,<}$ methods, we expect that data to exhibit the same general behavior as we observe for the 3d method.

\begin{figure}[h]
	\begin{minipage}{0.49\textwidth}
        \includegraphics[width=\textwidth]{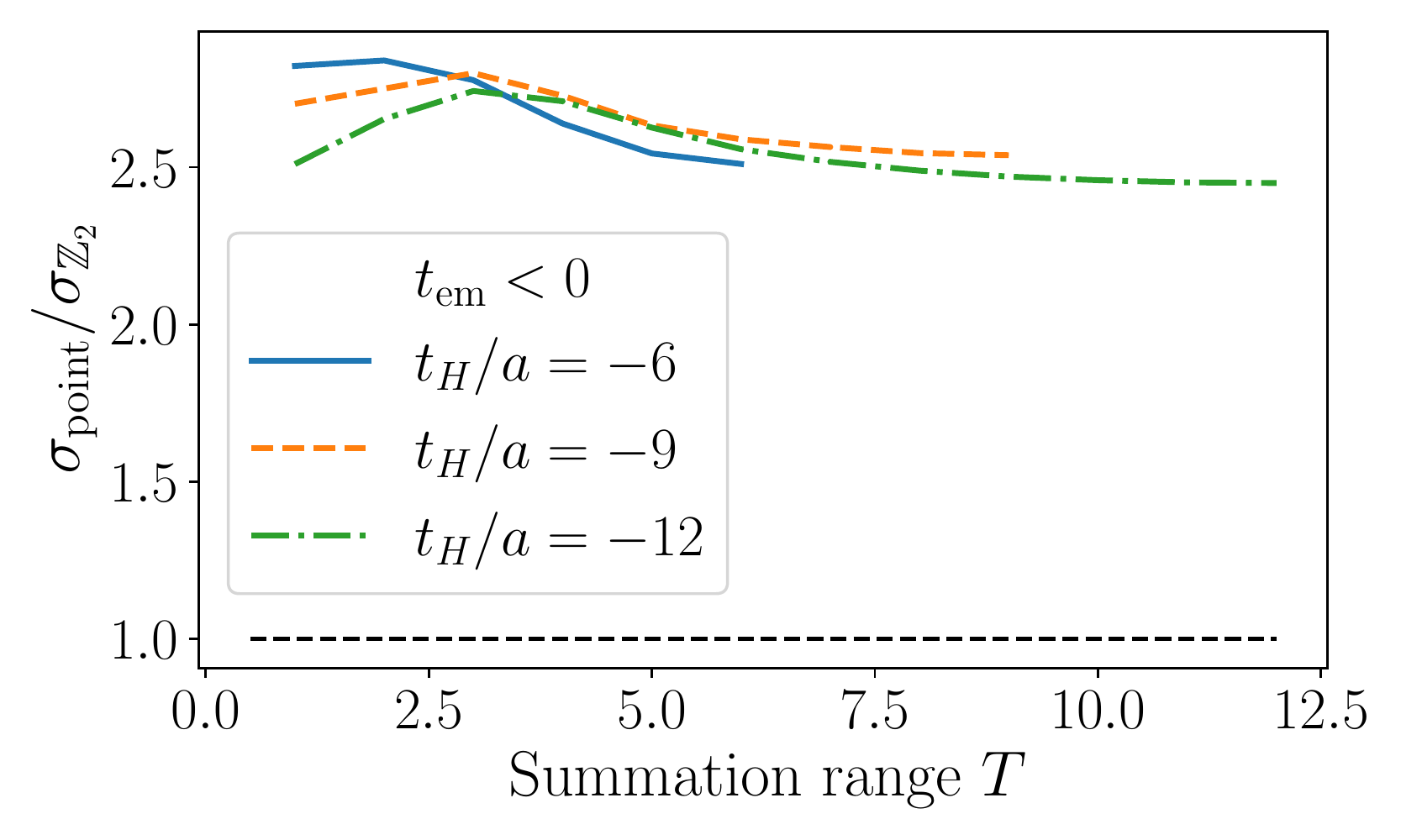}
    \end{minipage}
    \begin{minipage}{0.49\textwidth}
        \includegraphics[width=\textwidth]{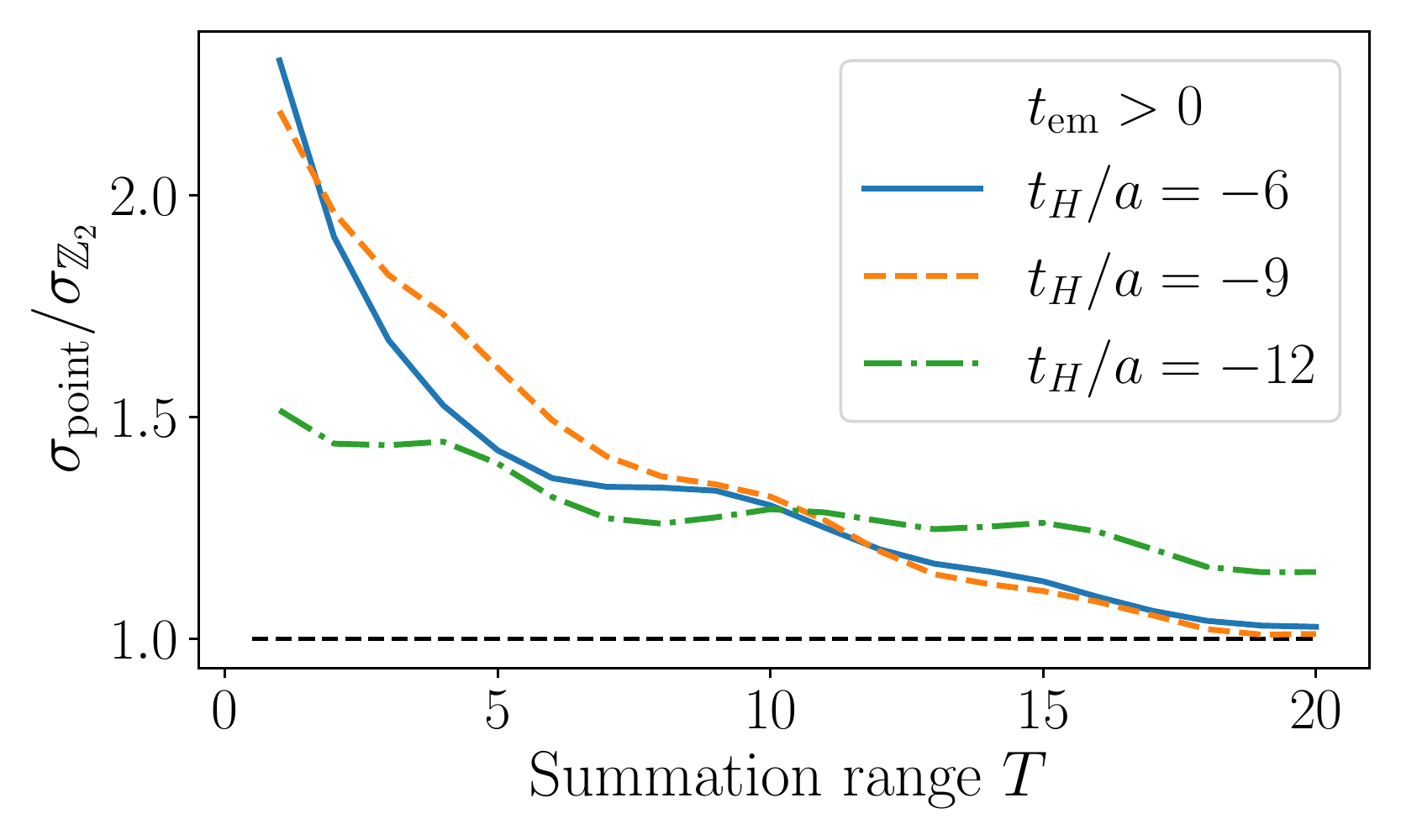}
    \end{minipage}
    \caption{Ratio of statistical uncertainties of point to noise sources as a function of summation range. The left and right plots show $F_V^<(T,t_H)$ and $F_V^>(T, t_H)$, respectively. Lines with different colors and line-styles indicate different source-sink separations. For both noise and point sources, calculations were performed on the same $N_{\rm cfg}=25$ configurations, both using one exact and 16 sloppy samples per configuration.}
    \label{fig:compare_noise_to_point}
\end{figure}

Figure~\ref{fig:compare_noise_to_point} compares the statistical uncertainty of $F_V^<(T,t_H)$ and $F_V^>(T,t_H)$ calculated using the 3d method for both point and noise sources. Specifically, Fig.~\ref{fig:compare_noise_to_point} shows the ratio of the statistical uncertainty from using point sources to using noise sources, as a function of summation range $T$. For the $t_\text{em}<0$ time ordering, the ratio approaches a constant value of $\sim 2.5$ as $T$ approaches $-t_H$. The ratio for the $t_\text{em}>0$ time ordering decreases as the summation range is increased. The data shown in Fig.~\ref{fig:compare_noise_to_point} was calculated in the rest frame of the meson with $\vec{p}_\gamma=2\pi/L(0,1,1)$; we observed that these general trends also hold for other values of $\vec{p}_\gamma$ given in Table~\ref{tab:parameter_details}. 

The differences in behavior of the two time orderings can be understood by considering the maximum Euclidean time separation between any of the three operators in the correlation function. The maximum time separation between any two operators in the $t_\text{em}<0$ time ordering is equal to a constant given by the source-sink separation $t_H$. For the $t_\text{em}>0$ time ordering on the other hand, the maximum separation is given by $-t_H+T$, which grows with summation range. The relative statistical uncertainty is generally observed to increase more quickly for noise sources than point sources as the maximum separation increases, leading to the behavior observed in Fig.~\ref{fig:compare_noise_to_point}.

To determine which source offers the best precision to computational cost ratio, we need to refer back to Table~\ref{tab:num_solves} and compare the number of solves required for the 3d method for both noise and point sources. Using point sources, for a given meson momentum one calculates all values of photon momentum for free. The reduction in the number of required solves per configuration can be used to perform the calculation on more configurations. Therefore, if the square root of the ratio of the number of solves for point to noise sources is larger than the ratio of their statistical uncertainties, point sources will be more cost effective. The number of photon momenta that should be used in this comparison is the number of momenta that provide physically allowed values of $x_\gamma$. In the rest frame of the $D_s$ meson with $L=24$ and $a^{-1} = 1.785(5)$ GeV, there are four values of $p_\gamma$ that are kinematically allowed. Plugging in $N_{p_{D_s}}=1$, $N_{p_\gamma}=4$, and $N_{t_H}=3$, one finds that noise sources require twice as many solves as point sources. Therefore, looking at Fig.~\ref{fig:compare_noise_to_point}, we observe that noise sources are more cost effective for the $t_\text{em}<0$ time ordering. For the $t_{\text{em}}>0$ time ordering, we find that noise sources are more cost effective for smaller values of $T$, and point sources become more cost effective for larger values of $T$. 

One additional factor to consider is that noise sources benefit from volume averaging, while point sources do not. Because our numerical test was performed on a relatively small lattice with $N_L=24$ spatial lattice sites, noise sources are expected to improve relative to point sources by a larger margin for lattices with more spatial sites.

\FloatBarrier

\section{Fit Method}
\label{section:fit_method}
In this section we describe the fit methods used to remove unwanted exponentials from the form factor results presented here and in Sec.~\ref{sec:compare_3d_4d}. Before proceeding, it will be useful to introduce the notation $I_{\mu \nu}(t_H,T) = I^A_{\mu \nu}(t_H, T) + I^V_{\mu \nu}(t_H, T)$, where $I^V_{\mu \nu}(t_H,T)$ and $I^A_{\mu \nu}(t_H, T)$ are the weak vector and axial-vector current components of $I_{\mu \nu}(t_H,T)$, respectively.

We start by studying, in continuum QCD, the quantum numbers of the states that have a non-zero contribution to the sum over states in the spectral decompositions of $I_{\mu \nu}^<(t_H, T)$ and $I_{\mu \nu}^>(t_H, T)$\footnote{Note that because we neglect disconnected diagrams in this present work, certain states will not contribute to the spectral decomposition of $I_{\mu \nu}^<(t_H, T)$ and $I_{\mu \nu}^>(t_H, T)$ that would otherwise. This is expected to have a more significant effect for the $t_\text{em}>0$ time ordering, where, for example, a $\pi \pi$-like state will not contribute as a result of neglecting the disconnected diagrams for the $D_s$ decay.}. For the $t_{\text{em}}<0$ time ordering, the states $\ket{n(\vec{p}_H-\vec{p}_\gamma)}$ must have the same quark-flavor quantum numbers as the initial pseudoscalar meson $H$. Additionally, parity constrains the $J^P$ quantum numbers of the states that contribute, which are in general different for the weak axial-vector and weak vector current components. For $I_{\mu \nu}^{<,A}(t_H, T)$ one finds that the allowed values are $J^P \in \{0^-,1^+,2^\pm, \dots\}$, which implies that the lowest-energy state that contributes to $I_{\mu \nu}^{<,A}(t_H, T)$ is the pseudoscalar meson itself. Moving on to $I_{\mu \nu}^{<,V}(t_H, T)$, one finds that the allowed values are $J^P \in \{0^+,1^-,2^\pm, \dots\}$, which implies that the lowest-energy state that contributes to $I_{\mu \nu}^{<,V}(t_H, T)$ is the vector meson ($H^*$) associated with the initial-state pseudoscalar meson, \textit{e.g.} for $H=D_s$ it would be a ($D_s^*$)-like state. We calculate the energies of the $D_s$ and $D_s^*$ states for all necessary combinations of $\vec{p}_{D_s}-\vec{p}_\gamma$ by performing single-exponential fits of the associated two-point functions projected to definite momenta. The results of these fits are then used as Gaussian priors in the form-factor fits, where the prior value and prior width are set as the central value and uncertainty of the fit results, respectively. For the $t_{\text{em}}>0$ time ordering, the states $\ket{m(\vec{p}_\gamma)}$ are flavorless and we leave their energies as fit parameters. Parity constrains the quantum numbers of the states that contribute to the sum over states in $I_{\mu \nu}^{>,A}(t_H,T)$ and $I_{\mu \nu}^{>,V}(t_H,T)$ to be $J^P \in \{0^+, 1^-, 2^\pm, \dots\}$ and $J^P \in \{1^-, 2^\pm, \dots\}$, respectively. 

From this discussion we also learn that in general, for a particular momentum and a given time ordering, the same states contribute to all $\mu, \nu$ components of $I^A_{\mu \nu}(t_H, T)$, and similarly for $I^V_{\mu \nu}(t_H, T)$. Therefore, while the matrix elements multiplying the unwanted exponentials will in general be different for different $\mu,\nu$ indices, the energies appearing in the exponents will be the same. Because only $I^V_{\mu \nu}(t_H, T)$ contributes to $F_V(t_H, T)$ and only $I^A_{\mu \nu}(t_H, T)$ contributes to $F_A(t_H, T), F_{A,SD}(t_H, T)$, and $f_H(t_H, T)$, one can fit the form factors directly without mixing unwanted exponentials. Fitting the form factors directly offers two advantages over fitting $I_{\mu \nu}(t_H, T)$. First, fitting the form factors requires fewer fit parameters, which helps stabilize the fits. Second, consider the scenario where taking linear combinations of $I_{\mu \nu}(t_H, T)$ results in cancellations that reveal features in the form factors that $I_{\mu \nu}(t_H, T)$ is not sensitive to. If one fits $I_{\mu \nu}(t_H, T)$ first, these features could be missed by the fit, and would then propagate as a source of systematic uncertainty in the form factors. This possibility is eliminated by fitting the form factors directly.

To help constrain the energy gap $\Delta E$ between the ground state and the first excited state created by the interpolating field, we first perform two-exponential fits to the pseudoscalar two-point function. The result of the fit for $\Delta E$ is then used as a Gaussian prior in the form factor fits, with the prior width equal to the statistical uncertainty scaled by a factor 1.5. We extract $F_V(t_H, T), F_A(t_H, T)$ and $f_H(t_H, T)$ from the time-integrated correlation function. We then calculate the structure-dependent axial form factor by $F_{A,SD}(t_H, t)=F_A(t_H,T)-(-Q_l f_H(t_H,T)/E_\gamma^{(0)})$. We perform simultaneous fits to the form factors $F_V(t_H, T), F_A(t_H, T), f_H(t_H, T)$ and $F_{A,SD}(t_H, t)$ for all kinematic points. This takes advantage of the fact that data on a given ensemble will have common energies, including the excited-state energy gap, as well as energies that appear in unwanted exponentials from intermediate states.

We fit our data as a function of both source-sink separation $t_H$ and integration range $T$. Because each successive value of $T$ is a sum of the previous values of $T$, the data for a given value of $t_H$ is highly correlated. These large correlations manifest as small eigenvalues in the correlation matrix, which makes correlated fits to this data unstable. We instead perform uncorrelated fits and use jackknife to estimate uncertainties of the fit parameters. To replace the $\chi^2$ as a goodness-of-fit, we check that the fit result of an individual form factor at a given momentum is stable under variations of the fit range. For the 3d method, we perform simultaneous fits to all values of $t_H$ while searching for stability in $T$. For the 4d and 4d$^{>,<}$ methods, we perform simultaneous fits to all values of $T$ searching for stability in $t_H$. The global fits are then performed using these chosen stable fit ranges.
\begin{figure}
    \centering
    \includegraphics[width=\textwidth]{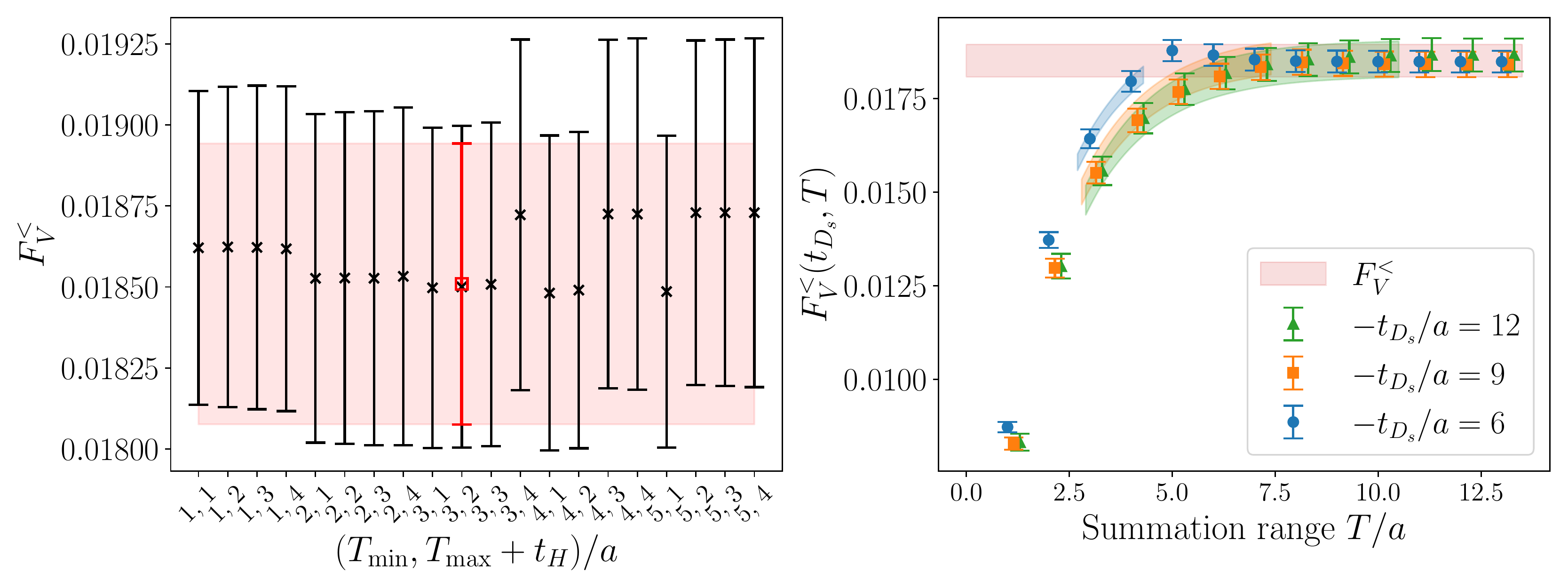}
    \caption{$F_V^<$ data calculated using the 3d method. The left plot shows $F_V^<$ resulting from a fit for different fit ranges $(T_\text{min}, T_\text{max}+t_{D_s})/a$. The red square is the chosen stable fit range and is the result of the global fit to all 3d method data. Fit ranges where $-t_{D_s}/a=6$ has no data points indicates that data set was left out of the fit. The right plot shows $F_V^<(t_{D_s}, T)$ calculated using the 3d method as a function of $T$. The three differently colored, shaped sets of data points correspond to different values of $t_{D_s}$. The red horizontal band is the one sigma extrapolated value of $F_V^<$, and corresponds to the red band in the left plot. The blue, orange, and green bands are the one sigma global fit results for $-t_{D_s}/a=6, 9, 12$, respectively. The error bands are only shown for data included in the fit. The data was calculated with $\vec{p}_{D_s}=0$ and $\vec{p}_\gamma=2\pi/L(1,1,1)$.}
    \label{fig:stability_old_3d}
\end{figure}

The fit form used for the 3d method data includes one exponential to account for the unwanted exponential from the lowest-energy excited state created by the interpolating field, and one exponential for the unwanted exponential that comes with the lowest-energy intermediate state. The fit forms used for the $t_{\text{em}}<0$ and $t_{\text{em}}>0$ time orderings of the 3d method data for a form factor $F=F_V, F_{A,SD}, F_A, f_H$ are given by
\begin{align}
    F^{<}(t_H, T) &= F^< + B_{F}^{<} e^{-(E_\gamma - E_{H} + E^<)T} + C^{<}_{F} e^{\Delta E t_H},
    \label{eq:3d_fit_form_neg}
    \\
    F^{>}(t_H, T) &= F^> + B_{F}^{>} e^{(E_\gamma - E^>)T} + C^{>}_{F} e^{\Delta E t_H}.
    \label{eq:3d_fit_form_pos}
\end{align}
The fit form used for the $t_\text{em}>0$ time ordering of the 4d$^{>,<}$ data is the same as the 3d method, and the fit form for the $t_\text{em}>0$ time ordering of the 4d$^{>,<}$ data is the same as the 3d method except with $C^<_F=0$. Recall from Sec.~\ref{ssec:Euclidean} that for $t_\text{em}<0$, excited-state effects become larger as one integrates towards the interpolating field. For this reason, stability tests for the $t_\text{em}<0$ time ordering are done by varying both the minimum fit range as well as the distance from the interpolating field. For $t_{\text{em}}>0$ on the other hand, we only need to check for stability in the minimum fit range.

Because the 4d data is a sum of both time orderings, one possible fit form would be a sum of those in Eqs.~\eqref{eq:3d_fit_form_neg} and \eqref{eq:3d_fit_form_pos}. However, we perform fits to regions of the data that has plateaued in $t_H$ and therefore use the fit form
\begin{align}
    F(T) &= F + B_{F}^{<} e^{-(E_\gamma - E_{H} + E^<)T} + B_{F}^{>} e^{(E_\gamma - E^>)T}.
    \label{eq:4d_fit_form}
\end{align}
To help stabilize the fits to the 4d-method data, we put a Gaussian prior on the fit parameter $E^>$ centered at the $\phi$-meson mass with a width of 200 MeV. 

\begin{figure}
    \centering
    \includegraphics[width=\textwidth]{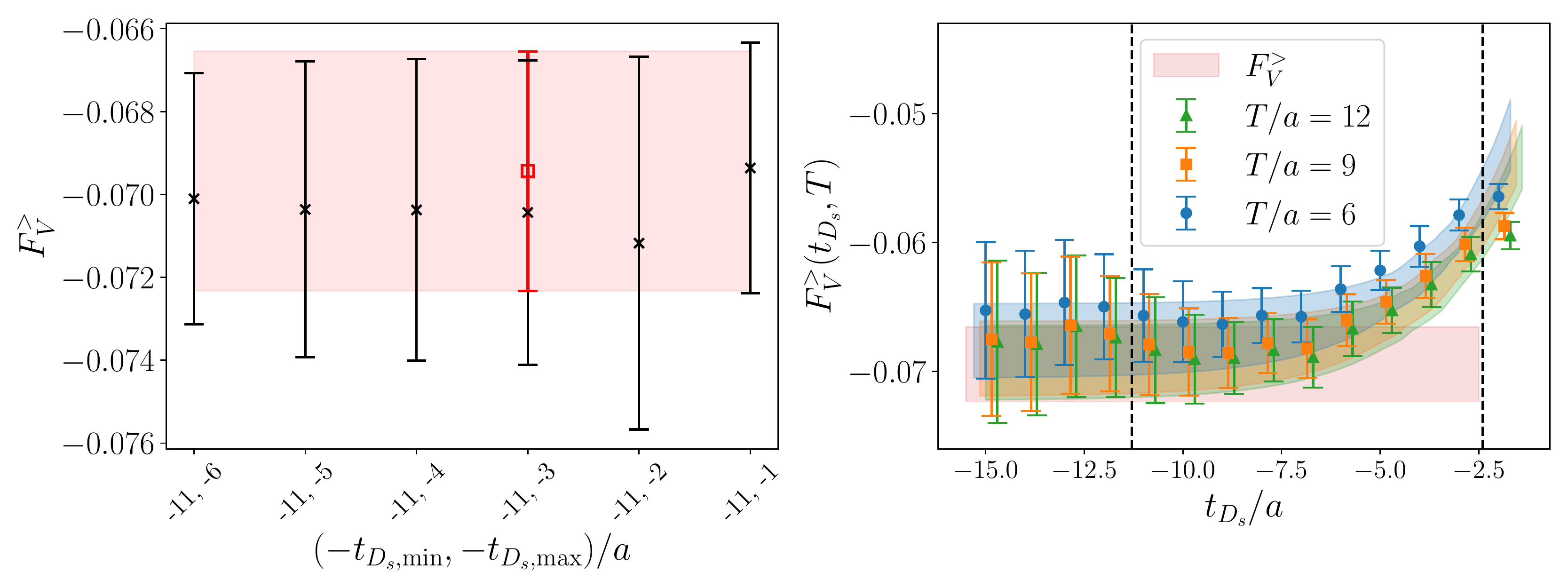}
    \caption{$F_V^>$ data calculated using the 4d$^{>,<}$ method. The left plot shows $F_V^>$ resulting from a fit for different fit ranges $(-t_{D_s, \text{min}}, t_{D_s, \text{max}})/a$ for a fixed choice of $-t_{D_s, \text{min}}/a=11$. The red square is the chosen stable fit range and is the result of the global fit to all 4d$^{>,<}$ method data. The right plot shows $F_V^>(t_{D_s}, T)$ calculated using the 4d$^{>,<}$ method as a function of $t_{D_s}$. The three different colored, shaped data points correspond to different values of $T$. The red horizontal band is the one sigma extrapolated value of $F_V^>$, and corresponds to the red band in the left plot. The blue, orange, and green bands are the one sigma global fit results for $T/a=6, 9, 12$, respectively. The vertical black dashed lines indicate the data included in the fit. The data was calculated with $\vec{p}_{D_s}=2\pi/L(0,0,1)$ and $\vec{p}_\gamma=2\pi/L(0,0,1)$.}
    \label{fig:stability_4d_pos}
\end{figure}

Figure \ref{fig:stability_old_3d} shows example $D_s$ stability-test-fit plots for the 3d method as well as the result of the fit on top of the data. For the 3d method we find that, in general, the global fit does not significantly reduce the statistical errors. Similar plots for the 4d$^{>,<}$ method are shown in Fig.~\ref{fig:stability_4d_pos}. The global fit to this 4d$^{>,<}$ method data improves the statistical precision by a larger factor than for the 3d method data. One possible explanation for this improvement is that all 4d$^{>,<}$ method data was calculated using the same value of $\vec{p}_\gamma$. The fit forms for the different momentum combinations of $F_V^>(t_{D_s}, T)$ included in the global fit therefore all have the fit parameter $E^>$ in common. This can be seen by looking at the spectral decomposition in Eq.~\eqref{eq:Imunu_spectral_greaterthan}, which indicates that the value of the energy $E^>$ for a given component of the weak current depends only on $\vec{p}_\gamma$.

\section{Comparing 3d and 4d methods}
\label{sec:compare_3d_4d}
In this section, we compare the 3d, 4d, and 4d$^{>,<}$ methods. Before proceeding, recall that the 3d method offers better control over taking $T \to \infty$, while the 4d and 4d$^{>,<}$ methods offer better control over the $t_H \to -\infty$ limit. The 3d and 4d/4d$^{>,<}$ methods therefore complement each other with regard to control over the two types of unwanted exponentials appearing in the calculation. To test if this complementarity can be exploited to improve the quality of the fits, we also perform simultaneous fits to the 3d and 4d$^{>,<}$ method. As a metric we will compare the vector form factor as a function of $x_\gamma$. Note that some data at different $x_\gamma$ values are in different little groups of the cubic group, and therefore can have different discretization errors.

\begin{figure}[h]
    \centering
    \begin{minipage}{0.49\textwidth}
        \centering
        \includegraphics[width=\textwidth]{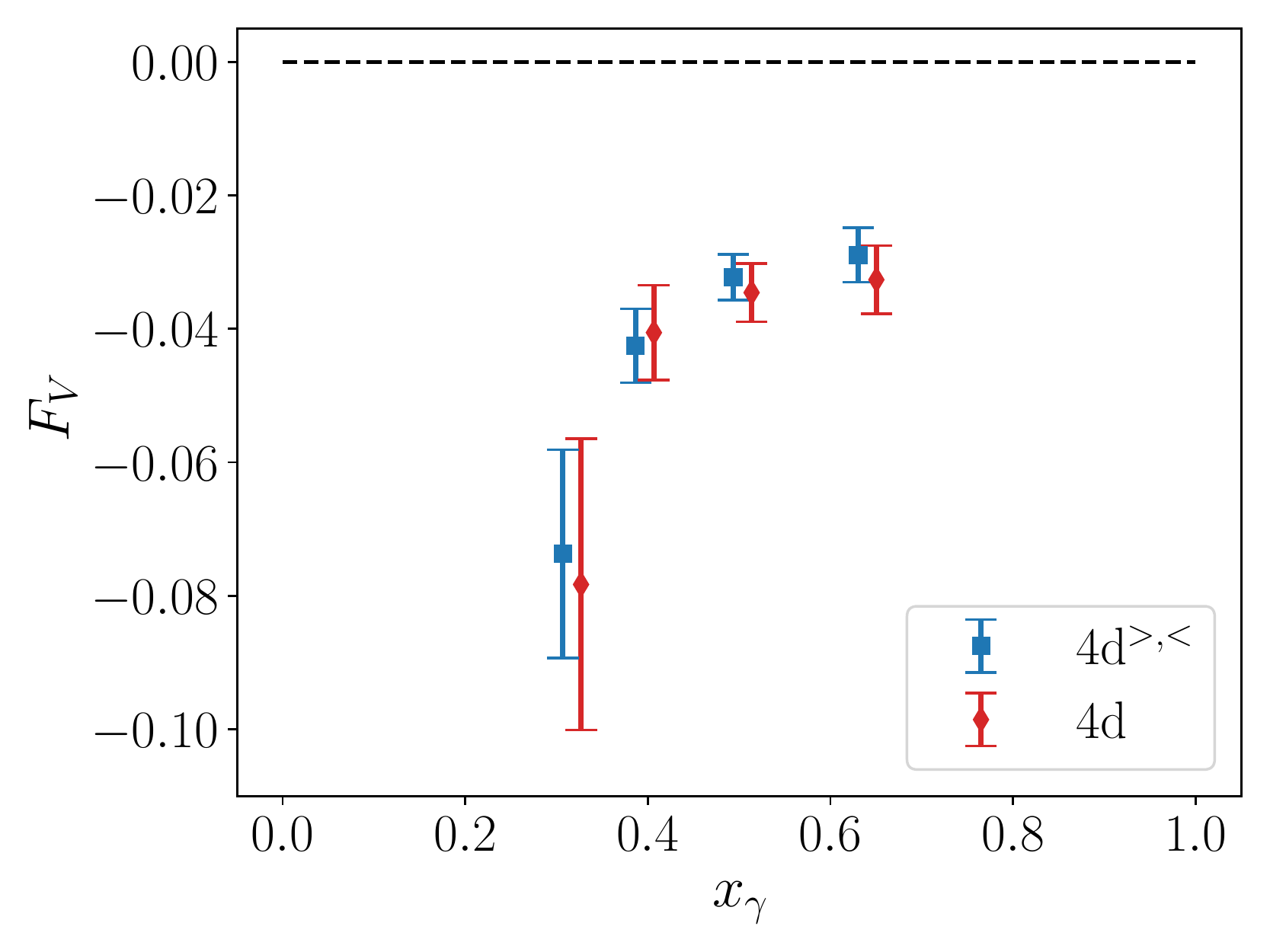}
    \end{minipage}
    \begin{minipage}{0.49\textwidth}
        \centering
        \includegraphics[width=\textwidth]{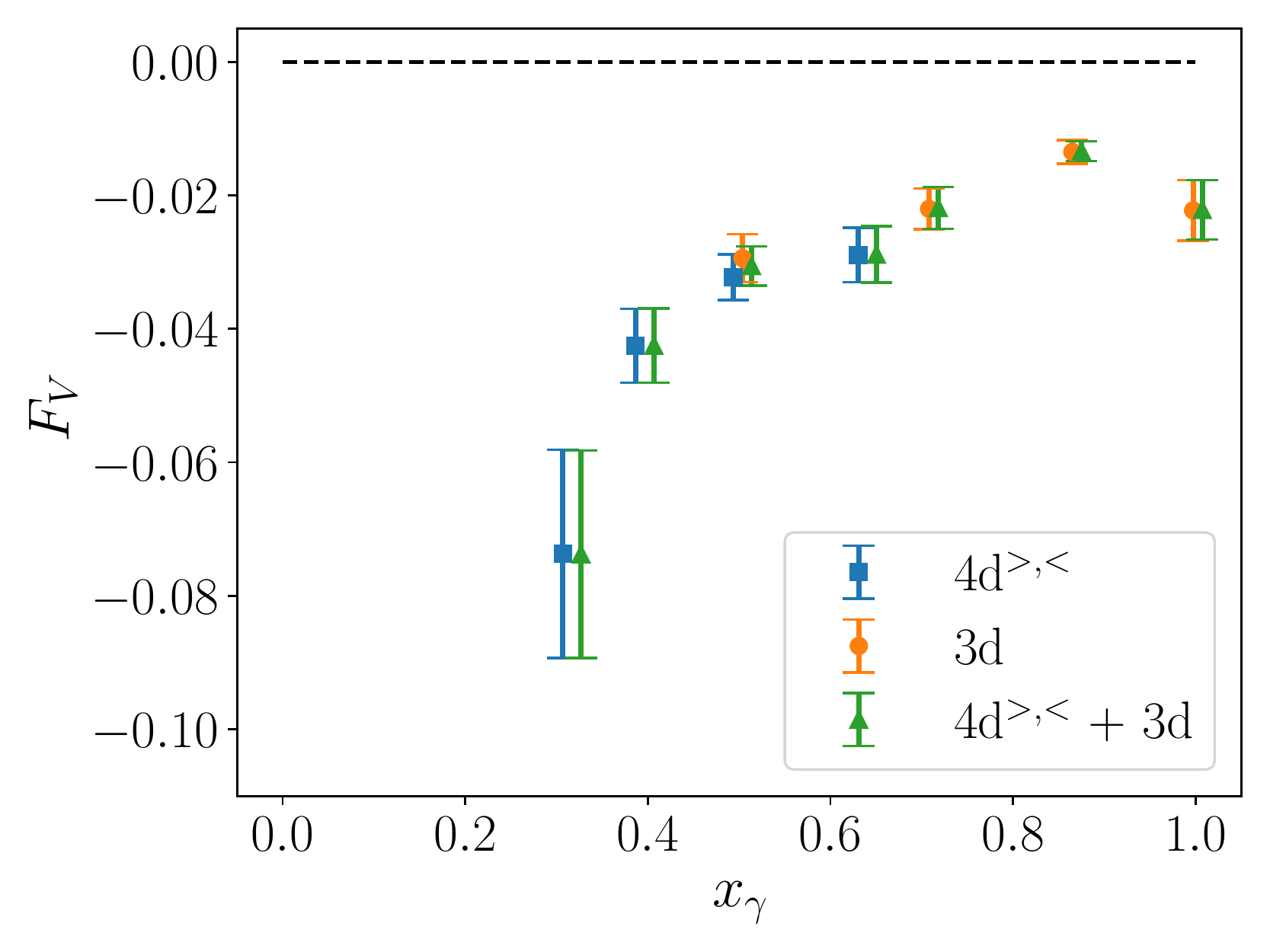}
    \end{minipage}
    \caption{Comparison of the 3d, 4d, and 4d$^{>,<}$ methods for $F_V$, plotted as a function of $x_\gamma$ (note that these are not our final results for the form factor; see Fig.~\ref{fig:3dprime_FF_vs_xgamma} for the final results with all improvements). Left: The red diamonds(blue squares) were calculated using the 4d(4d$^{>,<}$) method. Right: The blue squares(orange circles) show results using only 4d$^{>,<}$(3d) method data. The green triangles show results of simultaneous fits to both the 4d$^{>,<}$ and 3d data (since all fits include data at multiple $x_\gamma$ values, we can obtain results from the combination of methods even at $x_\gamma$ where we do not have both 3d and 4d$^{>,<}$ correlation functions). Points at the same $x_\gamma$ are shifted slightly for clarity.}
    \label{fig:FV_vs_xgamma_compare_seq_props}
\end{figure}

We start by comparing the 4d and 4d$^{>,<}$ methods. The left plot in Fig.~\ref{fig:FV_vs_xgamma_compare_seq_props} shows the results of $F_V$ as a function of $x_\gamma$ calculated using the 4d and 4d$^{>,<}$ methods. We observe that for all values of $x_\gamma$, the 4d$^{>,<}$ method yields  smaller statistical uncertainties than the 4d method. Recall that, when using the 4d$^{>,<}$ method, the different time orderings can be resolved. This allows the use of more detailed fit forms and the fits can be done at earlier values of $t_H$, resulting in the smaller uncertainties. Looking at Table~\ref{tab:num_solves}, the computational cost of the 4d$^{>,<}$ method is roughly twice as much as the 4d method. However, the ability to resolve the time orderings using the 4d$^{>,<}$ method allows for a more robust control over systematic uncertainties from unwanted exponentials. For this reason, we choose to compare the 4d$^{>,<}$ to the 3d method moving forward.

The right plot in Fig.~\ref{fig:FV_vs_xgamma_compare_seq_props} shows $F_V$ as a function of $x_\gamma$ calculated using the 3d and 4d$^{>,<}$ methods, as well as simultaneous fits to all data from both data sets. Focusing first on the individual fits, we see that for the value of $x_\gamma$ where we have data for both, the fit results agree and are of similar precision. We find that performing simultaneous fits to both data sets results in a factor $\sim 2$ improvement for this particular $x_\gamma$ value. For values where we have only 3d or 4d$^{>,<}$ method data, we find little to no improvement in statistical precision. Additionally, performing combined fits to the 3d and 4d$^{>,<}$ method data did not have a significant improvement in the stability of the global fit. Moving on to the computational cost, looking at Table~\ref{tab:num_solves}, the 3d method generally requires less solves than the 4d$^{>,<}$ method. This is due to the number of sequential solves required, which for the 3d method is $N_{t_H} N_{p_H}$, and for the 4d$^{>,<}$ method is $8 N_T N_{p_\gamma}$. As explained in Sec.~\ref{section:sequential_propagators}, the factor 8 results from having to do a solve for each $\gamma_\mu$ matrix associated with the electromagnetic current, for each time ordering. From this, if one uses only a single method, the 3d method offers similar precision and control over the unwanted exponentials compared to the 4d$^{>,<}$ method, but for a significantly cheaper computational cost. 

If one uses both methods however, our results suggest that a factor of $\sim 2$ improvement in precision could be achieved by performing calculations using both methods for each $x_\gamma$. This would also allow for more robust control of both sources of systematic uncertainties from unwanted exponentials. However, even if one keeps $N_T$ and $N_{p_\gamma}$ small, the additional solves required for the 4d$^{>,<}$ method will be significant relative to using only the 3d method. One could instead perform the calculation using the 3d method for more values of $t_H$ and more configurations, improving both the precision and control over unwanted exponentials, for less computational cost than using both methods. For these reasons, we proceed using the 3d method.

\FloatBarrier

\section{Improved three-point function calculation}
\label{sec:improved_C3}
In the following, we describe our improved methods of calculating lattice correlators that will be used to extract the form factors using the 3d method. We begin by discussing the infinite-volume approximation, which allows us to calculate the three-point functions at arbitrary photon momentum (i.e., not subject to the usual restriction from the periodic boundary conditions) with errors exponentially small in the lattice volume. Then, in Sec.~\ref{ssec:em_curr_at_origin} we introduce an alternate three-point function that can be used to extract the form factors. We demonstrate how it can be extracted for free by reusing propagators required to calculate the original three-point correlation function in Eq.~\eqref{eq:three_point}.

\subsection{Infinite-volume approximation}
\label{ssec:infinite_volume_approximation}
In this section we describe our approach to estimate momentum-projected correlation functions at arbitrary momenta (i.e., not restricted to integer multiples of $2\pi/L$) with exponentially small errors in the finite volume.  We simplify the discussion without loss of generality and consider the case of one spatial dimension with even integer extent $L$ (here we use lattice units). Let $C^{L}(x)$ be a finite-volume correlator and $C^{\infty}(x)$ the corresponding correlator in the $L\to \infty$ limit.  We assume there exist $c,d,\Lambda,\Lambda^\prime \in \mathbb{R}^+$ and $L_0 \in \mathbb{N}$ for which
\begin{align}\label{eqn:infinite_volume_approximation_condition_a}
\vert C^{\infty}(x) - C^{L}(x)\vert \leq c e^{-\Lambda L}
\end{align}
for all $x$ with $-L/2 \leq x \leq L/2$ and $L \geq L_0$ and
\begin{align}\label{eqn:infinite_volume_approximation_condition_b}
 \vert C^{\infty}(x) \vert \leq d e^{-\Lambda^\prime \vert x \vert}
\end{align}
for all $x$ with $\vert x\vert > L/2$. We now define
\begin{align}\label{eqn:fvsymmsum}
\tilde{C}^L(q) &\equiv \sum_{x=-L/2}^{L/2-1} C^L(x) e^{iqx} 
\end{align}
and
\begin{align}
\tilde{C}^\infty(q) &\equiv \sum_{x=-\infty}^{\infty} C^{\infty}(x) e^{iqx} \,.
\end{align}
Under the above assumptions, it then follows that there is a $\tilde{c} \in \mathbb{R}^+$ for which
\begin{align}
\vert \tilde{C}^\infty(q) - \tilde{C}^L(q) \vert  \leq \tilde{c} e^{-\Lambda_0 L}
\end{align}
for all $q \in [-\pi,\pi]$ and all $L \geq L_0$, with $\Lambda_0 \equiv \min(\Lambda,\Lambda^\prime/2)$.
In other words, $\tilde{C}^L(q)$ is exponentially close to the infinite-volume version $\tilde{C}^\infty(q)$.
In practice, the coordinate $x$ is often the relative distance between vertices $y$ and $z$, i.e., we are interested in
\begin{align}
 \tilde{C}(q) \equiv \sum_{y,z} C(y,z) e^{i q(y-z)} \,.
\end{align}
The constraint of Eq.~\eqref{eqn:infinite_volume_approximation_condition_b} can be satisfied if $C(y,z)$ eventually decreases exponentially as coordinates $y$ and $z$ are separated.  The implementation of
Eq.~\eqref{eqn:fvsymmsum}, however, requires the truncation of the double sum over $y$ and $z$ to $-L/2 \leq y-z < L/2$.  We are able to do this with point sources for either $y$ or $z$ but not if sequential solves and wall sources are used for both $y$ and $z$.  We therefore develop a method in Sec.~\ref{ssec:ratio_method} that combines the statistical benefits of a sequential solve with the improved momentum resolution offered by a point-source-based setup.

In practice, $\Lambda=m_\pi$ but $\Lambda^\prime$ can be substantially larger such that $\Lambda_0=m_\pi$ often holds.

Finally, we would like to point out that the method is similar in spirit to the QED$_\infty$ method \cite{RBC:2018dos}.  In Ref.~\cite{Feng:2018qpx} an extension of this method was presented that allows for the calculation of QED self energies with only exponentially small finite-volume errors.

\subsection{Three-point function with electromagnetic current at origin}
\label{ssec:em_curr_at_origin}
The three-point function in Eq.~\eqref{eq:three_point} used to extract the hadronic tensor has the weak current fixed to the origin. In this section, we show how $T_{\mu \nu}$ can be extracted from a similar correlation function, except with the electromagnetic current fixed to the origin, given by
\begin{align}
    C_{3,\mu \nu}^{\text{EM}}(t_W, t_H) = e^{E_H t_W}\int d^3x \int d^3y \, e^{i(\vec{p}_\gamma-\vec{p}_H) \cdot \vec{x}} e^{i \vec{p}_H \cdot \vec{y}} \langle J_{\mu}^\text{em}(0) J_\nu^\text{weak}(t_W, \vec{x}) \phi^\dagger_H(t_H, \vec{y}) \rangle.
    \label{eq:C3_EM}
\end{align}
The superscript $\text{EM}$ is used throughout this work to differentiate between the correlation function with the weak current at the origin in Eq.~\eqref{eq:three_point}. The additional factors $e^{E_H t_W}$ and $e^{-i \vec{p}_H \cdot \vec{x}}$ are required to shift the interpolating field in Euclidean time and space, respectively, relative to the other operators. Note that the phase to project to definite photon momentum is flipped relative to the three-point function in Eq.~\eqref{eq:three_point}. When using point sources, this correlation function can be calculated for free by reusing propagators used to calculate the three-point function in Eq.~\eqref{eq:three_point}. In particular, two sets of propagator solves must be performed, one for each component of the electromagnetic current $J^{\text{em}}_\mu$. For example, when $H = D_s$, the sequential propagator calculated for the strange(charm) quark contribution of $J^{\text{em}}_\mu$ when the weak current is at the origin is the same sequential propagator needed for the charm(strange) quark contribution of $J^{\text{em}}_\mu$ when the electromagnetic current is at the origin. 

We define the time integrals of this correlation function for the different time orderings as
\begin{align}
    &I^{\text{EM}, >}_{\mu \nu}(t_H, T) = \int_0^T dt_W \, e^{-E_\gamma t_W} C^\text{EM}_{3,\mu \nu}(t_W, t_H),
    \\
    &I^{\text{EM}, <}_{\mu \nu}(t_H, T) = \int_{-T}^0 dt_W \, e^{-E_\gamma t_W} C^\text{EM}_{3,\mu \nu}(t_W, t_H).
\end{align}
By inserting two complete sets of states and performing the integrals over time, we find the spectral decompositions
\begin{align}
    I^{\text{EM}, >}_{\mu \nu}(t_H, T) &= \sum_{n,l} \frac{\bra{0} J^\text{weak}_\nu(0) \ket{n(\vec{p}_H-\vec{p}_\gamma)} \bra{n(\vec{p}_H-\vec{p}_\gamma)} J^\text{em}_\mu(0) \ket{l(\vec{p}_H)} \bra{l(\vec{p}_H)} \phi^\dagger_H \ket{0}}{2E_{n,\vec{p}_H-\vec{p}_\gamma} 2E_{l, \vec{p}_H}(E_\gamma + E_{n,\vec{p}_\gamma - \vec{p}_H} - E_H)} \nonumber
    \\
    &\hspace{0.4in} \times e^{E_{l,\vec{p}_H} t_H} \Big[1- e^{-(E_\gamma + E_{n,\vec{p}_\gamma - \vec{p}_H} - E_H) T} \Big]
    \label{eq:Imunu_spectral_decomp_EM_greaterthan}
\end{align}
and
\begin{align}
    I^{\text{EM}, <}_{\mu \nu}(t_H, T) &= \sum_{m,l} \frac{\bra{0} J^\text{em}_\mu(0) \ket{m(\vec{p}_\gamma)} \bra{m(\vec{p}_\gamma)} J^\text{weak}_\nu(0) \ket{l(\vec{p}_H)} \bra{l(\vec{p}_H)} \phi^\dagger_H \ket{0}}{2E_{m,\vec{p}_\gamma} 2E_{l,\vec{p}_H}(E_\gamma-E_{m,\vec{p}_\gamma} + \Delta E_{l, \vec{p}_H})} \nonumber
    \\
    &\hspace{0.4in} \times e^{E_{l,\vec{p}_H} t_H} \Big[ e^{(E_\gamma - E_{n,\vec{p}_\gamma} + \Delta E_{l, \vec{p}_H})T} - 1 \Big],
    \label{eq:Imunu_spectral_decomp_EM_lessthan}
\end{align}
where $\Delta  E_l = E_{l,\vec{p}_H}-E_{H,\vec{p}_H}$ is the excited-state energy gap for the $l^\text{th}$ excited state created by the interpolating field. Using similar arguments as in Sec.~\ref{ssec:Euclidean}, we find that, for $\vec{p}_\gamma \neq 0$, the hadronic tensor can be extracted by
\begin{align}
    T^<_{\mu \nu} &= - \lim_{T \to \infty} \lim_{t_H \to -\infty} \frac{2 E_{H, \vec{p}_H} e^{-E_{H, \vec{p}_H} t_H}}{\bra{H(\vec{p}_H)} \phi^\dagger_H(0) \ket{0}} I^{\text{EM}, >}_{\mu \nu}(t_H, T),
    \\
    T^>_{\mu \nu} &= - \lim_{T \to \infty} \lim_{t_H \to -\infty} \frac{2 E_{H, \vec{p}_H} e^{-E_{H, \vec{p}_H} t_H}}{\bra{H(\vec{p}_H)} \phi^\dagger_H(0) \ket{0}} I^{\text{EM}, <}_{\mu \nu}(t_H, T).
\end{align}
The largest possible integration ranges for the $t_W<0$ and $t_W>0$ time orderings are $-t_H$ and $aN_T+t_H$, respectively, where $N_T$ is the number of temporal lattice sites. As before, as one integrates closer to the interpolating field for $t_W<0$, excited state effects become larger. 

\begin{figure}
    \centering
    \includegraphics[width=0.7\textwidth]{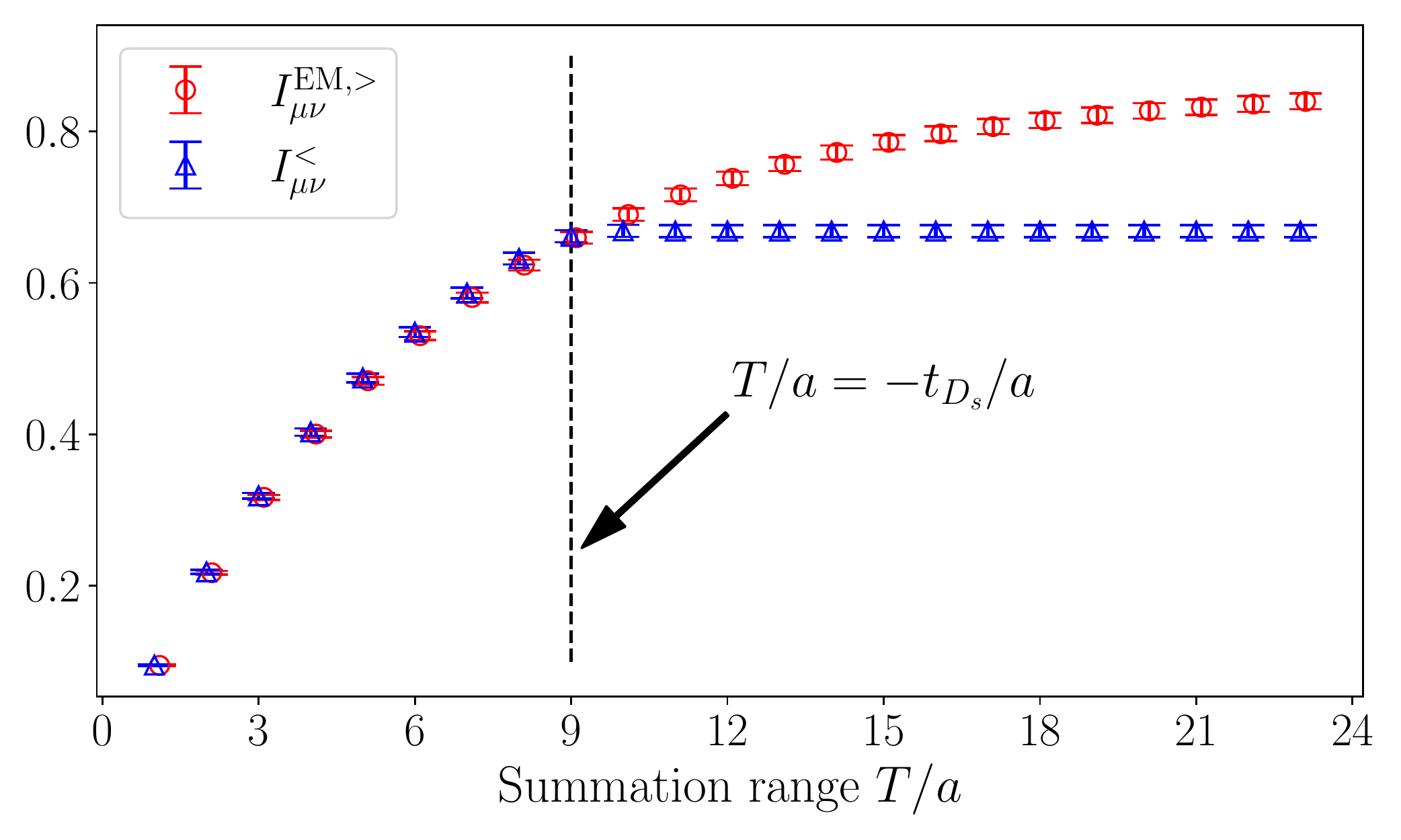}
    \caption{$I_{\mu \nu}^{<}(T, t_H)$ and $I_{\mu \nu}^{\text{EM},>}(T,t_H)$ as a function of $T$ for $-t_H/a=9$. Notice that $I_{\mu \nu}^{<}(T, t_H)$ can only be evaluated up to $T=-t_H$, while $I_{\mu \nu}^{\text{EM},>}(T, t_H)$ can be evaluated for larger values of $T$. Both were calculated in the rest frame of the meson with $\vec{p}_\gamma=2\pi/L(0,0,0.6)$. The indices shown are $\mu=\nu=0$.}
    \label{fig:C3_compare_em_weak}
\end{figure}

Notice that the spectral decompositions of the $t_W>0$ time ordering of $I^{\text{EM}}_{\mu \nu}$ and the $t_\text{em}<0$ time ordering of $I_{\mu \nu}$ are equal up to excited state effects; the same is true for the $t_W<0$ and $t_\text{em}>0$ time orderings of $I^{\text{EM}}_{\mu \nu}$ and $I_{\mu \nu}$. This implies that one can perform simultaneous fits to the $I^{\text{EM},>}_{\mu \nu}$ and $I^<_{\mu \nu}$ data using common fit parameters, and similarly for the $I^{\text{EM},<}_{\mu \nu}$ and $I^>_{\mu \nu}$ data. 
As an example of the different behavior of the two data sets with $T$, Fig.~\ref{fig:C3_compare_em_weak} shows the weak axial-vector component of both $I_{00}^{<}(t_H, T)$ and $I_{00}^{\text{EM},>}(T,t_H)$, calculated in the rest frame of the meson with $\vec{p}_\gamma=2\pi/L(0,0,0.6)$ using $-t_H/a=9$.  
Looking at the blue triangles, for $T<|t_H|$, the $I_{\mu \nu}^{<}(t_H, T)$ data begins to plateau as $T$ is increased up until the maximum value of $T=-t_H$. For the $I_{\mu \nu}^{\text{EM},<}(t_H, T)$ data on the other hand, it is possible to integrate past $T=-t_H$ because one is integrating away from the interpolating field in this case. 

This example leads to a clear scenario where having both sets of data would be crucial for the analysis. 
In particular, consider the possibility where the results of the fits to the $I_{\mu \nu}^{<}(t_H, T)$ data were not stable for all allowed values of $T < |t_H|$. 
To get around this problem, one option would be to extend the allowed values of $T$ by increasing the source-sink separation $t_H$. 
However, because increasing $t_H$ generally results in noisier data, it might not be practical to extend $t_H$ large enough to observe stability in the fit range for $T$.
Another option would be to add additional exponential terms in the fit form, but because fits with multiple exponentials are generally unstable, this might not be possible without introducing priors on the energies of the intermediate state, which could bias the results. 
A better solution to the problem would be to perform the calculation for $I_{\mu \nu}^{\text{EM},>}(t_H, T)$, which would allow one to extend the fit range in $T$ while keeping $t_H$ constant. A similar situation can occur for $t_W<0$ and $t_\text{em}>0$ data, except that now 
$I_{\mu \nu}^{\text{EM},<}(t_H, T)$ has a limited range of $T<|t_H|$, and 
$I_{\mu \nu}^{>}(t_H, T)$ can be evaluated for larger values of $T$. 

Considering instead the opposite scenario where one observes stability for both time orderings of both $I_{\mu \nu}^{\text{EM}}(t_H, T)$ and $I_{\mu \nu}(t_H, T)$, having both sets of data provides improvements to the extracted form factors beyond simply extra statistics. This can be understood by considering the maximum Euclidean time separation of the three operators in the correlation functions. The maximum separation for the $t_\text{em}<0$ time ordering is $t_H$, while for $t_W>0$ the maximum separation is $T+|t_H|$. It is therefore expected that, for the same value of $t_H$, the $I^{<}_{\mu \nu}$ data will be more precise than the $I^{\text{EM},>}_{\mu \nu}$ data; using similar arguments, the $I^{\text{EM},<}_{\mu \nu}$ data is expected to be more precise than the $I^{>}_{\mu \nu}$ data. 

In Sec.~\ref{ssec:compare_EM_WEAK} we compare the form factors extracted the individual data sets, as well as the improvements achieved by performing simultaneous fits to both data sets.

\section{Improved form factor determination}
\label{sec:improved_FF}
In this section we describe improvements to the original 3d method calculation presented in Secs.~\ref{section:fit_method} and \ref{sec:compare_3d_4d}. In particular, the four improvements we make are
\begin{itemize}
    \item Sec.~\ref{ssec:ratio_method} shows improvements by taking ratios of the point source data to noise source data
    \item Sec.~\ref{ssec:ave_pgamma} shows improvements from averaging over positive and negative photon momentum
    \item Sec.~\ref{ssec:sub_pgamma_zero_FA_SD} shows improvements by extracting $F_{A,SD}$ using a method that removes contact terms that diverge at small $x_\gamma$
    \item Sec.~\ref{ssec:compare_EM_WEAK} shows improvements by doing a combined analysis of both data calculated using the original three-point function in Eq.~\eqref{eq:three_point} and the alternate three-point function in Eq.~\eqref{eq:C3_EM}.
\end{itemize}
As each improvement is presented, we also present numerical studies that demonstrate the level of the improvements. Note that, when doing comparisons of different methods, we still implement all the other improvements not being studied. So, in any given analysis, three of the four improvements are being used. 

Another modification to the analysis is that we now fit the contributions to the form factors from the separate quark components of the electromagnetic current separately. This is done for two reasons. First, these separate contributions are well defined QCD form factors and are therefore of phenomenological interest. The second is that, in general, the intermediate states that contribute to the different quark contributions are different. Fitting them separately therefore reduces the possible number of exponential that contribute at finite integration range $T$, which stabilizes the fits. We denote the charm and strange quark components of the form factors with superscripts $(c)$ and $(s)$, respectively, such that \textit{e.g.} $F_V = F^{(c)}_V + F^{(s)}_V$. The details of the final analysis methods are presented in Sec.~\ref{sec:final_procedure}.

\subsection{Ratio methods}
\label{ssec:ratio_method}

To be able to use the infinite-volume approximation in Sec.~\ref{ssec:infinite_volume_approximation}, one must use point sources. In our analysis in Sec.~\ref{sec:noise_vs_point}, we found that noise sources generally perform better than point sources for the same statistics. Because of this, we would like a way to improve the precision of point sources to be similar to that of noise sources. In this section, we present a method that achieves this which works by taking ratios of correlation functions calculated using both point and noise sources. 

Specifically, suppose we calculate a three-point function (either in Eq.~\eqref{eq:three_point} or Eq.~\eqref{eq:C3_EM}) using point sources in the rest frame of the meson at photon momentum $\vec{p}_\gamma$, denoted as $C^{\rm point}_{3,\mu \nu}(\vec{p}_\gamma, t, t_H)$. Here $t$ could be either $t_{\text{em}}$ or $t_W$. Using noise sources, we calculate the same correlation function, but at a photon momentum $\vec{p}^*$ that is allowed by periodic boundary conditions, denoted as $C^{\mathbb{Z}_2}_{3, \mu \nu}(\vec{p}^*, t, t_H)$. The improved estimator is calculated using the ratio
\begin{align}
    C^{\rm improved}_{3, \mu \nu}(\vec{p}_\gamma, t, t_H) = C^{\rm point}_{3, \mu \nu}(\vec{p}_\gamma, t, t_H) \frac{C^{\mathbb{Z}_2}_{3, \mu \nu}(\vec{p}^*, t, t_H)}{C^{\rm point}_{3, \mu \nu}(\vec{p}^*, t, t_H)},
    \label{eq:ratio_C3}
\end{align}
where it is understood that one must first calculate the expectation values of the individual correlation functions \textit{before} taking the ratio. Note that the value of $\vec{p}^*$ must be chosen such that the expectation value of the denominator is non-zero. In our analysis, we perform calculations for $\vec{p}_\gamma$ in the $z$-direction, and calculate two values of $\vec{p}^* = 2\pi/L(0, 0, p^*_z)$ with $p^*_z\in \{0, 1\}$.

When deciding what ratio to take for the weak axial-vector component of the three-point function, for each value of $\vec{p}_\gamma, t, t_H$, we try four ratios, calculate the statistical uncertainty for each of the four possible ratios, and choose to implement the method with the smallest statistical uncertainty. The four ratios we consider are 
\begin{enumerate}
    \item No ratio,
    \item Ratio using $p^*_z = 0$,
    \item Ratio using $p^*_z = 1$,
    \item Ratio using the two values of $p^*_z$ linearly interpolated to the value of $\vec{p}_\gamma$.
\end{enumerate}
Because the expectation value of the weak vector component of the three-point correlation function is zero when both the meson and photon momentum are zero, we only consider methods 1 and 3 in this case. Additionally, when calculating $F_{A,SD}$ by subtracting the correlation function at zero photon momentum as described in Sec.~\ref{ssec:sub_pgamma_zero_FA_SD}, we take ratios after subtracting. 

\begin{figure}[h]
    \centering
    \begin{minipage}{0.48\textwidth}
        \centering
        \includegraphics[width=\textwidth]{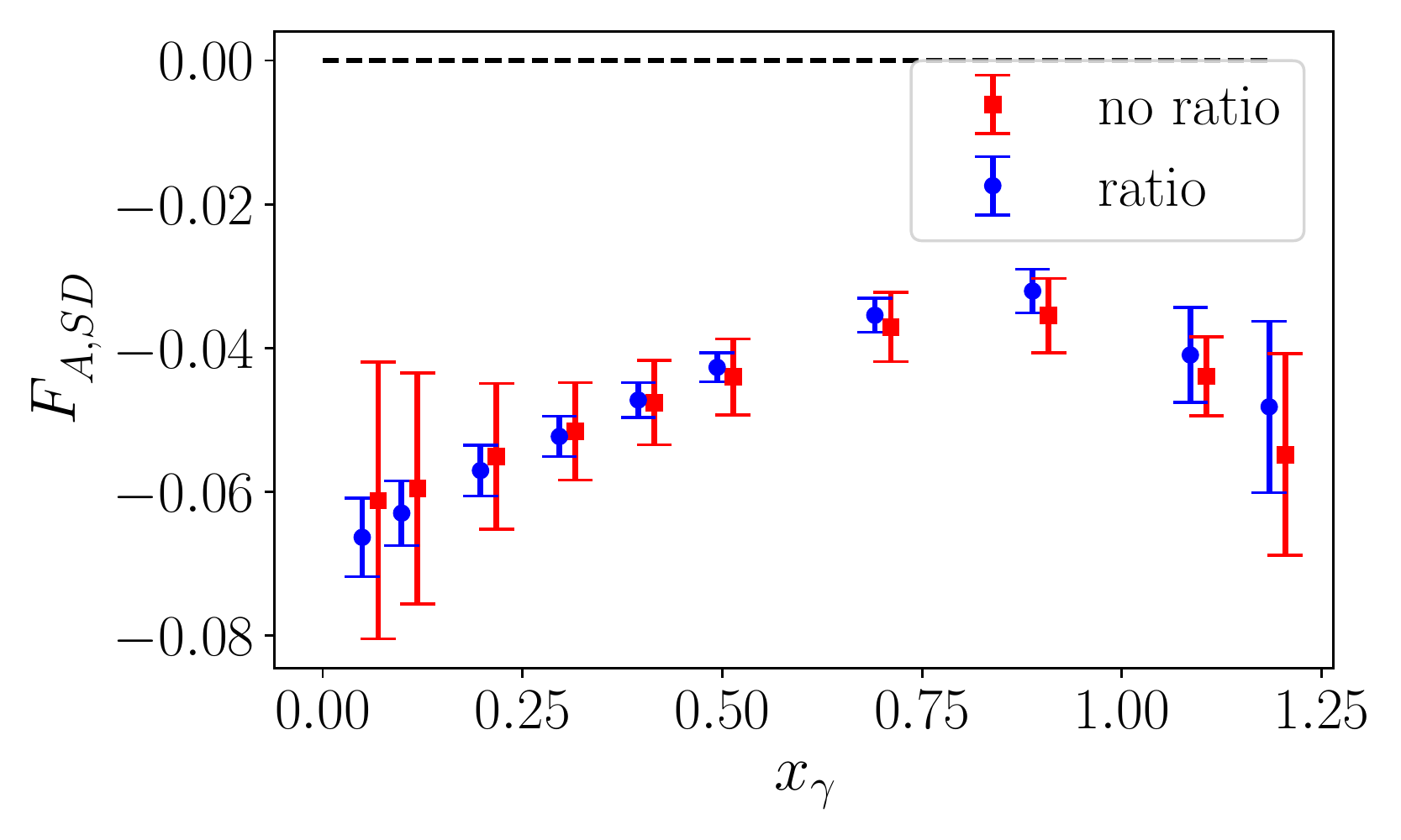}
    \end{minipage}
    \begin{minipage}{0.48\textwidth}
        \centering
        \includegraphics[width=\textwidth]{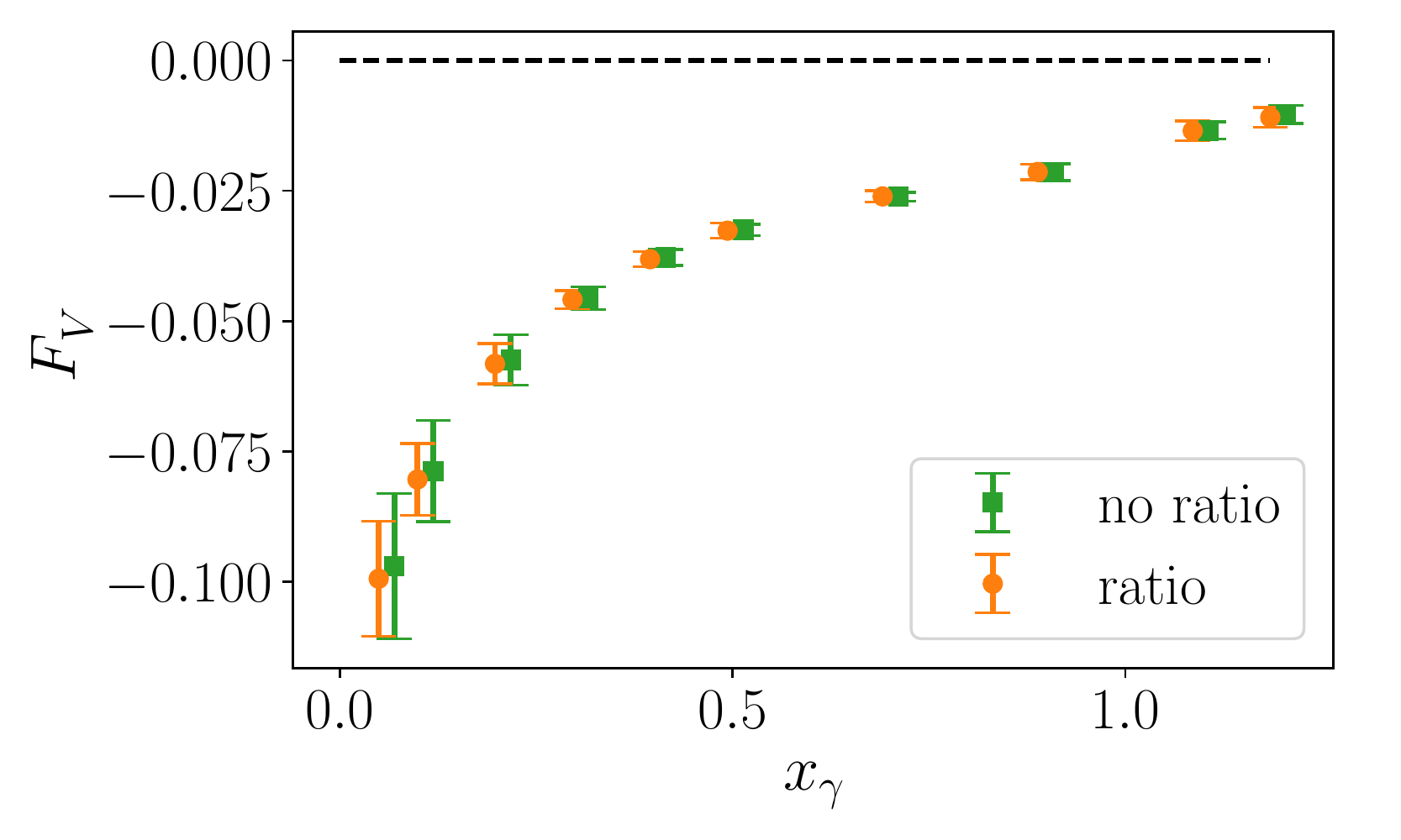}
    \end{minipage}
    \caption{Left(right) compares $F_{A,SD}(F_V)$ as a function of $x_\gamma$ using the ratio method and without using the ratio method. The ratio method results in a more significant improvement for $F_{A,SD}$. Points at the same $x_\gamma$ have been shifted slightly for clarity.}
    \label{fig:compare_ratio}
\end{figure}

To test the improvements gained using this ratio method, we compare the form factors as a function of $x_\gamma$ both using the improved correlation function in Eq.~\eqref{eq:ratio_C3}, and using the original correlation function without multiplying by the ratio. The analysis of a specific component of a given form factor was performed using the same fit forms and fit ranges for both the original and improved data. The fit forms and fit ranges were chosen by performing a stability analysis to the improved data. The results for both $F_{A,SD}$ and $F_V$ are shown in Fig.~\ref{fig:compare_ratio}. 

Looking first at $F_{A,SD}$, we observe a $\sim 4$ times reduction in the statistical error for small $x_\gamma$, with the improvements generally decreasing as $x_\gamma$ increases. More specifically, the time orderings $t_{\text{em}}<0$ and $t_W>0$ see the greatest improvement in precision. The time orderings $t_{\text{em}}>0$ and $t_W<0$ for the charm-quark component of the EM current sees a factor $\sim 2$ improvement, while the strange-quark component of the EM current sees little to no improvement. We observe only modest reductions in statistical noise for the vector form factor $F_V$.

\FloatBarrier

\subsection{Averaging over \texorpdfstring{$\pm \vec{p}_\gamma$}{+-pgamma}}
\label{ssec:ave_pgamma}
One advantage of our improved method is the ability to average over the positive and negative photon momenta for free. In this section, we compare the precision of the form factors calculated by performing this average to form factors calculated using only positive photon momentum. As in Sec.~\ref{ssec:ratio_method}, the analysis of a specific component of a given form factor was performed using the same fit forms and fit ranges for the analysis of both data. The fit forms and fit ranges were chosen by performing a stability analysis to the data averaged over photon momentum.

\begin{figure}[h]
    \centering
    \begin{minipage}{0.48\textwidth}
        \centering
        \includegraphics[width=\textwidth]{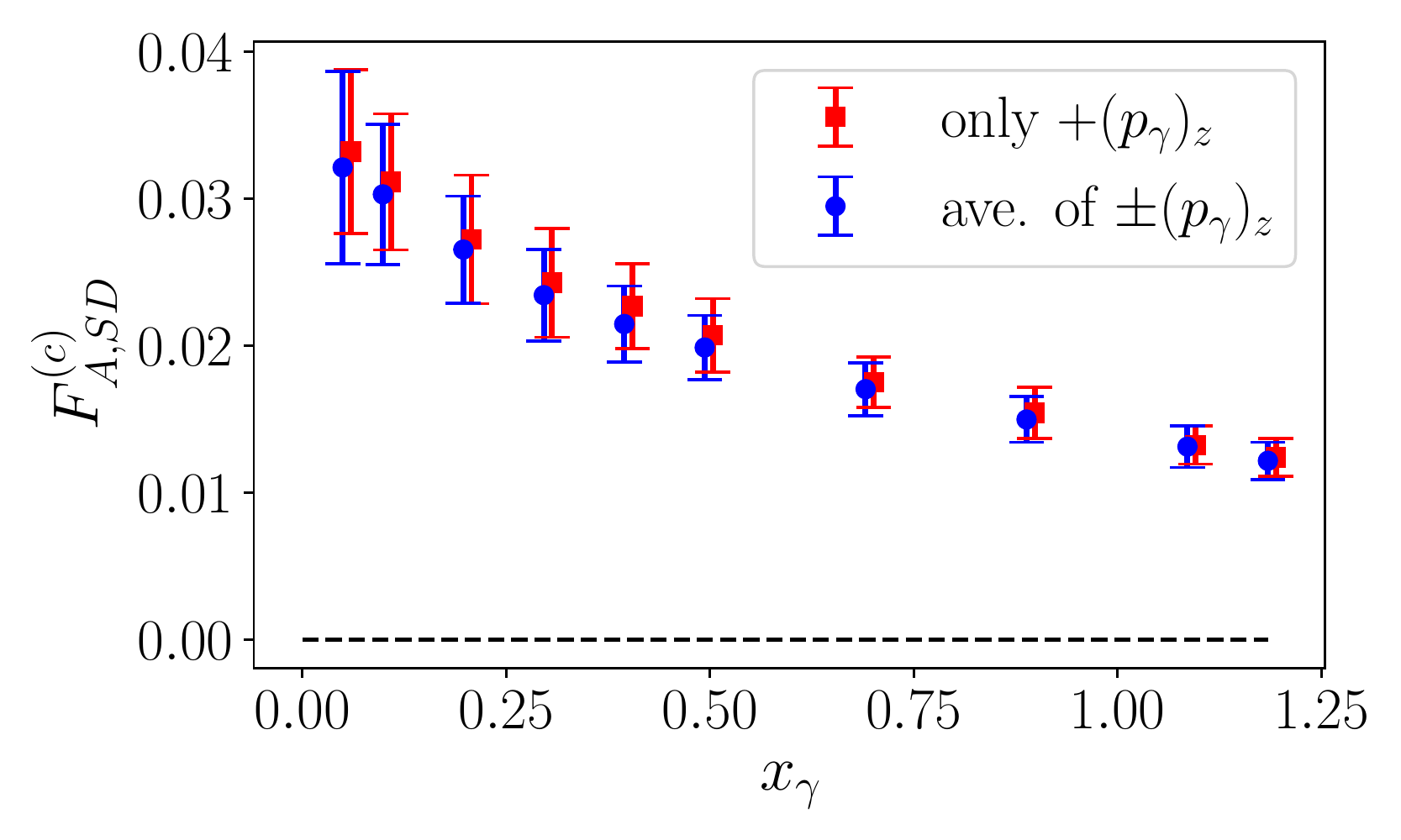}
    \end{minipage}
    \begin{minipage}{0.48\textwidth}
        \centering
        \includegraphics[width=\textwidth]{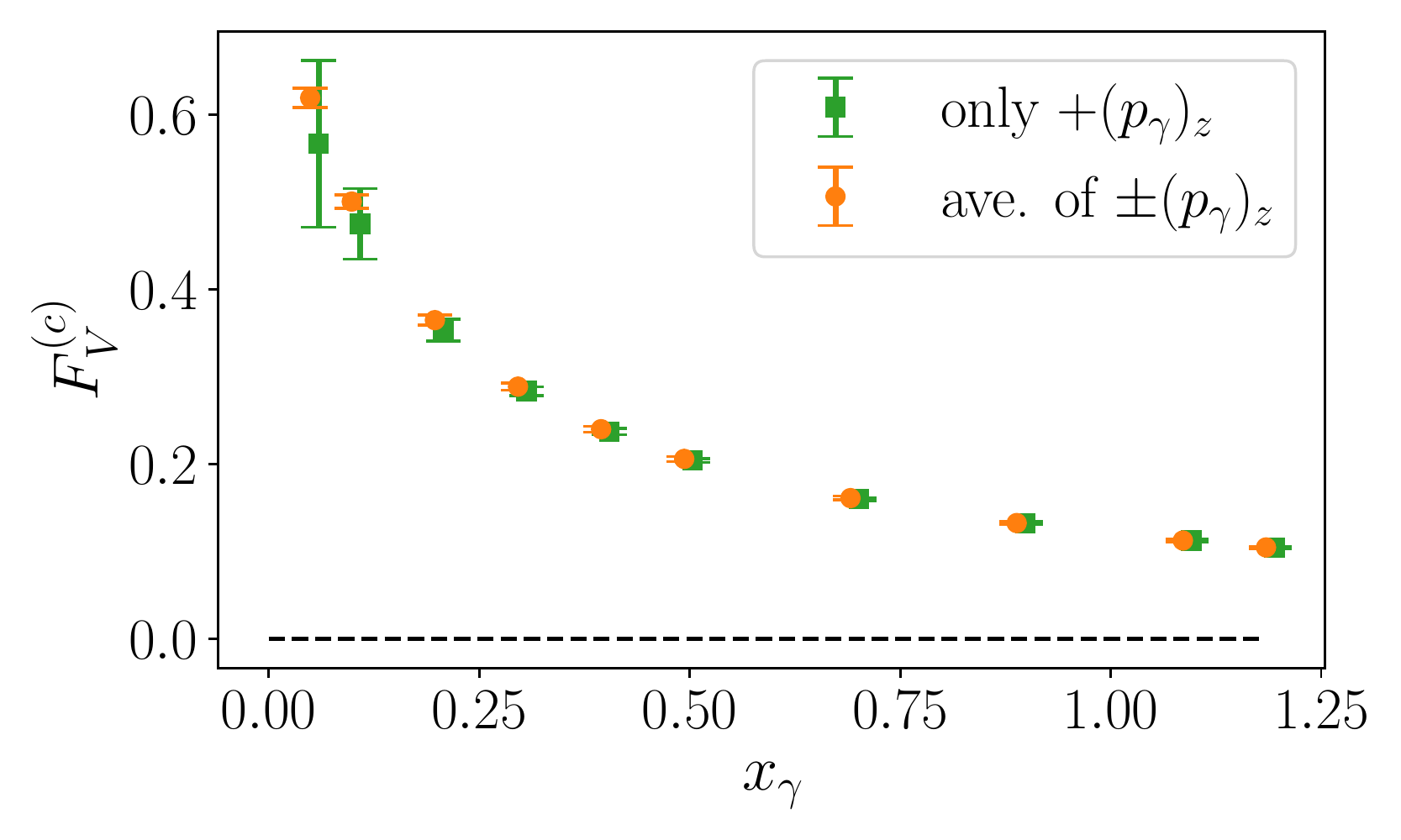}
    \end{minipage}

    \begin{minipage}{0.48\textwidth}
        \centering
        \includegraphics[width=\textwidth]{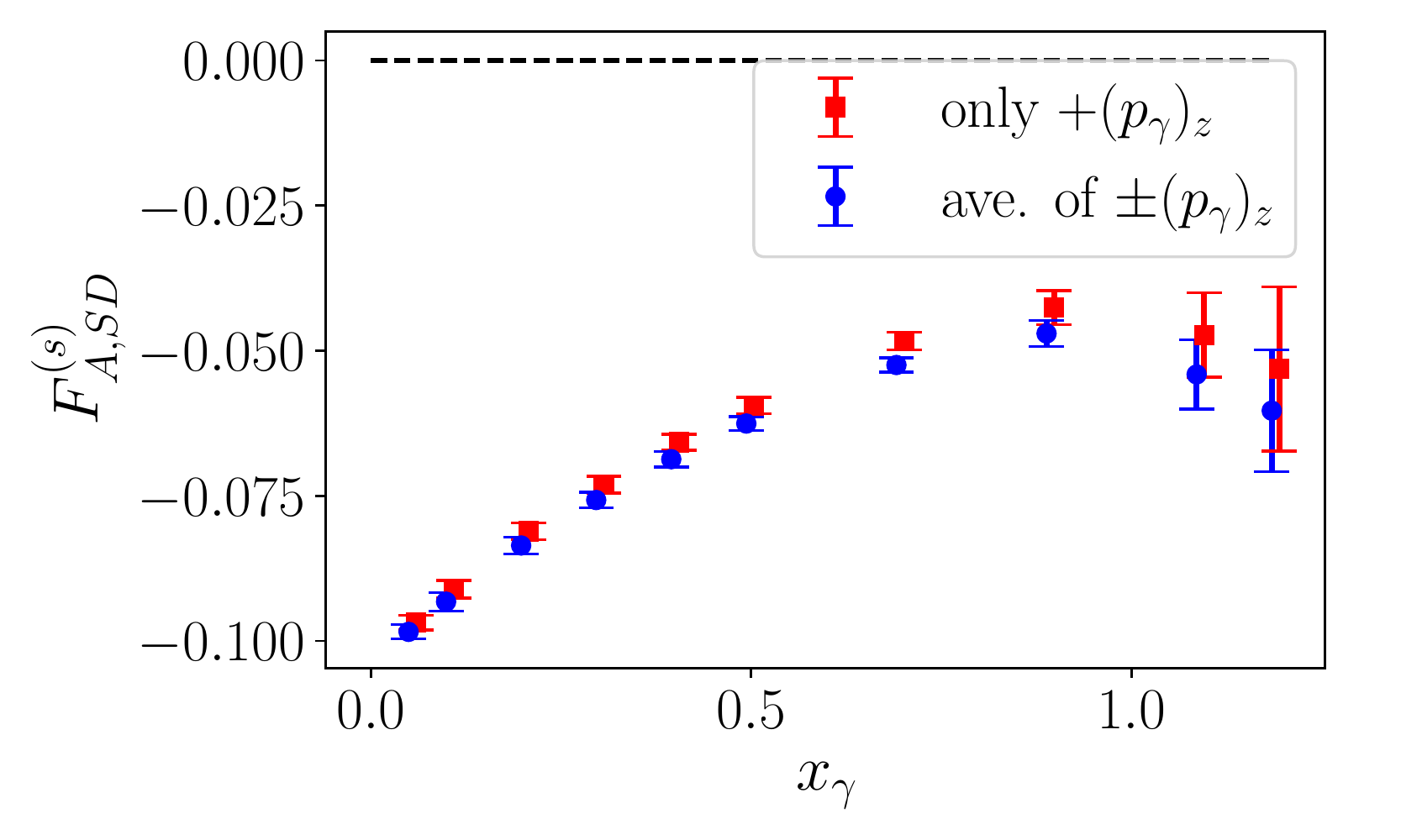}
    \end{minipage}
    \begin{minipage}{0.48\textwidth}
        \centering
        \includegraphics[width=\textwidth]{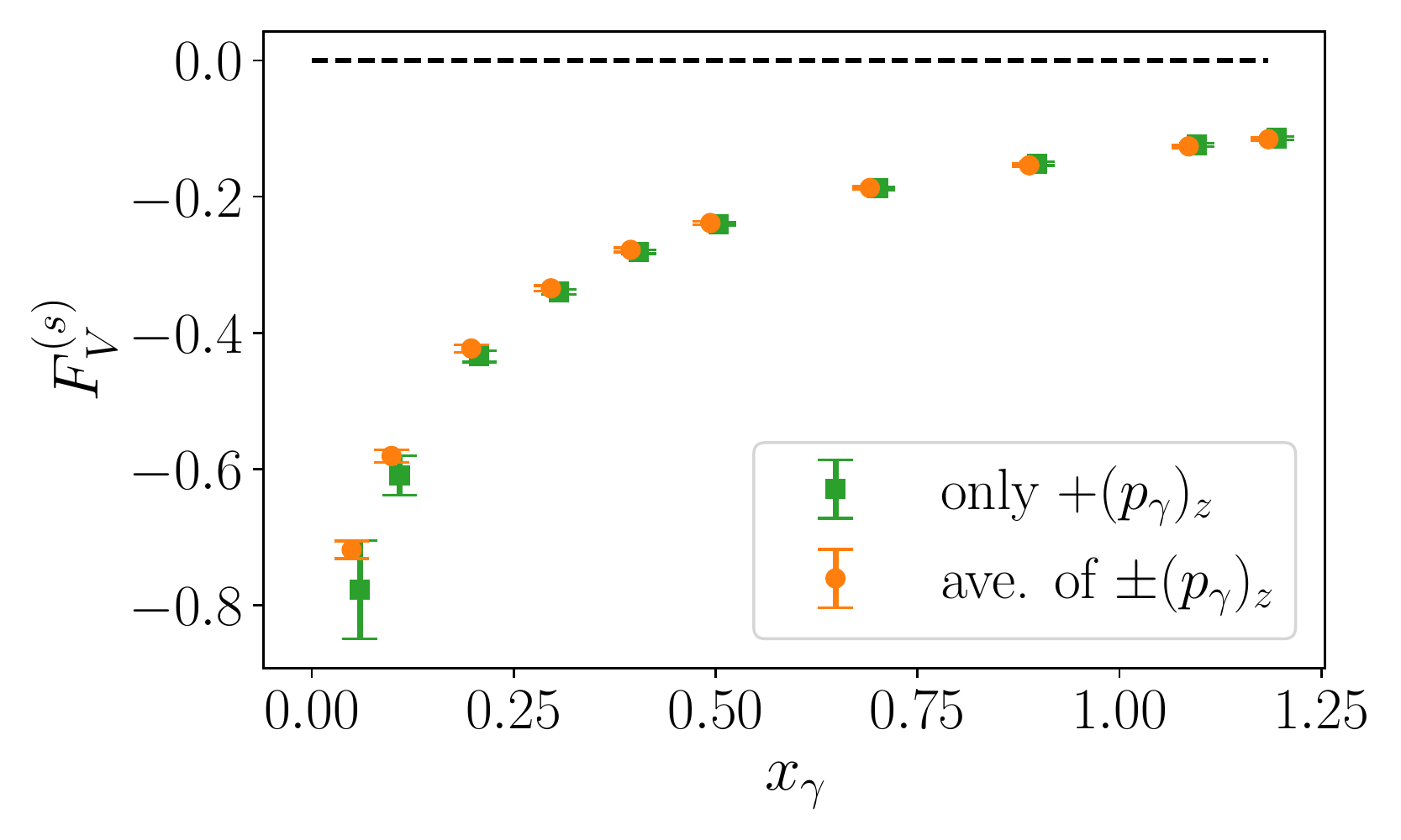}
    \end{minipage}

    \begin{minipage}{0.48\textwidth}
        \centering
        \includegraphics[width=\textwidth]{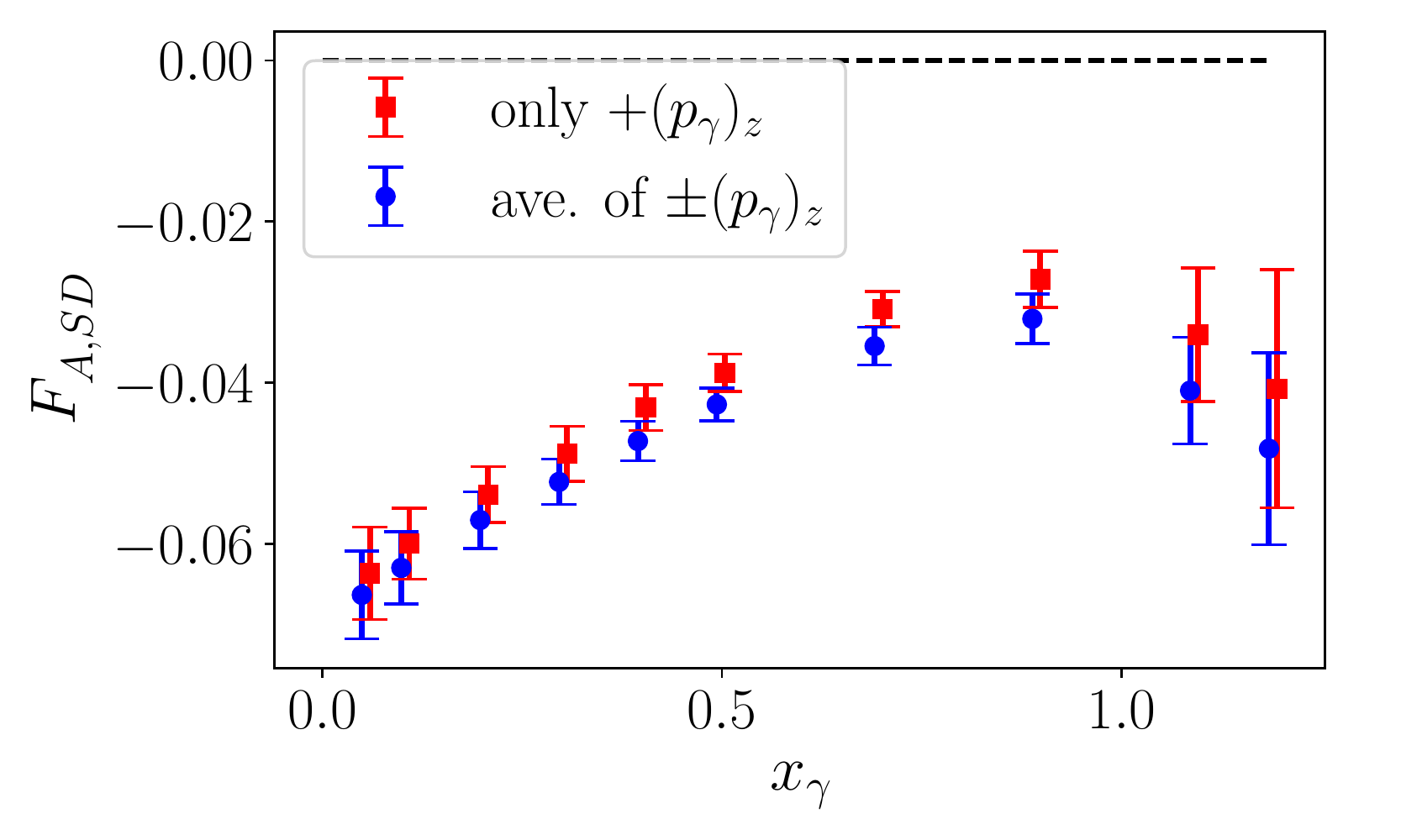}
    \end{minipage}
    \begin{minipage}{0.48\textwidth}
        \centering
        \includegraphics[width=\textwidth]{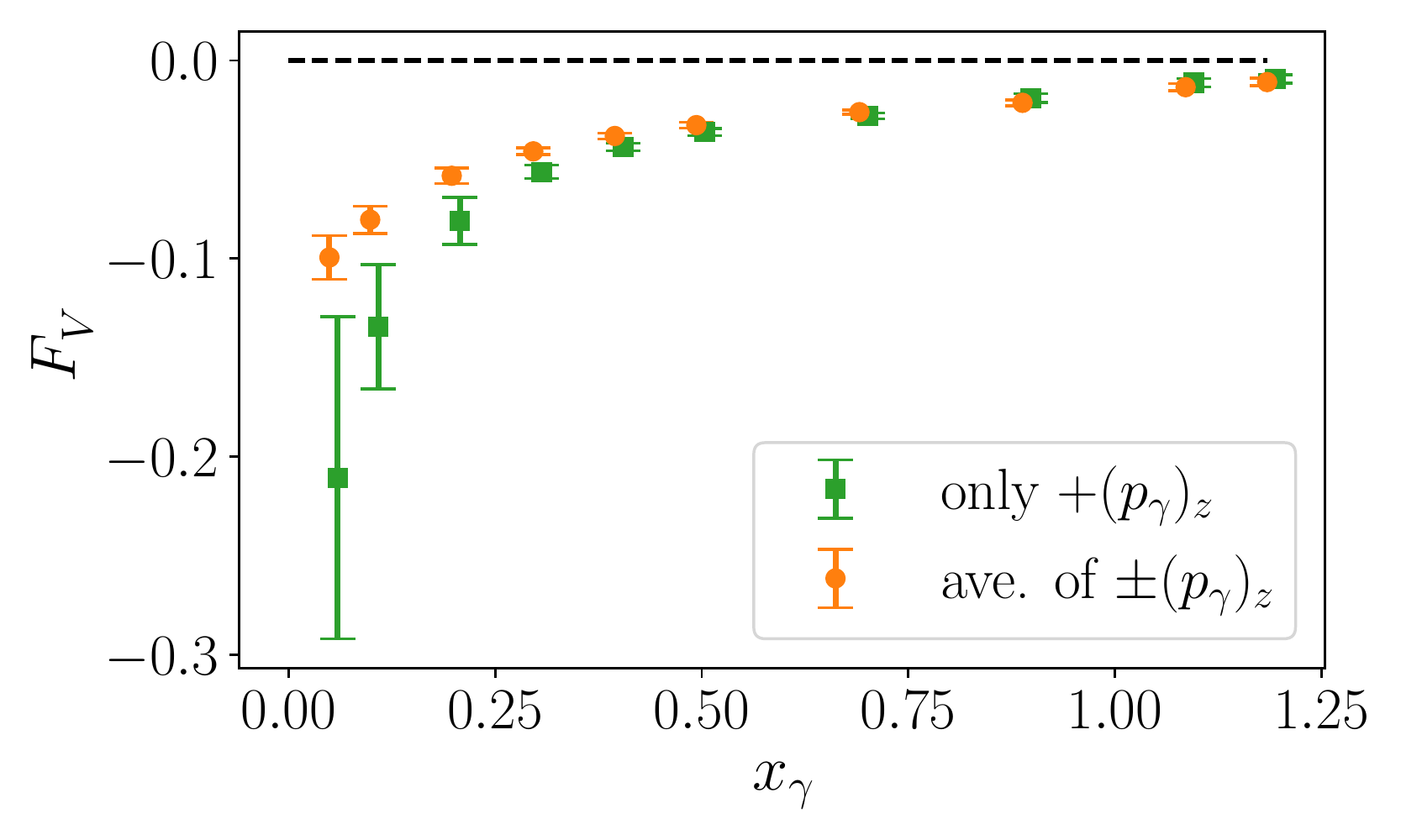}
    \end{minipage}
    \caption{Left(right) column compares $F_{A,SD}(F_V)$ calculated using only positive $(p_\gamma)_z$ to $F_{A,SD}(F_V)$ calculated averaging over positive and negative $(p_\gamma)_z$. The different rows show the full form factors, as well as the individual charm and strange quark current components. For $F_V, F_V^{(c)},$ and $F_V^{(s)}$, dramatic improvements in precision are observed for small $x_\gamma$, with the improvement generally decreasing with $x_\gamma$. Only modest improvements are observed for all $F_{A,SD}, F_{A,SD}^{(c)},$ and $F_{A,SD}^{(c)}$ data.  Points at the same $x_\gamma$ have been shifted slightly for clarity.}
    \label{fig:compare_ave}
\end{figure}

Looking at Fig.~\ref{fig:compare_ave}, we see that at small $x_\gamma$, averaging over $\pm \vec{p}_\gamma$ results in anywhere from a factor 3 to factor 9 improvement in precision for $F_V^{(c)}$, $F_V^{(s)}$ and $F_V$. The dramatic improvement in precision at small $x_\gamma$ can be understood by first noting that the form factor decomposition of $T_{\mu \nu}$ in Eq.~\eqref{eq:Tmunu_FF_decomp} implies the weak vector component of the three-point correlation function is purely real. This information can be used to show that $F_V$ receives a pure noise contribution, which is exactly canceled out by averaging over positive and negative photon momentum, leading to the dramatic improvement. On the other hand, the weak axial-vector component of the three-point correlation function is purely imaginary and does not receive a pure noise contribution. For this reason, averaging over $\pm \vec{p}_\gamma$ has only a modest improvement in precision for $F_{A,SD}$.

Another observation from Fig.~\ref{fig:compare_ave} is that there is a strong cancellation between the strange and charm quark contributions of $F_V$ (similar cancellations were also observed in the $D_s D_s^*\gamma$ couplings \cite{Donald:2013sra,Pullin:2021ebn}, which correspond to pole residues in the $D_s\to\ell\bar{\nu}\gamma$ form factors). Additionally, although results for $F_V^{(c)}$ and $F_V^{(s)}$ agree between averaging and not averaging, there is a slight tension for $F_V$ at small $x_\gamma$. 
Recall that our updated analysis method involves first fitting the $F_V^{(c)}(t_H, T)$ and $F_V^{(s)}(t_H, T)$ data and then taking linear combinations of the fit results to extract $F_V$. 
To ensure that fitting $F_V^{(c)}(t_H, T)$ and $F_V^{(s)}(t_H, T)$ first and then taking linear combinations does not introduce systematic uncertainties in the results for $F_V$, we also did the analysis performing fits to $F_V(t_H, T)$ directly. We found that the results for $F_V$ between the two analysis methods were consistent within errors, and from this conclude that the tension in Fig.~\ref{fig:compare_ave} is the result of a statistical fluctuation. Furthermore, we also found that fitting $F_V^{(c)}(t_H, T)$ and $F_V^{(s)}(t_H, T)$ first resulted in slightly smaller statistical uncertainties than fitting $F_V(t_H, T)$ directly.
\FloatBarrier

\subsection{Comparing different methods to calculate \texorpdfstring{$F_{A,SD}$}{F_A,SD}}
\label{ssec:sub_pgamma_zero_FA_SD}

In this section, we compare three different methods for calculating $F_{A,SD}$, denoted using the superscripts I, II, and III, as in $F_{A,SD}^\text{I}, F_{A,SD}^\text{II},$ and $F_{A,SD}^\text{III}$, to differentiate between them. We use method III in the final analysis presented in Sec.~\ref{sec:final_procedure}.

\begin{figure}[h]
    \centering
    \includegraphics[width=0.8\textwidth]{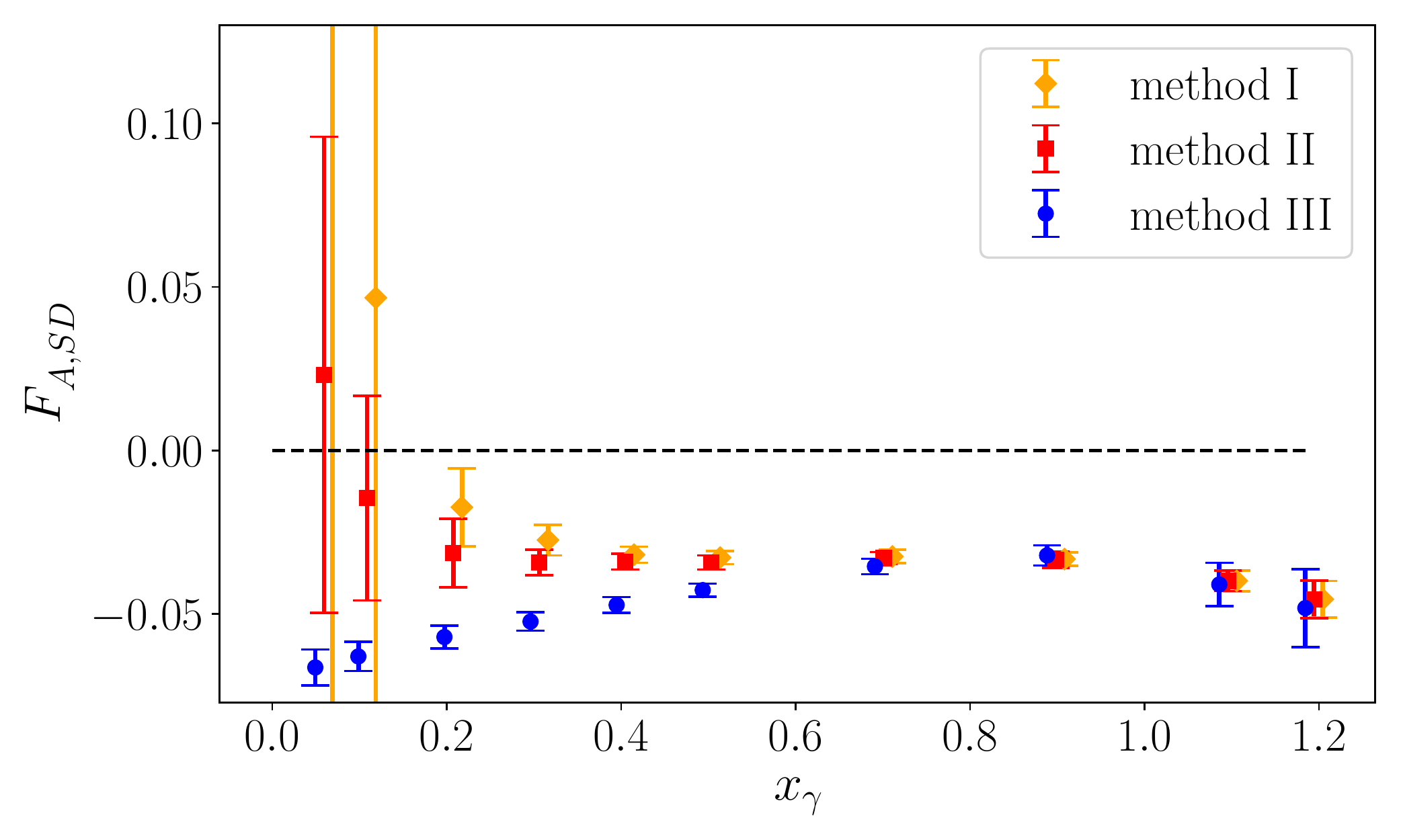}
    \caption{$F_{A,SD}$ as a function of $x_\gamma$ calculated using methods I, II, and III. Method III is significantly more precise at small $x_\gamma$. Methods I and II disagree with method III at smaller $x_\gamma$, due to $\mathcal{O}(a^n/x_\gamma)$ discretization effects.}
    \label{fig:compare_subtraction_FA_SD}
\end{figure}

Using method I, the structure-dependent part of the axial form factor is first calculated as a function of $t_H$ and $T$, denoted $F_{A,SD}(t_H, T)$. It is calculated as $F_{A,SD}(t_H, T) = F_A(t_H, T) - (-Q_\ell f_H(t_H, T)/E_\gamma^{(0)})$, where $F_A(t_H, T)$ and $f_H(t_H, T)$ are extracted from appropriate linear combinations of the time-integrated correlation function. We then fit $F_{A,SD}(t_H, T)$ to take the $T \to \infty$ and $t_H \to -\infty$ limits. To understand how the fits to $F_{A,SD}(t_H, T)$ are performed, we must first understand what intermediate states contribute in the spectral decomposition. In particular, subtleties appear when considering the time ordering $t_{\text{em}}<0$ and $t_W>0$. As explained in Sec.~\ref{section:fit_method}, the lowest-energy intermediate state that contributes to the weak axial-vector component of the time-integrated correlation function for these time orderings is the initial-state pseudoscalar meson $H$. Looking at the form of the spectral decomposition in Eqs.~\eqref{eq:Imunu_spectral_lessthan} and~\eqref{eq:Imunu_spectral_decomp_EM_greaterthan}, for $\vec{p}_H=\vec{0}$, each term in the sum is proportional to $\bra{0} J_\nu^\text{weak}(0) \ket{n(-\vec{p}_\gamma)}$. In this case the state is $n=H$, and this matrix element is the definition of the pseudoscalar decay constant, \textit{i.e.} $\bra{0} J_\nu^\text{weak}(0) \ket{H(-\vec{p}_\gamma)} \sim i (p_\gamma)_\nu f_H$. This implies that the state $n=H$ only contributes for indices $\nu$ with $(p_\gamma)_\nu \neq 0$. Our analysis uses $\vec{p}_\gamma$ in the $\hat{z}$-direction, and so this state only contributes in the $\nu=0,3$ matrix elements. The axial form factor $F_A$ is extracted using the indices $(\mu, \nu) \in \{(1,1), (2,2)\}$, and the decay constant is extracted using the the indices $(\mu, \nu) \in \{(0,0), (3,0)\}$. Therefore, when fitting $F_{A,SD}(t_H, T) = F_A(t_H, T) - (-Q_\ell f_H(t_H, T)/E_\gamma^{(0)})$, the lowest-energy state that appears for the $t_{\text{em}}<0$ and $t_W>0$ time orderings is $n=H$. The unwanted exponential in this case decays in $T$ according to the combination of energies $E_\gamma + E_{H,\vec{p}_\gamma} - m_B$, which approaches zero as $\vec{p}_\gamma \to 0$. An example of the behavior in $T$ for $p_{\gamma,z}=0.6 \times 2\pi/L$ is given in Fig.~\ref{fig:C3_compare_em_weak}. Looking at Fig.~\ref{fig:compare_subtraction_FA_SD}, $F_{A,SD}^\text{I}$ is precise at large $x_\gamma$, and the error increases dramatically for $x_\gamma \lesssim 0.1$. Part of the increase in error is due to the long extrapolation in $T$ performed for the $t_\text{em}<0$ and $t_W>0$ data. Another factor is that, for smaller $x_\gamma$, we did not observe stability in the $t_\text{em}<0$ data for any of the possible values of $T<|t_H|$. Specifically, stability was not observed for $F_{A,SD}^{(c)}$ with $p_{\gamma,z} \in 2\pi/L\{0.1, 0.2\}$, and for $F_{A,SD}^{(s)}$ with $p_{\gamma,z}\in 2\pi/L\{0.1, 0.2, 0.4, 0.6, 0.8, 1.0\}$. For these cases, we used only the $t_W>0$ data. Lastly, we observe that as $x_\gamma$ decreases, there are cancellations as large as 98\% between the two time orderings for the smallest $x_\gamma$. 

Method II improves upon the first by exactly subtracting the unwanted exponential contribution from the $n=H$ state to the $t_\text{em}<0$ and $t_W>0$ time orderings using a technique put forth in Ref.~\cite{Tuo:2021ewr}. The procedure follows from the observation that, when the energy of the lowest-energy intermediate state contributing to the spectral decomposition of the time-integrated correlation function is known, \textit{e.g.}~when determined by fitting to a two-point function, this information can be combined with the three-point correlation function to exactly subtract the unwanted exponential associated with that state. As a concrete example, consider the spectral decomposition for $I^{A,<}_{\mu 0}$ with $\vec{p}_H=\vec{0}$ and $ \vec{p}_\gamma = (0, 0, p_{\gamma,z})$. In this case, assuming ground-state saturation has been achieved for the interpolating field, the lowest-energy unwanted exponential when $n=H$ takes the form 
\begin{align}
    -\frac{\bra{0} J^A_0(0) \ket{H(-\vec{p}_\gamma)} \bra{H(-\vec{p}_\gamma)} J^\text{em}_\mu(0) |H(\vec{0})\rangle \langle H(\vec{0})| \phi^\dagger_H(0) \ket{0}}{2E_{H, \vec{p}_\gamma} 2m_H(E_\gamma + E_{H, \vec{p}_\gamma} - m_H)} e^{m_H t_H} e^{-(E_\gamma - m_H + E_{H, \vec{p}_\gamma})T},
    \label{eq:unwante_exp}
\end{align}
where $J_\nu^A(0)$ is the axial-vector component of the weak current. Because $I_{30}^{A,<}$ is the integral of $C^A_{3,\mu 0}(t_H, t_\text{em})$ over $t_\text{em}$, the spectral decompositions are equal up to the factor $(E_\gamma + E_{H\vec{p}_\gamma}-m_H)$ in the denominator. Therefore, the unwanted exponential in Eq.~\eqref{eq:unwante_exp} can be exactly subtracted by taking the combination
\begin{align}
    I_{\mu 0}^{A,<}(T, t_H) + C_{3,\mu \nu}^A(-T, t_H) \frac{e^{-E_\gamma T}}{E_\gamma + E_{H, \vec{p}_\gamma} - m_H}.
\end{align}
A similar procedure can be done for data with the electromagnetic current at the origin. Note that, for $t_W>0$, this combination subtracts the unwanted exponential corresponding to $n=H$ for the ground state as well as excited states created by the interpolating field. For $t_\text{em}<0$ however, the cancellation only occurs for the ground-state contribution. One must use a modified fit form to account for this given by
\begin{align}
    F_{A,SD}^<(T, t_H) = F^<_{A,SD} + B e^{-(E_\gamma-m_H+E^<)T} + A_{\rm exc} e^{-\Delta E(T+t_H)} e^{-(E_\gamma-m_H+E_{H,\vec{p}_\gamma})T} + C e^{\Delta E t_H},
\end{align}
where the term proportional to $A_{\rm exc}$ accounts for the imperfect cancellation when excited states contribute. Because the unwanted exponential with the smallest energy has been subtracted, the data plateaus more quickly with $T$. This allows one to fit earlier in $T$ to the more precise data, and also results in a shorter extrapolation in $T$. Additionally, fits to all $t_\text{em}<0$ data were stable for the allowed values of $T<|t_H|$. The results for $F_{A,SD}^\text{II}$ are shown in Fig.~\ref{fig:compare_subtraction_FA_SD}. As expected, the data agree for larger $x_\gamma$, and the error bars are significantly reduced for small $x_\gamma$. However, because the large cancellation between the two time orderings is still present, the error bars using this method also increase dramatically as one goes to smaller $x_\gamma$.

The third method, originally put forth in Ref.~\cite{Desiderio:2020oej}, calculates $F_{A,SD}$ by exploiting properties of the weak axial-vector three-point function at zero photon momentum. Further details and comments are given in Appendix~\ref{app:WI}. To summarize, one can extract the structure-dependent part of the axial form factor by replacing the original correlation function
\begin{equation}
\label{eq:3pt_unsub}
\int d^3x \int d^3y \,  e^{-i\vec p_\gamma \cdot \vec x}\, 
 \langle J_\mu^{\text{em}}(t_{\text{em}},\vec x) J^A_\nu(0)  \phi^\dagger_H(t_H,\vec y) \rangle
\end{equation}
by
\begin{equation}
\label{eq:3pt_submaintext}
\int d^3x \int d^3y \, \left( e^{-i\vec p_\gamma \cdot \vec x}-1\right)\, 
 \langle J_\mu^{\text{em}}(t_{\text{em}},\vec x) J^A_\nu(0)  \phi^\dagger_H(t_H,\vec y) \rangle, 
\end{equation}
for $\mu=\nu \in \{1,2\}$ (here we set $\vec{p}_H=0$), and similarly for the case in which the electromagnetic current is fixed at the origin. By applying the same steps previously used to extract $F_A(t_H, T)$ from Eq.~(\ref{eq:3pt_unsub}) to Eq.~(\ref{eq:3pt_submaintext}) instead, one directly obtains $F_{A,SD}(t_H, T)$. One advantage of this method is that $F_{A,SD}$ is extracted only using the $(\mu, \nu)\in \{(1,1), (2,2)\}$ indices, and so the state $n=H$ does not contribute to $t_\text{em}<0$ and $t_W>0$ data. This implies that the data for these time orderings will plateau more quickly in $T$, and a shorter extrapolation is required. Additionally, using this method results in at most a $50\%$ cancellation between the time orderings at the smallest $x_\gamma$. Looking at Fig.~\ref{fig:compare_subtraction_FA_SD}, these factors lead to significantly more precise results at small $x_\gamma$. One downside to this method was that stability was not observed for the $t_W<0$ time ordering for any photon momentum, and we only used the $t_\text{em}>0$ data. For this reason, the results for $F_{A,SD}^\text{III}$ at large $x_\gamma$ are less precise than the other two methods.

Another advantage of method III has to do with discretization effects. It was shown in Ref.~\cite{Desiderio:2020oej} that subtracting the point-like contribution to $F_A$ using the decay constant $f_H$ calculated in the usual way from a two-point function results in $\mathcal{O}(a^n/x_\gamma)$ discretization effects, in spite of the naive expectations based on the lattice Ward identity. Calculating $F_{A,SD}$ using method III, however, was shown to avoid this problem \cite{Desiderio:2020oej}, with only discretization errors of the form $\mathcal{O}(a^2)$. While methods I and II extract $f_H$ using the time-integrated three-point function and not the two-point function, we observe that those procedures still result in discretization errors of the form $a^n/x_\gamma$. In fact, since the axial form factor, $F_A$, and the decay constant, $f_H$, are computed from different combinations (see above) of the time-integrated-correlation-function components, which, we note, are not related by $H(3)$ symmetry and have their own lattice artifacts, residual discretization effects that scale as $\sim 1/x_\gamma$ will survive in $F_{A,SD}$ once the point-like part of the axial form factor is subtracted. On the contrary, in method III a unique combination of the time-integrated-correlation-function components is involved, leading to a complete cancellation of the unphysical, infrared-divergent contribution to $F_{A,SD}$ at finite cutoff. This is corroborated by the findings of Fig.\ref{fig:compare_subtraction_FA_SD}, where the results for $F_{A,SD}^\text{III}$ agree with  methods I and II at large $x_\gamma$, but disagree for $x_\gamma < 0.6$. In the light of the above considerations we choose to use method III in our final analysis presented in Sec.~\ref{sec:final_procedure}. 

\FloatBarrier

\subsection{Comparing the different three-point function analyses}
\label{ssec:compare_EM_WEAK}
In this section, we compare form factor results calculated using the three-point functions in Eq.~\eqref{eq:three_point} and Eq.~\eqref{eq:C3_EM}. For the comparison, we perform fits to the individual data sets, as well as simultaneous fits to both sets of data. Note that, for the strange-quark electromagnetic-current contribution to the $t_W<0$ time ordering of $F_{A,SD}$, we found that the fit results were not stable for any allowed values of integration range $T$. For that particular data set we therefore only used $t_\text{em}>0$ data. We begin this section by providing theoretical arguments for which method will be more precise at extreme values of $x_\gamma$, and conclude by discussing the form-factor results.

Starting with the $t_{\text{em}}<0$ data, as the integration range $T$ is increased, the maximum distance between any of the three operators is fixed by the source-sink separation $t_H$. 
For the $t_W>0$ data however, the maximum distance is given by $T+|t_H|$, which increases with $T$. 
This implies that the signal for the $t_W>0$ data will decrease with $T$, while the signal for the $t_\text{em}<0$ data will be relatively constant with $T$. 
Because the unwanted exponentials for the $t_\text{em}<0$ and $t_W>0$ data decay more quickly as $\vec{p}_\gamma$ is increased, one must fit larger values of $T$ for small $\vec{p}_\gamma$. 
Taken together, these facts imply that the $t_\text{em}<0$ data will be more precise than the $t_W>0$ data, with a larger relative improvement for small $p_\gamma$. 
Similar arguments can be made for the $t_{\text{em}}>0$ and $t_W<0$ data, except that as $\vec{p}_\gamma$ is increased, the unwanted exponentials decay more slowly with $T$, and the roles of the $t_{\text{em}}>0$ and $t_W<0$ are flipped with regards to the minimum distance between the operators. Therefore, the improvement in precision of the $t_W<0$ data over the $t_{\text{em}}>0$ data will be more significant at large $x_\gamma$.

\begin{figure}[h]
    \centering
    \begin{minipage}{0.48\textwidth}
        \centering
        \includegraphics[width=\textwidth]{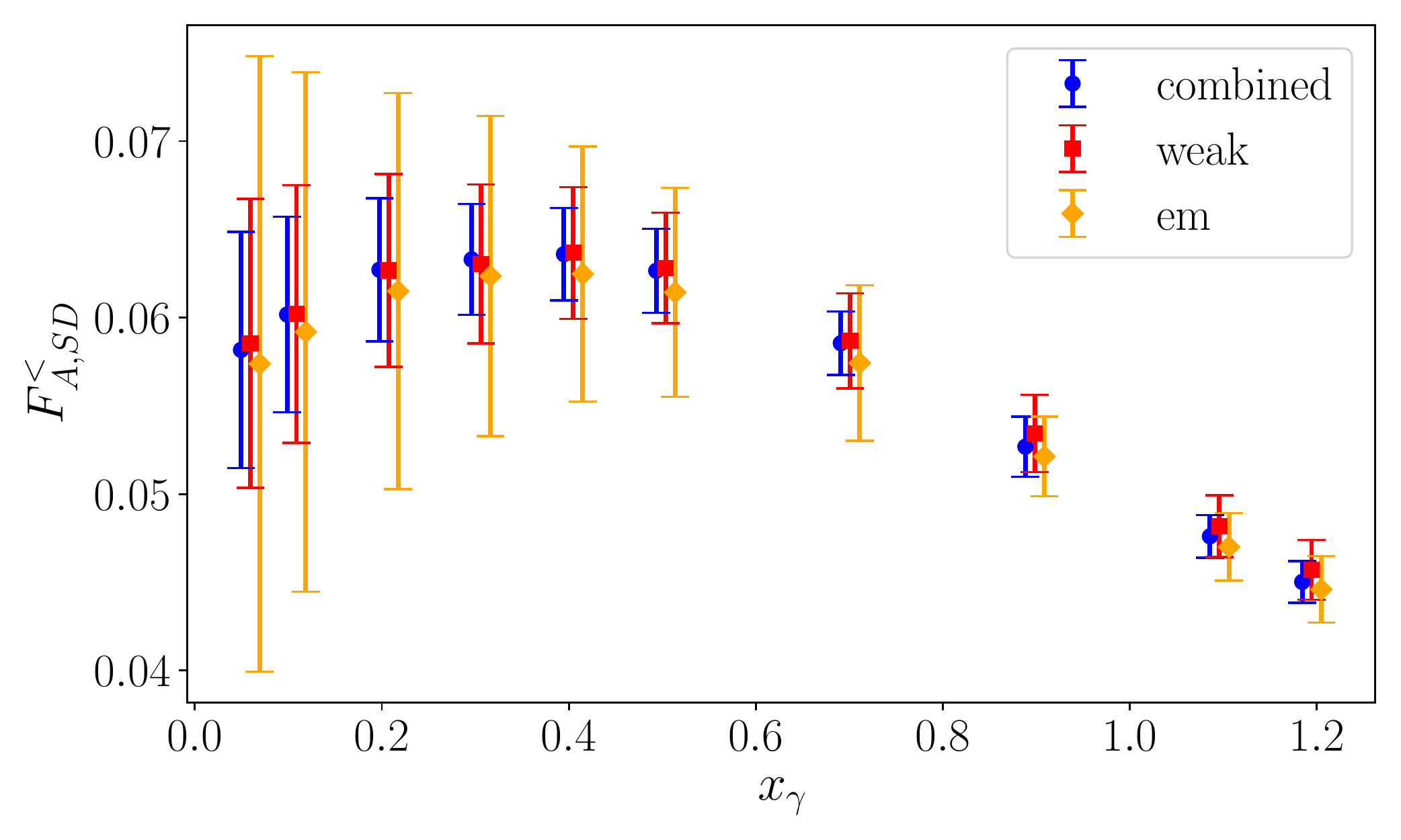}
    \end{minipage}
    \begin{minipage}{0.48\textwidth}
        \centering
        \includegraphics[width=\textwidth]{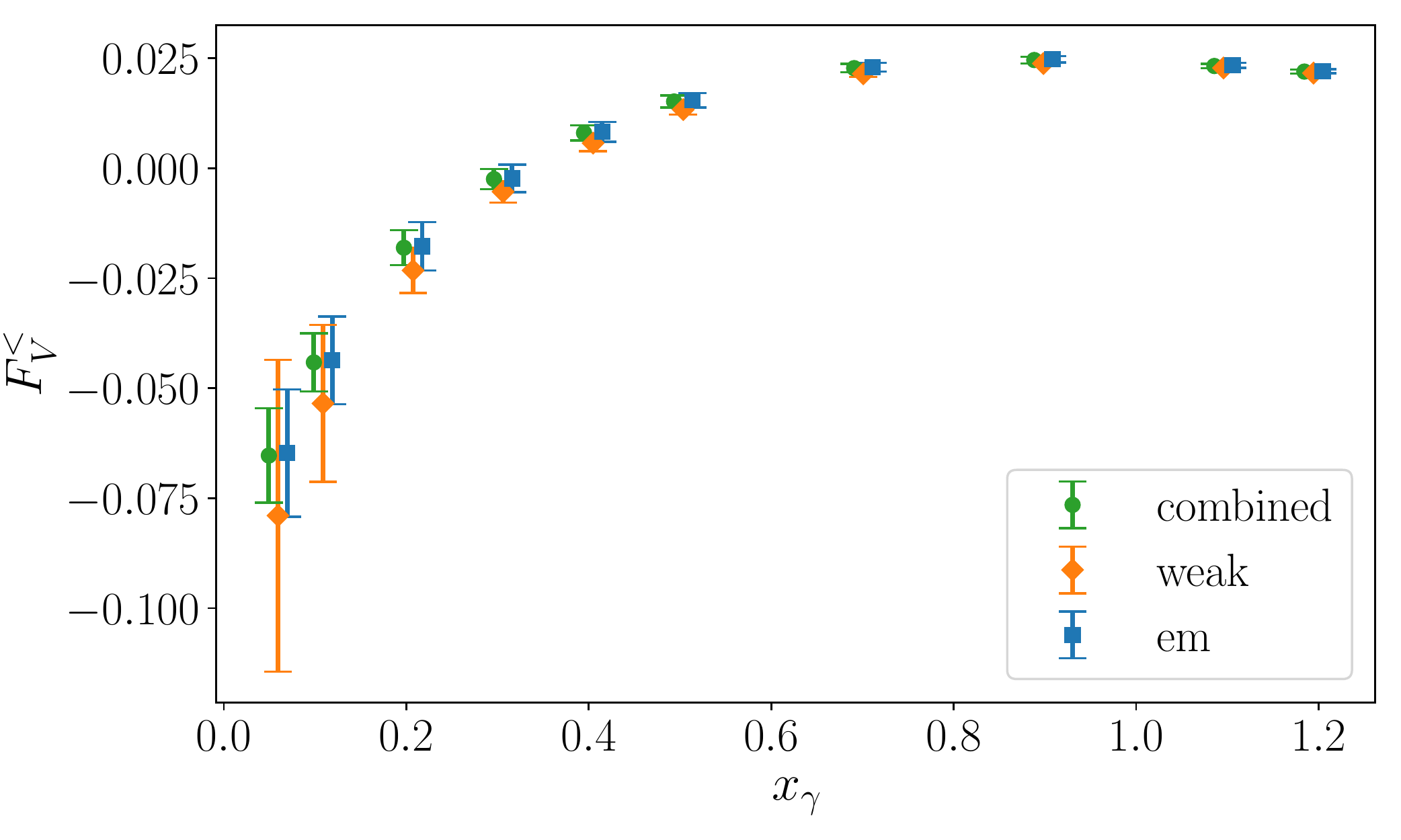}
    \end{minipage}

    \begin{minipage}{0.48\textwidth}
        \centering
        \includegraphics[width=\textwidth]{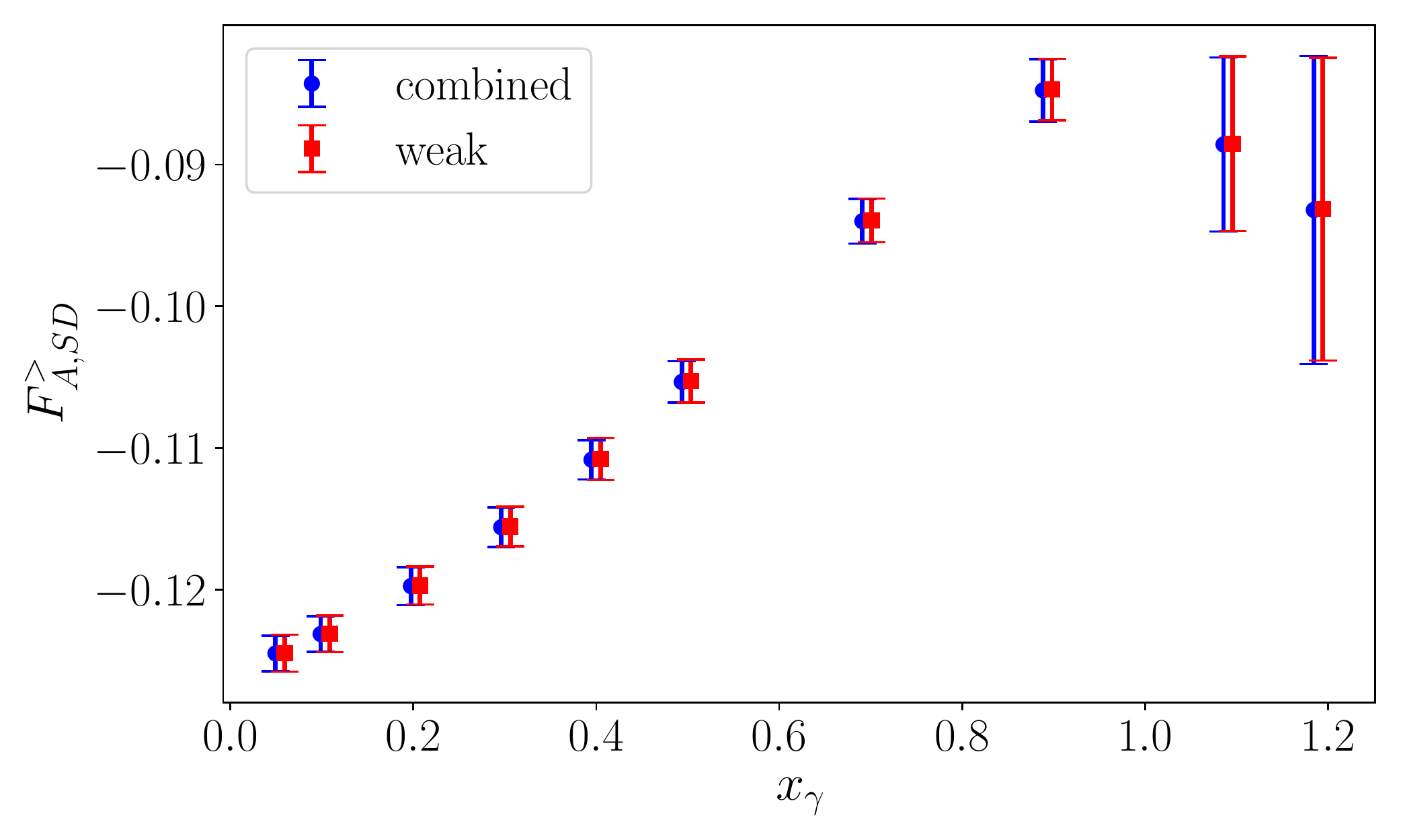}
    \end{minipage}
    \begin{minipage}{0.48\textwidth}
        \centering
        \includegraphics[width=\textwidth]{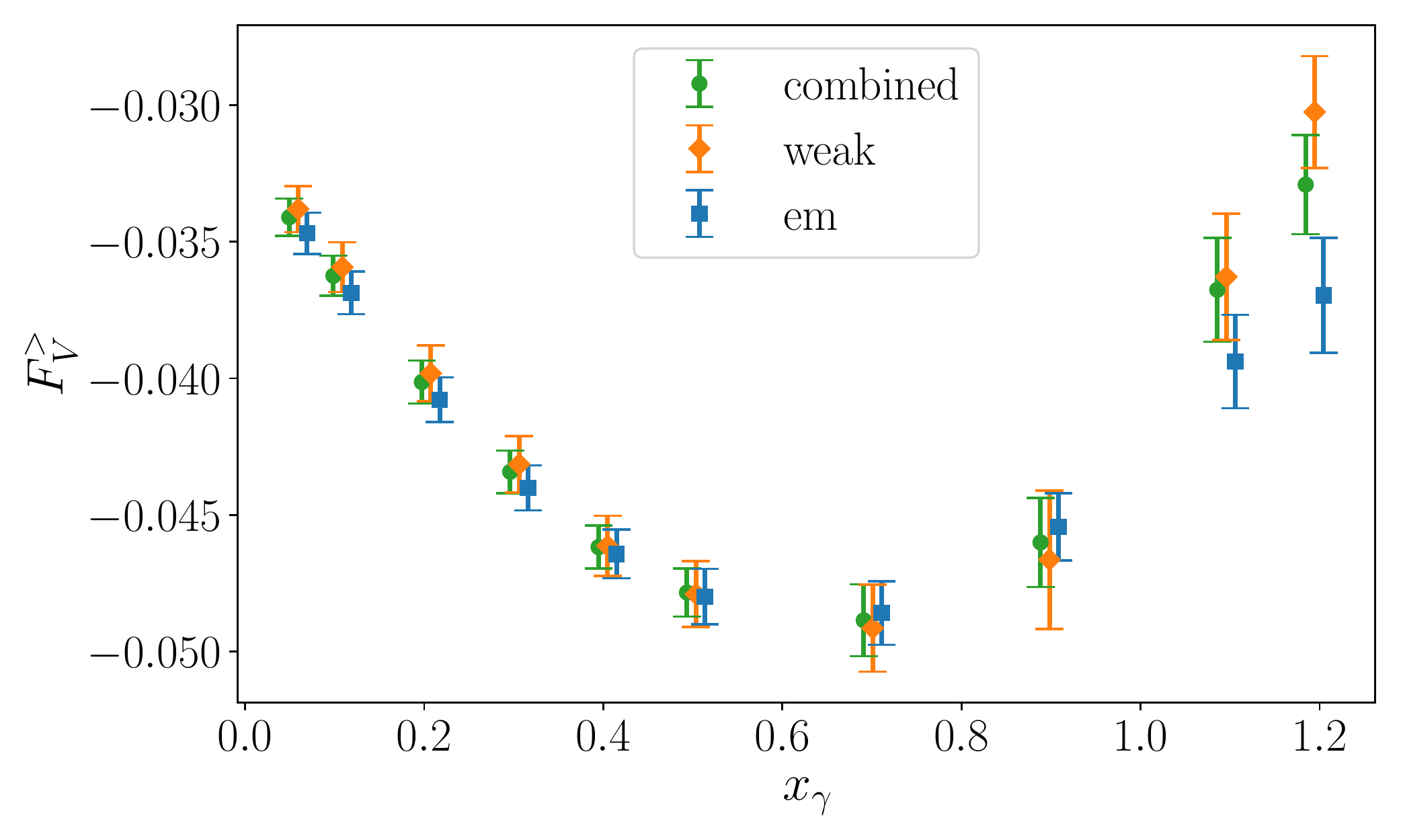}
    \end{minipage}

    \begin{minipage}{0.48\textwidth}
        \centering
        \includegraphics[width=\textwidth]{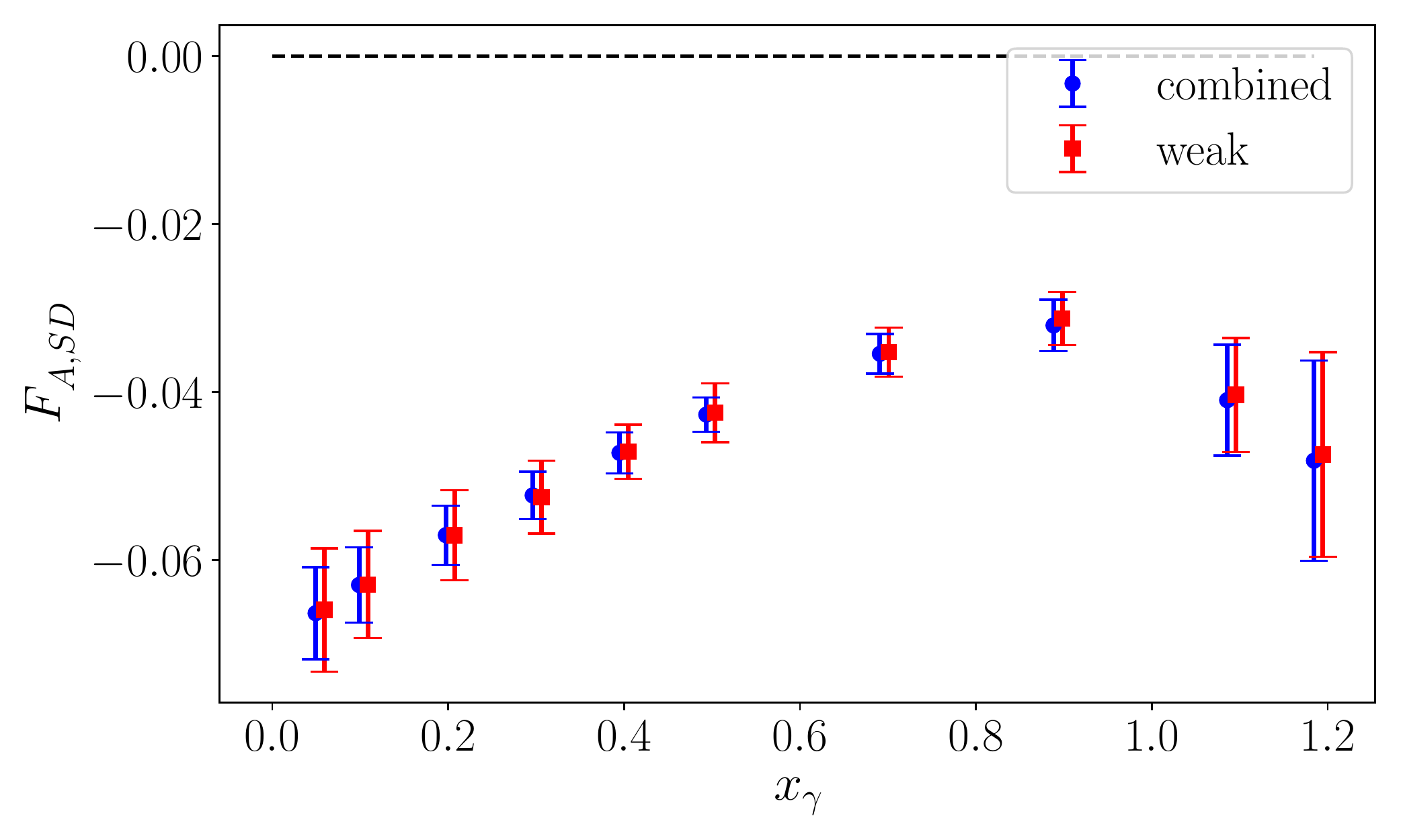}
    \end{minipage}
    \begin{minipage}{0.48\textwidth}
        \centering
        \includegraphics[width=\textwidth]{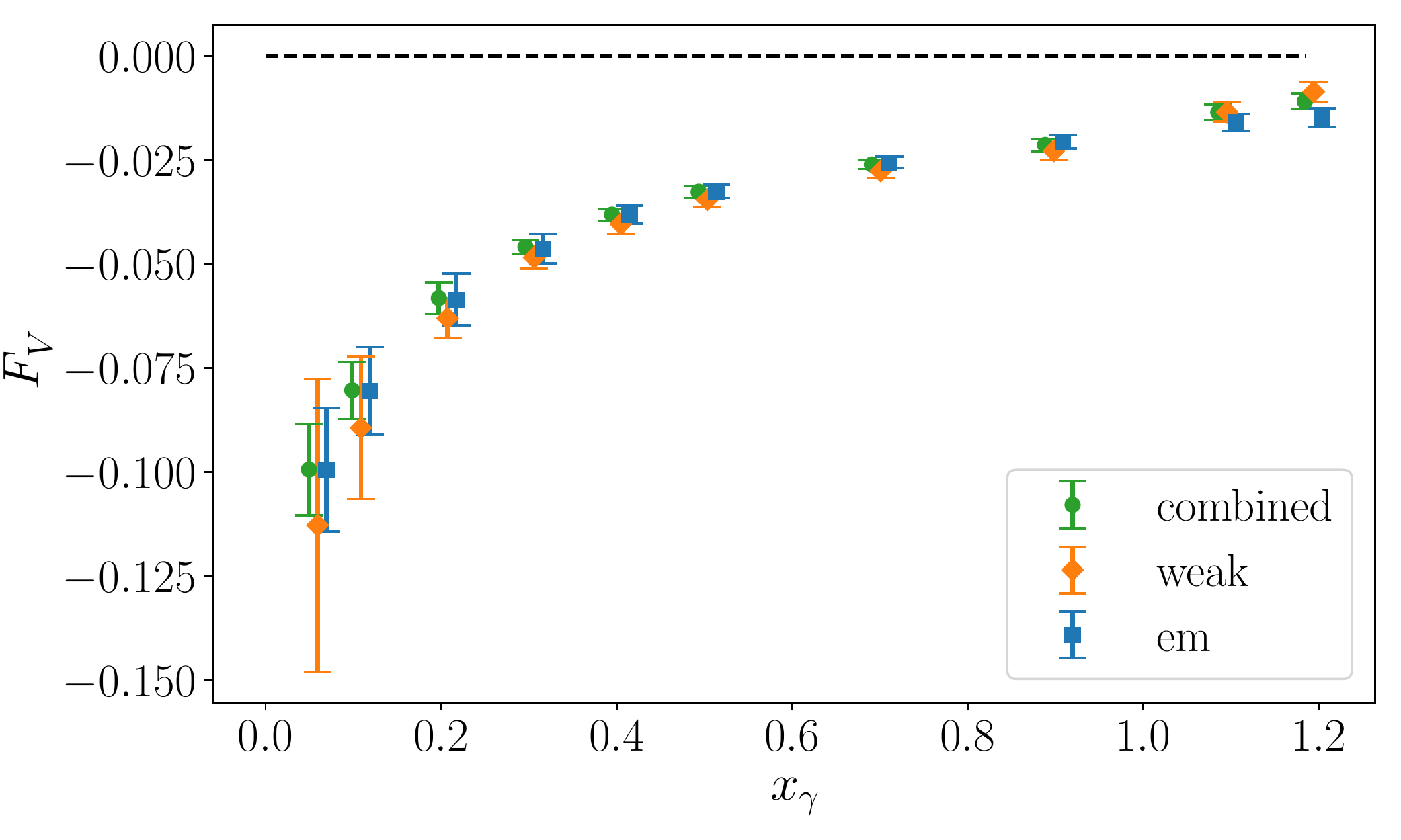}
    \end{minipage}
    \caption{Left(right) column compares $F_{A,SD}(F_V)$ calculated using only the three-point function in Eq.~\eqref{eq:three_point} (labeled ``weak''),  only the three-point function in Eq.~\eqref{eq:C3_EM} (labeled ``em''), and a simultaneous fit of both data sets (labeled ``combined''). The different rows show the full form factors, as well as the individual-time-ordering contributions. The vertical axis label $F_V^<$ indicates the $t_\text{em}<0$ and $t_W>0$ data, and the label $F_V^<$ indicates the $t_\text{em}>0$ and $t_W<0$ data; similar labels are used for $F_{A,SD}$. The $t_\text{em}<0$ data is more precise than the $t_W>0$ data for small $x_\gamma$, and the $t_W<0$ data is more precise than the $t_\text{em}>0$ data for large $p_\gamma$.}
    \label{fig:FA_SD_vx_xgamma_compare_weak_vs_em}
\end{figure}

Figure~\ref{fig:FA_SD_vx_xgamma_compare_weak_vs_em} shows the different time orderings of $F_{A,SD}$ and $F_V$ as a function of $x_\gamma$ determined using each method separately, as well as from a combined analysis. As expected, at small $x_\gamma$, results using $t_\text{em}<0$ data are more precise than using $t_W>0$ data for both $F_V$ and $F_{A,SD}$. Looking at the vector form factor, for larger $x_\gamma$, we observe that the $t_W<0$ result is more precise than the $t_\text{em}>0$ result. While we cannot perform the same comparison for $F_{A,SD}$, we observed a similar trend for the charm-quark-current contribution to the $t_W<0$ and $t_\text{em}>0$ time ordering of $F_{A,SD}$.

To summarize, data calculated using either of the three-point functions in Eq.~\eqref{eq:three_point} or Eq.~\eqref{eq:C3_EM} has inherent limitations to the precision that can be achieved at the extreme values of $x_\gamma$. However, we found that performing combined fits to both sets of data allows us to achieve a high precision for both small and large $x_\gamma$. Additionally, for intermediate $x_\gamma$ values, we see an overall improvement compared to a single method. Lastly, because we had to discard the $t_W<0$ data for the strange-quark-current contribution for $F_{A,SD}$, it was crucial to the analysis that we performed the calculation using both methods.

\FloatBarrier

\section{Form factor results using all improvements}
\label{sec:final_procedure}

\begin{figure}
    \includegraphics[width=\textwidth]{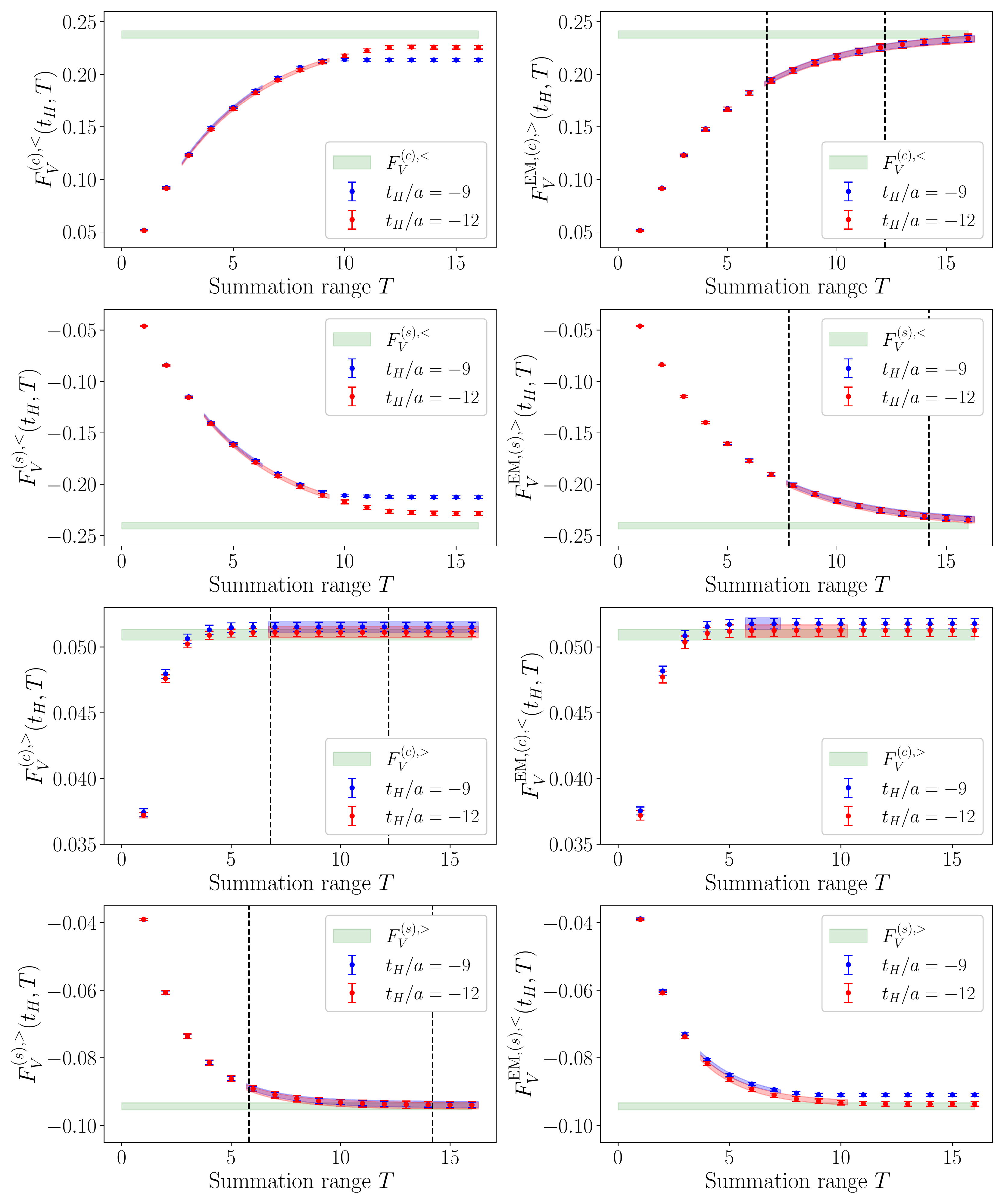}
    \caption{Fit results for $F_V$ with $\vec{p}_\gamma = 2\pi/L(0, 0, 0.6)$. The green horizontal band is the one-sigma region of the desired constant term in the fit form. The blue and red bands are the one-sigma bands of the fits as a function of $T$ for $t_H/a=-9$ and $t_H/a=-12$, respectively. For the $t_\text{em}>0$ and $t_W>0$ time orderings, the black vertical dashed lines indicate the fit range used. For the $t_\text{em}<0$ and $t_W<0$ time orderings, the error bands are only shown for data included in the fit.}
    \label{fig:3dprime_FV_stability}
\end{figure}

\begin{figure}
    \includegraphics[width=\textwidth]{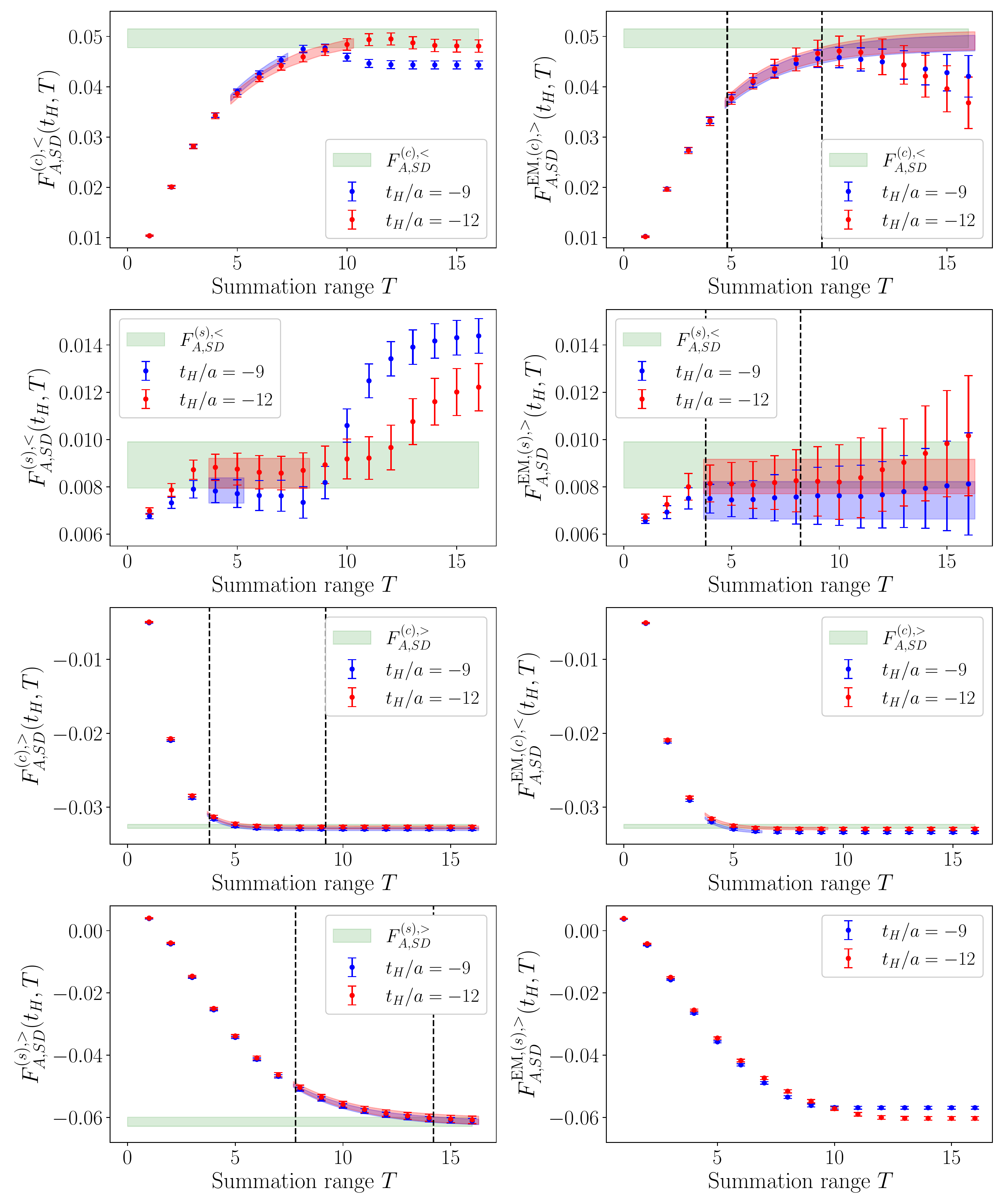}
    \caption{Fit results for $F_{A,SD}$ with $\vec{p}_\gamma = 2\pi/L(0, 0, 1.4)$. The green horizontal band is the one-sigma region of the desired constant term in the fit form. The blue and red bands are the one-sigma bands of the fits as a function of $T$ for $t_H/a=-9$ and $t_H/a=-12$, respectively. For the $t_\text{em}>0$ and $t_W>0$ time orderings, the black vertical dashed lines indicate the fit range used. For the $t_\text{em}<0$ and $t_W<0$ time orderings, the error bands are only shown for data included in the fit. Fits to $F_{A,SD}^{\text{EM}, (s), >}(t_H, T)$ were not stable for any allowed fit ranges, and this data was not included in the final analysis.}
    \label{fig:3dprime_FA_SD_stability}
\end{figure}

\begin{figure}
    \centering
    \begin{minipage}{0.48\textwidth}
        \centering
        \includegraphics[width=\textwidth]{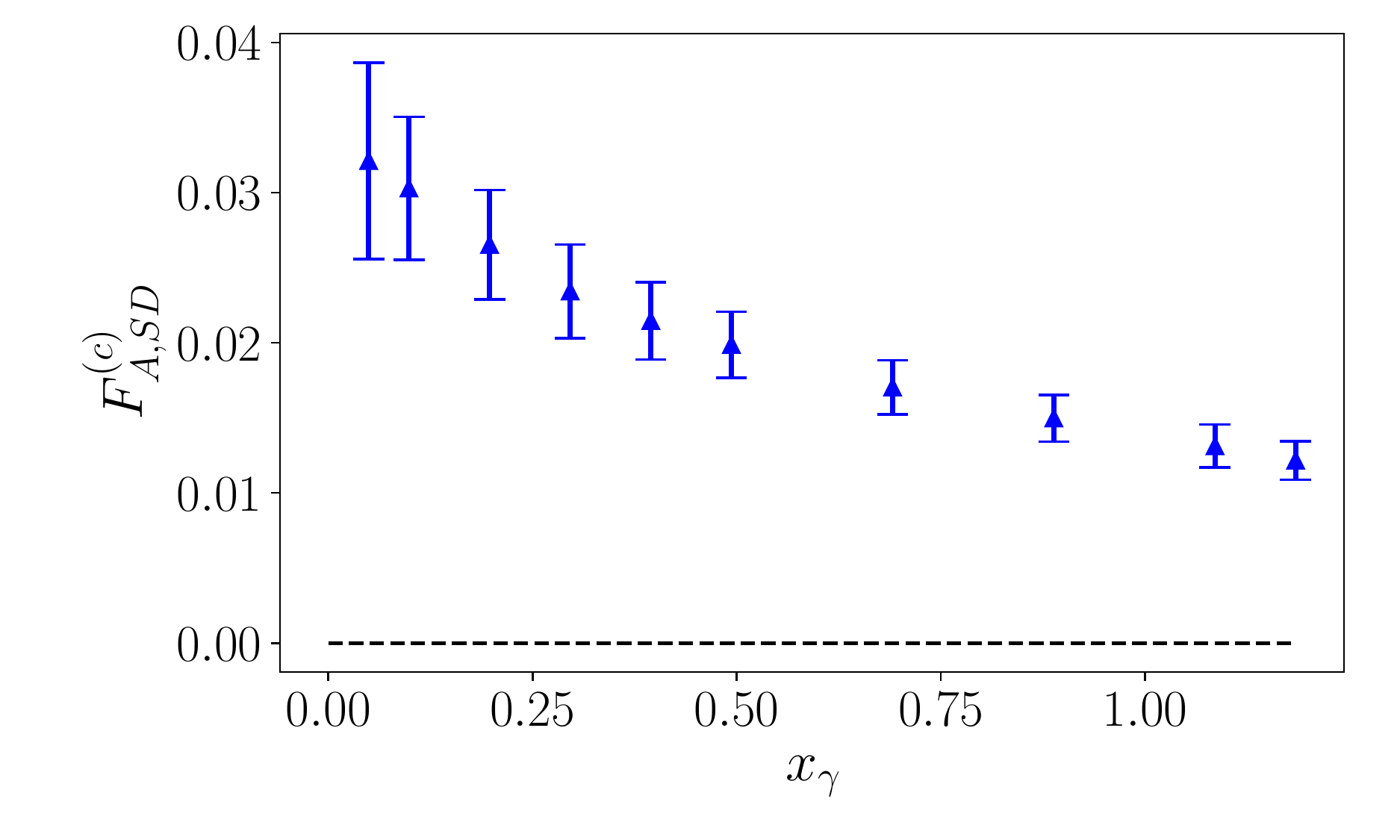}
    \end{minipage}
    \begin{minipage}{0.48\textwidth}
        \centering
        \includegraphics[width=\textwidth]{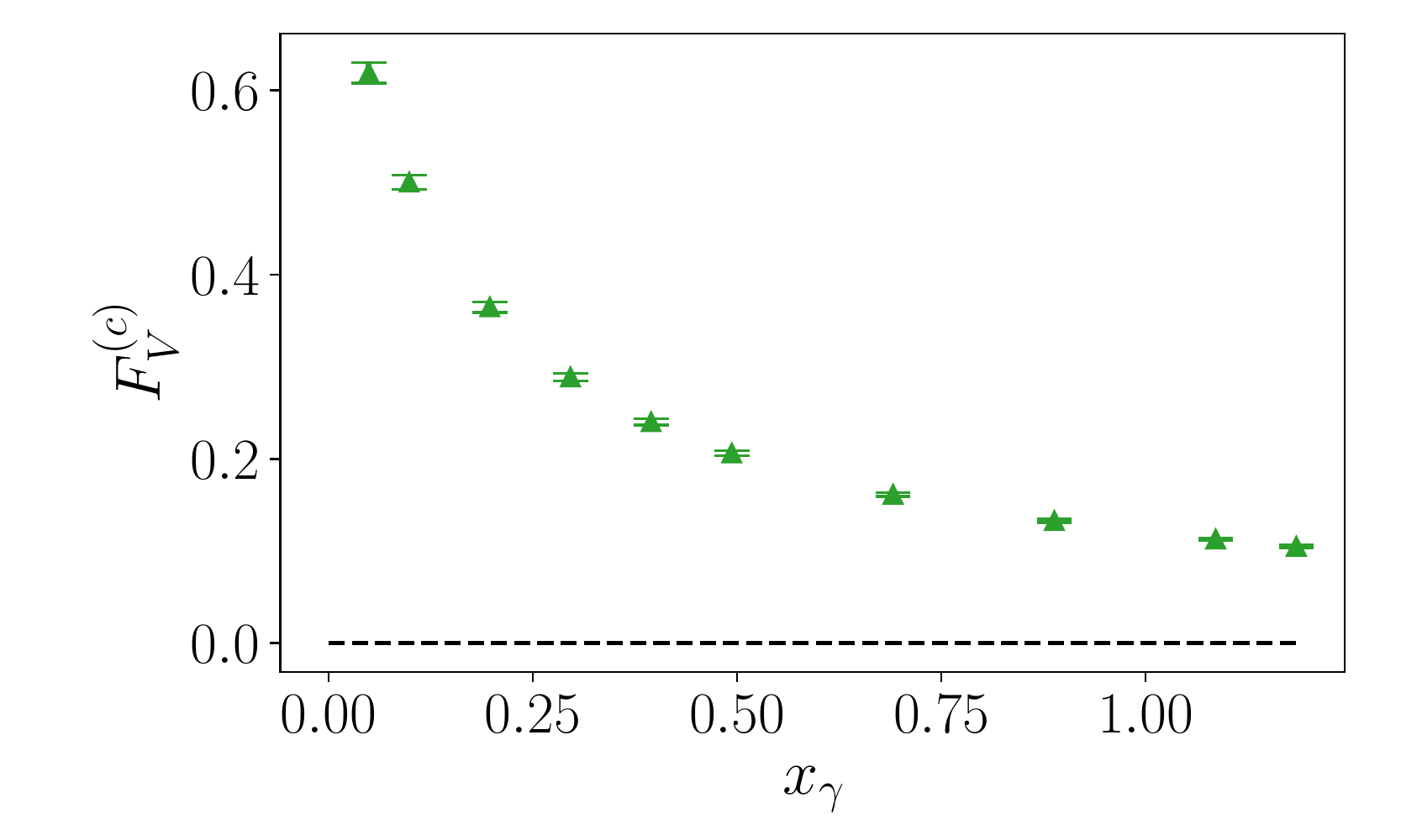}
    \end{minipage}
    
    \begin{minipage}{0.48\textwidth}
        \centering
        \includegraphics[width=\textwidth]{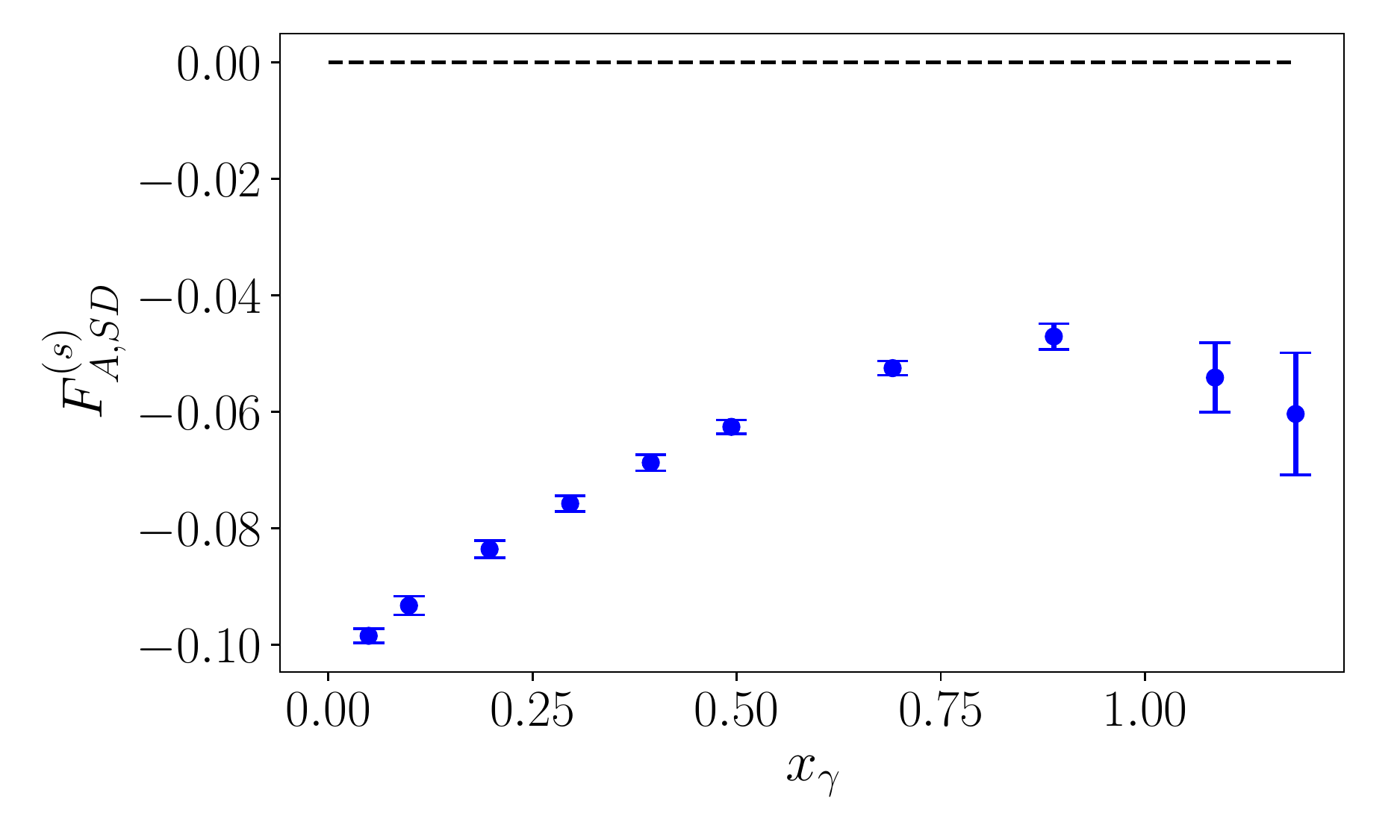}
    \end{minipage}
    \begin{minipage}{0.48\textwidth}
        \centering
        \includegraphics[width=\textwidth]{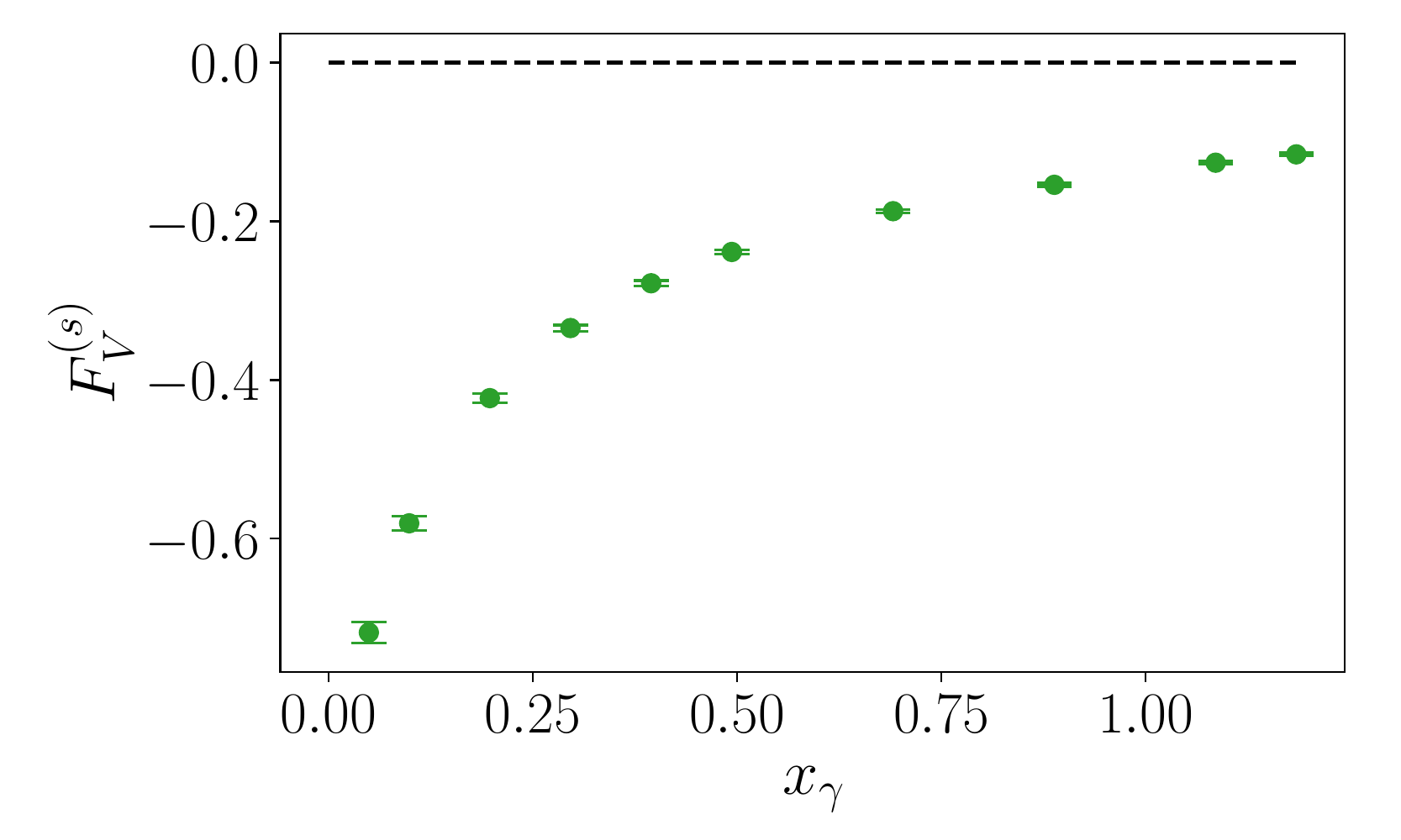}
    \end{minipage}
    
    \begin{minipage}{0.48\textwidth}
        \centering
        \includegraphics[width=\textwidth]{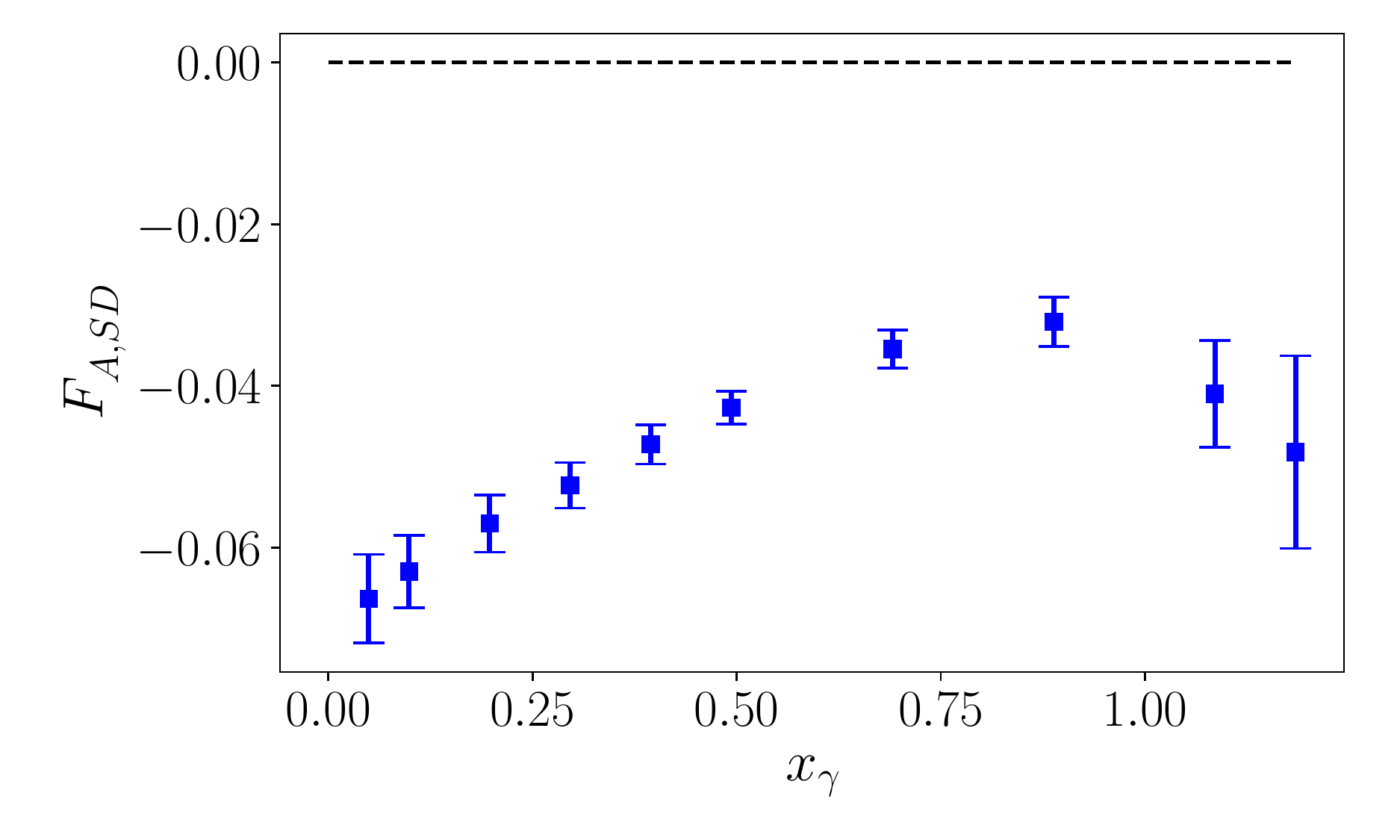}
    \end{minipage}
    \begin{minipage}{0.48\textwidth}
        \centering
        \includegraphics[width=\textwidth]{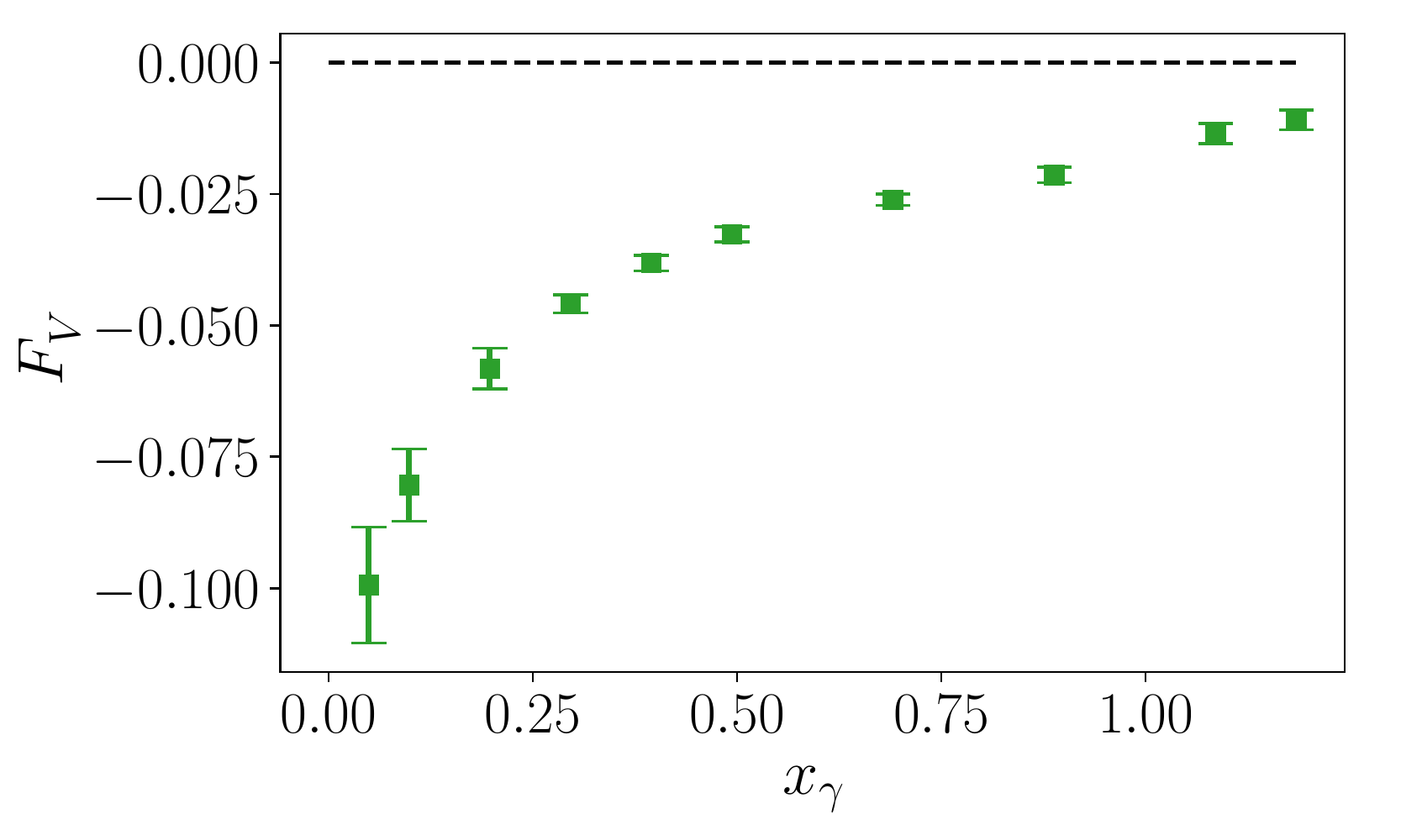}
    \end{minipage}
    \caption{Results of $F_{A,SD}$ and $F_V$, as a function of $x_\gamma$ calculated using the complete analysis method. Also shown are the contributions from the individual quark flavors in the electromagnetic current. The data shown in these plots are also provided in machine-readable files \cite{supplemental}. Note that these results are from a single gauge-field ensemble, and thus not yet extrapolated to the continuum limit and physical pion mass.}
    \label{fig:3dprime_FF_vs_xgamma}
\end{figure}

In this section we summarize the improved methods used to extract the form factors. 

Using the 3d method, we calculate the three-point functions with the weak and electromagnetic currents at the origin using the infinite-volume approximation method described in Sec.~\ref{ssec:infinite_volume_approximation}. Calculations are performed in the rest frame of the meson for photon momenta in the $\hat{z}$ direction $p_{\gamma,z} \in 2\pi/L\{0.1, 0.2, 0.4, 0.6, 0.8, 1.0, 1.4, 1.8, 2.2, 2.4\}$. The three-point functions are averaged over positive and negative photon momentum. We also calculate the three-point function using $\mathbb{Z}_2$ random-wall sources for photon momentum $p_{\gamma,z} \in 2\pi/L\{0, 1\}$, and take ratios with the point source data as explained in Sec.\ref{ssec:ratio_method}. We then extract the form factors as a function of $t_H$ and $T$ by taking linear combinations of the improved time-integrated correlation functions. 

Contrary to the analysis methods outlined in Sec.~\ref{section:fit_method}, in the final analysis we perform stability-test fits to the two individual quark components of the electromagnetic-current contributions to the form factors; the full form factors are obtained by summing the fit results of these contributions. As before, we check that the fit result of an individual data set is stable under variations of the fit range. This stability analysis is first performed to the data with the weak and EM currents at the origin separately. For the $t_\text{em}<0$ and $t_W<0$ data, we search for stability under variations of the minimum fit ranges $T^{<}_\text{min}$ and $T^{<,\text{EM}}_\text{min}$, as well as the distances from the interpolating field $T^{<}_\text{max}+t_H$ and $T^{<,\text{EM}}_\text{max}+t_H$. For the $t_\text{em}>0$ and $t_W>0$ data on the other hand, stability is only checked under variations of the minimum fit ranges $T^>_\text{min}$ and $T_\text{min}^{>,\text{EM}}$. The stable fit ranges determined from these individual fits are then used to perform simultaneous fits to the $t_{\text{em}}<0 (t_{\text{em}}>0)$ and $t_W>0 (t_W<0)$ data. 
To check that the combined fits are also stable, we perform fits to a number of different fit ranges varied about these chosen fit ranges. In particular, we vary each of the three possible fit ranges individually by $-1, 0,$ and $+1$, resulting in 27 total fits. For simultaneous fits to the $t_\text{em}<0$ and $t_W>0$ data, the three fit ranges we vary are $T^{<}_\text{min}$, $T_\text{min}^{>,\text{EM}}$, and $T_\text{max}^{<}+t_H$, and for simultaneous fits to the $t_\text{em}>0$ and $t_W<0$ data, the three fit ranges we vary are $T_\text{min}^>, T_\text{min}^{<,\text{EM}}$, and $T_\text{max}^{<, \text{EM}}+t_H$.
Because we found that performing global fits to all $x_\gamma$ did not significantly improve the precision of the extrapolated values for the form factors when using the 3d method, we now take the stable fit range to the fits at a single $x_\gamma$ as the final value. The most detailed fit forms used in the analysis for the $t_{\text{em}}<0$ and $t_W>0$ data are given by
\begin{align}
    F^{(q),<}(t_H, T) &= F^{(q),<} + B_{F^{(q)}}^{<}(1+ B^{<}_{F^{(q)}, \text{exc}} e^{\Delta E (T+t_H)}) e^{-(E_\gamma - E_{H} + E_{F^{(q)}}^<)T} + C^{<}_{F^{(q)}} e^{\Delta E t_H},
    \\
    F^{(q), >, \text{EM}}(t_H, T) &= F^{(q),<} + B_{F^{(q)}}^{<} e^{-(E_\gamma - E_{H} + E_{F^{(q)}}^<)T} + C^{<,\text{EM}}_{F^{(q)}} e^{\Delta E t_H},
    \label{eq:3dprime_fit_form_neg}
\end{align}
where $F = F_V, F_{A,SD}$. Note that these fit forms have the parameters $F^{(q),<}, B_{F^{(q)}}^<, E_H, E_{F^{(q)}}^<$ and $\Delta E$ in common. The fit forms for the $t_{\text{em}}>0$ and $t_W<0$ data are given by
\begin{align}
    F^{(q),>}(t_H, T) &= F^{(q),>} + B_{F^{(q)}}^{>} e^{(E_\gamma - E_{F^{(q)}}^>)T} + C^{>}_{F^{(q)}} e^{\Delta E t_H}
    \\
    F^{(q),<,\text{EM}}(t_H, T) &= F^{(q),>} + B_{F^{(q)}}^{>}\left(1+ B^{>}_{F^{(q)}, \text{exc}} e^{\Delta E (T+t_H)} \right)e^{(E_\gamma - E_{F^{(q)}}^>)T} + C^{>,\text{EM}}_{F^{(q)}} e^{\Delta E t_H},
    \label{eq:3dprime_fit_form_pos}
\end{align}
which have the parameters $F^{(q),>}, B_{F^{(q)}}^>, E_{F^{(q)}}^>$ and $\Delta E$ in common. In the cases where the data plateaus quickly in $T$ we use a fit form with $B_{F^{(q)}}^{(<),(>)}=0$. 

For fits to $F_V$, we calculate the vector-meson $(D_s^*)$ energies for all values of $\vec{p}_{D_s}-\vec{p}_\gamma$ using a lattice dispersion relation of the form
\begin{equation}
    E_{{H^*}} = m_{H^*} + \alpha |\vec{p}|^2 + \beta |\vec{p}|^4.
\end{equation}
The parameters $m_{H^*}, \alpha$ and $\beta$ are determined by performing fits to vector-meson energies determined from the associated vector-meson two-point functions for $|\vec{p}|^2 \in (2\pi/L)^2\{0, 1, 2, 3, 4\}$. For reasons explained in Sec.~\ref{ssec:infinite_volume_approximation}, this lattice dispersion relation is valid for momenta at non-integer multiples of $2\pi/L$ up to errors suppressed exponentially in the volume. The vector-meson energies are used as priors in fits for the $t_\text{em}<0$ and $t_W>0$ time orderings of $F_V$, with the prior equal to the central value and the prior width equal to the uncertainty of the fit result. The $D_s$ mass determined from fits to the associated two-point function is used as a prior in the fits, with the prior value and prior width equal to the central value and the uncertainty of the fit result, respectively. The excited-state energy gap $\Delta E$ between the ground state and the first excited state created by the interpolating field is extracted by performing two-exponential fits to the two-point function. The fit result for $\Delta E$ is used as a prior in the form-factor fits, with the prior equal to the central value and the prior width equal to the uncertainty of the fit result scaled by a factor 1.5.

The vector-form-factor fits are performed using the methods outlined in Sec.~\ref{section:fit_method}. Figure~\ref{fig:3dprime_FV_stability} shows examples of the error bands from fits to $F_V$ with $\vec{p}_\gamma = 2\pi/L(0, 0, 0.6)$.
We calculate the structure-dependent part of the axial form factor using the improved method in Sec.~\ref{ssec:sub_pgamma_zero_FA_SD}, and the fits to $F_{A,SD}(t_H, T)$ must be modified accordingly. The improved method involves taking combinations of correlation functions for the same $\mu, \nu$ indices of $I_{\mu \nu}$ at both non-zero photon momentum \textit{and} zero photon momentum. This combination will receive contributions from two sets of intermediate states, each with different momentum. Fitting each of these states is in general difficult. To get around this, we perform fits accounting only for the unwanted exponential that is expected to be dominant. For the $t_{\text{em}}>0$ and $t_W<0$ data, the unwanted exponentials decay more slowly as $p_\gamma$ is increased. The states with $p_\gamma=0$ are therefore sub-leading, and the fits to the $t_{\text{em}}>0$ and $t_W<0$ data only include the $p_\gamma \neq 0$ states. On the other hand, the unwanted exponentials for the $t_{\text{em}}<0$ and $t_W>0$ data decay more quickly as $p_\gamma$ is increased, and so the fits to this data only include the states with $p_\gamma=0$. Figure~\ref{fig:3dprime_FA_SD_stability} shows examples of the error bands from fits to $F_{A,SD}$ with $\vec{p}_\gamma = 2\pi/L(0, 0, 1.4)$. Note that fits to $F_{A,SD}^{\text{EM}, (s), <}(t_H, T)$ were not stable for any allowed values of the fit ranges for all values of $p_\gamma$; this data was therefore not included in the final analysis.

The results of $F_{A,SD}$ and $F_V$, as well as the individual quark electromagnetic current  current contributions to the form factors, are shown in Fig.~\ref{fig:3dprime_FF_vs_xgamma} as a function of $x_\gamma$. Note that these results are from a single ensemble, and still contain nonzero-lattice-spacing and unphysical-pion-mass systematic errors.

\section{Conclusions}
\label{sec:conclusions}
In this work, we presented a study of lattice-QCD data-generation and analysis methods to determine the form factors describing  radiative leptonic decays of pseudoscalar mesons. We calculated the relevant non-local matrix elements using the 3d, 4d, and 4d$^{>,<}$ methods, and performed fits to the data to remove unwanted exponentials in the sum over intermediate states and from excited states created by the meson interpolating field. We demonstrated that the 3d method offers good control over both types of unwanted exponentials for a significantly reduced number of propagator solves compared to the 4d and 4d$^{>,<}$ methods.

From there, we further improved upon the 3d method by calculating the three-point function using the infinite-volume approximation method, which allows us to access the full range of kinetically allowed photon momenta without having to perform calculations in the moving frame of the meson. We then showed that the hadronic tensor could be extracted using an alternate three-point function with the electromagnetic current at the origin, rather than the weak current at origin. The alternate three-point function can be calculated by reusing propagators required for the original three-point function. 
Performing simultaneous fits to both data sets resulted in reductions in statistical noise for both $F_{A,SD}$ and $F_V$, with the largest improvements at small and large $x_\gamma$. Furthermore, having both data sets increases the maximum possible fit range in $T$ for data used to calculate a given time ordering of the hadronic tensor. Calculating both data sets and exploiting this property was found to be crucial for extracting $F_{A,SD}$.

Further improvements in the statistical precision were achieved by multiplying the desired three-point function by ratios of three-point functions calculated using noise and point sources. This procedure resulted in significant improvements in precision for $F_{A,SD}$, and modest improvements for $F_V$. We also averaged the three-point functions over positive and negative photon momentum, which resulted in significant improvements in precision for $F_V$ at small $x_\gamma$, and modest improvements for $F_{A,SD}$. Lastly, we extracted $F_{A,SD}$ using a subtraction method that utilizes the properties of the three-point function as $p_\gamma \to 0$. This method has a number of advantages, including an increased precision at small $x_\gamma$, data  plateauing more quickly in $T$, and removal of $\mathcal{O}(a^n/x_\gamma)$ lattice artifacts that diverge for $x_\gamma \to 0$.
The optimal combination of methods yields results for the $D_s^+ \to \ell^+ \nu \gamma$ structure-dependent vector and axial form factors in the entire kinematic range with statistical plus fitting uncertainties of order 5\%, using 25 gauge configurations with 64 samples per configuration.

Using the improved lattice methods developed in this work, we plan to perform calculations on more ensembles and perform extrapolations to the physical pion mass and the continuum for the pion, kaon, $D_{(s)}$, and $B_{(s)}$ radiative-leptonic-decay form factors. For the $B_{(s)}$ decays, using the domain-wall action will require extrapolating in the mass. Alternatively, one could perform calculations at the physical $b$-quark mass using the ``relativistic heavy-quark action'' \cite{Christ:2006us}. In that context, a new non-perturbative method to tune the parameters of the relativistic heavy-quark action has been developed in Ref.~\cite{Giusti:2021rsf} from which extensions of the present study to $B$-meson physics could benefit. We also plan to calculate the contributions from the quark-disconnected diagrams.

Precise determinations of the QCD form factors for radiative leptonic decays are relevant for a number of phenomenological reasons. At small photon energies, a calculation of the radiative-leptonic decay rate is needed in order to include $\mathcal{O}(\alpha_\text{em})$ corrections to purely leptonic decays. At large photon energies, radiative leptonic decays are useful probes of the internal structure of the mesons as well as sensitive probes of physics beyond the Standard Model. Additionally, the methods and main outcomes presented in this study could be relevant for the lattice calculation of transition form factors describing the interaction between pseudoscalar mesons and two off-shell photons, since similar Euclidean correlation functions are involved. From such form factors, important information can be extracted on parton distribution amplitudes in hadrons (see, {\it e.g.}, Ref.~\cite{Bali:2018spj}), as well as on the hadronic light-by-light contribution to the muon anomalous magnetic moment (see, {\it e.g.}, Ref.~\cite{Gerardin:2019vio} and references therein).

\FloatBarrier

\section*{Acknowledgements}

S.M.~thanks Diego Guadagnoli for asking the question whether the form factors describing radiative leptonic decays are calculable on the lattice. We thank the RBC and UKQCD Collaborations for providing the gauge-field configurations. C.F.K.~is supported by the DOE Computational Science Graduate Fellowship under Award Number DE-SC0020347. S.M.~is supported by the U.S.~Department of Energy, Office of Science, Office of High Energy Physics under Award Number DE-SC0009913. A.S.~is supported in part by the U.S.~DOE contract \#DE-SC0012704. This research used resources provided by the National Energy Research Scientific Computing Center, a DOE Office of Science User Facility supported by the Office of Science of the U.S. Department of Energy under Contract No.~DE-AC02-05CH11231, and the Extreme Science and Engineering Discovery Environment (XSEDE) \cite{XSEDE}, which was supported by National Science Foundation grant number ACI-1548562.  We acknowledge Partnership for Advanced Computing in Europe for awarding us access to SuperMUC-NG at GCS@LRZ, Germany.

\appendix
\section{The weak axial-vector three-point function at zero photon momentum}
\label{app:WI}

In the following, we discuss the non-trivial limit of the three-point axial-vector correlation function $C_{3,\mu \nu}^A$ as the momentum of the photon, $p_\gamma$, goes to zero. This is a key element in the subtraction of the point-like contribution from the relevant hadronic matrix element.
To this end we retrace the main steps of the study put forth in Ref.~\cite{Desiderio:2020oej}.
The starting point is to scrutinize the electromagnetic Ward identity (WI) that connects the three-point axial correlation function $C_{3,\mu \nu}^A$ with the axial-pseudoscalar correlation function and, consequently, the matrix element $T^A_{\mu \nu}$ with the decay constant $f_H$ of the meson.
As discussed in Ref.~\cite{Desiderio:2020oej}, a careful analysis of the cutoff effects reveals that the WI does not exclude the possibility of different artifacts appearing in the decay constant extracted from the three-point function and that from the two-point function.
However, thanks to a proper change to the kernel of $C_{3,\mu \nu}^A$ it is possible to nonperturbatively subtract infrared-divergent, ${\cal O}(a^n/x_\gamma)$ discretization effects which can jeopardize the extraction of $F_{A,SD}$ at small values of $x_\gamma$.

In  general, the  lattice vector WI at finite lattice spacing reads
\begin{equation}
\left\langle \frac{ \delta O}{\delta\alpha_V(x)}\bigg\vert_{\alpha_V(x)=0} \right\rangle - \left\langle O \frac{\delta S_F}{\delta\alpha_V(x)}\bigg\vert_{\alpha_V(x)=0} \right\rangle=0\, ,\label{eq:ward}
\end{equation}
where $S_F$ is the lattice fermion action,  $O$ is a generic operator, and $\langle \dots\rangle$ represents the matrix element of the operators on the vacuum, which is invariant under vector-like rotations controlled by the continuous  parameter $\alpha_V(x)$.

In the case of the correlation function $C_{3,\mu \nu}^A$, defined as
\begin{flalign}
C_{3, \mu\nu}^A(t_{\text{em}},t_H;\vec p_\gamma,\vec p_H) = a^6\sum_{\vec x,\vec y}\, e^{-i\vec p_\gamma \cdot \vec x+i\vec p_H\cdot \vec y}\, 
 \langle {\cal J}^{\text{em}}_\mu(t_{\text{em}},\vec x) J^A_\nu(0) \phi^\dagger_H(t_H,\vec y)\rangle\; ,
\end{flalign}
the WI, at fixed lattice spacing, related to the conserved electromagnetic current ${\cal J}_{\text{em}}^\mu$ is given by 
\begin{eqnarray}
&&\sum_{\vec x,\vec y}\, e^{-i\vec p_\gamma \cdot \vec x+i\vec p_H\cdot \vec y}\, 
 \langle \nabla_\mu^*{\cal J}^{\text{em}}_\mu(t_{\text{em}},\vec x) J^A_\nu(0) \phi^\dagger_H(t_H,\vec y)\rangle
\nonumber \\
&&=
a^{-4}\sum_{\vec x,\vec y}\, e^{-i\vec p_\gamma \cdot \vec x+i\vec p_H\cdot \vec y}
\left\{\delta_{t_{em},0}\,\delta_{\vec x,\vec 0}-\delta_{t_{em},t_H}\,\delta_{\vec x,\vec y}\right\}
 \langle J^A_\nu(0) \phi^\dagger_H(t_H,\vec y)\rangle\;,\label{eq:ward1}
\end{eqnarray}
where $\nabla_\mu^*$ is the backwards discretized derivative and $\phi^\dagger_H$ a pseudoscalar interpolating operator having the flavor quantum numbers of the incoming meson.  

To implement method III described in Section \ref{ssec:sub_pgamma_zero_FA_SD}, we are interested in studying the limit $\vec p_\gamma\to \vec 0$. This can be done by using  the  exact WI satisfied by $C_{3,\mu \nu}^A(t_{\text{em}}, t_H;\vec p_\gamma, \vec p_H)$ at finite lattice spacing;  in particular we aim to understand the structure of the correlation function $C_{3,\mu \nu}^A(t_{\text{em}}, t_H;\vec p_\gamma, \vec p_H)$ at $\vec p_\gamma=\vec 0$. To this end, we consider the two-point correlation functions on the right-hand side of Eq.\,(\ref{eq:ward1}) when $\nu$ is a spatial index (the case $\nu = 0$ is similar). From the spectral decomposition we get
\begin{equation}
a^3 \sum_{\vec y}\, e^{i\vec p\cdot \vec y}\, \langle J^A_k(0) \phi^\dagger_H(t_H,\vec y) \rangle = \frac{p_k f_H(\vec p) G_H(\vec p)}{2 E_H(\vec p)} e^{-t_H E_H(\vec p)} + \dots\, ,
\end{equation}
where the dots represent sub-leading exponentials. In the previous expression $f_H(\vec p)$, $G_H(\vec p)$, and $E_H(\vec p)$ are respectively the decay constant, the matrix element of the pseudoscalar density used as interpolating operator and the energy of the meson.  

By differentiating Eq.\,(\ref{eq:ward1}) with respect to the component $(p_\gamma)_j$ of $\vec p_\gamma$, using the previous expression and the symmetries of the lattice hyper-cubic group and then setting $\vec p_\gamma=\vec 0$, one gets
\begin{equation}
C_{3,jk}^A(t_{\text{em}},t_H;\vec 0,\vec p_H)= a^{-1} \delta_{t_{\text{em}},t_H}\,\frac{f_H(\vec p_H) G_H(\vec p_H)}{2 E_H(\vec p_H)} e^{-t_H E_H(\vec p_H)}
\left\{
\delta_{jk} - \frac{(p_H)_j (p_H)_k}{E^2_H(\vec p_H)}\left[ 1+t_H E_H(\vec p_H) + {\cal O}(a^2) \right]
\right\}+\dots\;,
\label{eq:specdec}
\end{equation}
where the ellipsis represents sub-leading exponentials.

As can be seen, the structure of $C_{3,jk}^A(t_{\text{em}},t_H;\vec 0,\vec p_H)$ is highly non trivial. Note in particular the term linear in $t_H$ that arises as a manifestation of the singular behavior at large distances of the correlation function.
In the rest frame of the meson ($\vec p_H = \vec 0$), which we use in our study, the terms in square brackets of Eq.\,(\ref{eq:specdec}) disappear leading to
\begin{flalign}
C_{3,jk}^A(t_{\text{em}},t_H;\vec 0,\vec 0)=
\delta_{jk} \,a^{-1} \delta_{t_{\text{em}},t_H}\,\frac{ f_H(\vec 0) G_H(\vec 0)}{2 m_H} e^{-t_H m_H}
+ \dots \;.
\end{flalign}
Therefore, we conclude that $C_{3,jk}^A(t_{\text{em}},t_H;\vec 0,\vec 0)$ can be analyzed to extract the coefficient of the leading exponential, {\it viz.}~the decay constant appearing in the lattice matrix element of the axial current.

We observe that the use of a non-conserved electromagnetic current $J^{\text{em}}_\mu$ does not induce the presence of contact terms which could spoil the enforcement of the correct WI in the continuum limit.
By dimensional analysis, the coefficient of the leading term in the operator product expansion of $J^{\text{em}}_\mu(x)\, J_\nu^{A}(0)$ scales as $\sim 1/|x|^3$ at small distances, leading to discretization terms after summation over $x$.
The use of an improved estimator to extract the structure-dependent form factor $F_{A,SD} (E_\gamma)$ makes it possible to nonperturbatively subtract those lattice artifacts.
This can be achieved by computing the subtracted correlation function
\begin{equation}
\label{eq:3pt_sub}
a^6 \sum_{\vec x, \vec y}\, \left( e^{-i\vec p_\gamma \cdot \vec x}-1\right)\, 
 \langle J_j^{\text{em}}(t_{\text{em}},\vec x) J^A_k(0)  \phi^\dagger_H(t_H,\vec y) \rangle\quad {\rm for}~j=k \;, 
\end{equation}
whose kernel sufficiently suppresses short-distance contributions. In this way we are able to use less computationally costly, non-conserved, local lattice vector currents.
Note that, by construction, the estimator (\ref{eq:3pt_sub}) vanishes identically at $x_\gamma = 0$ with vanishing noise.

\providecommand{\href}[2]{#2}\begingroup\raggedright\endgroup


\begin{thebibliography}{10}

\bibitem{Bloch:1937pw}
F.~Bloch and A.~Nordsieck, ``{Note on the Radiation Field of the electron},''
  \href{http://dx.doi.org/10.1103/PhysRev.52.54}{Phys. Rev. {\bfseries 52}
  (1937) 54--59}.

\bibitem{Giusti:2017dwk}
D.~Giusti, V.~Lubicz, G.~Martinelli, C.~T. Sachrajda, F.~Sanfilippo, S.~Simula,
  N.~Tantalo, and C.~Tarantino, ``{First lattice calculation of the QED
  corrections to leptonic decay rates},''
  \href{http://dx.doi.org/10.1103/PhysRevLett.120.072001}{Phys. Rev. Lett.
  {\bfseries 120} no.~7, (2018) 072001},
  \href{http://arxiv.org/abs/1711.06537}{{\ttfamily arXiv:1711.06537
  [hep-lat]}}.

\bibitem{DiCarlo:2019thl}
M.~Di~Carlo, D.~Giusti, V.~Lubicz, G.~Martinelli, C.~T. Sachrajda,
  F.~Sanfilippo, S.~Simula, and N.~Tantalo, ``{Light-meson leptonic decay rates
  in lattice QCD+QED},''
  \href{http://dx.doi.org/10.1103/PhysRevD.100.034514}{Phys. Rev. D {\bfseries
  100} no.~3, (2019) 034514}, \href{http://arxiv.org/abs/1904.08731}{{\ttfamily
  arXiv:1904.08731 [hep-lat]}}.

\bibitem{ParticleDataGroup:2022pth}
{\bfseries Particle Data Group} Collaboration, R.~L. Workman {\em et~al.},
  ``{Review of Particle Physics},''
  \href{http://dx.doi.org/10.1093/ptep/ptac097}{PTEP {\bfseries 2022} (2022)
  083C01}.

\bibitem{Kruger:2002gf}
F.~Kruger and D.~Melikhov, ``{Gauge invariance and form-factors for the decay
  $B\to \gamma l^+ l^-$},''
  \href{http://dx.doi.org/10.1103/PhysRevD.67.034002}{Phys. Rev. D {\bfseries
  67} (2003) 034002}, \href{http://arxiv.org/abs/hep-ph/0208256}{{\ttfamily
  arXiv:hep-ph/0208256}}.

\bibitem{Melikhov:2004mk}
D.~Melikhov and N.~Nikitin, ``{Rare radiative leptonic decays $B_{d,s}\to
  l^+l^- \gamma$},'' \href{http://dx.doi.org/10.1103/PhysRevD.70.114028}{Phys.
  Rev. D {\bfseries 70} (2004) 114028},
  \href{http://arxiv.org/abs/hep-ph/0410146}{{\ttfamily arXiv:hep-ph/0410146}}.

\bibitem{Dettori:2016zff}
F.~Dettori, D.~Guadagnoli, and M.~Reboud, ``{$B^{0}_{s} \to
  \mu^{+}\mu^{-}\gamma$ from $B^{0}_{s} \to \mu^{+}\mu^{-}$},''
  \href{http://dx.doi.org/10.1016/j.physletb.2017.02.048}{Phys. Lett. B
  {\bfseries 768} (2017) 163--167},
  \href{http://arxiv.org/abs/1610.00629}{{\ttfamily arXiv:1610.00629
  [hep-ph]}}.

\bibitem{Albrecht:2019zul}
J.~Albrecht, E.~Stamou, R.~Ziegler, and R.~Zwicky, ``{Flavoured axions in the
  tail of $B_{q}\to\mu^{+}\mu^{-}$ and $B \to\gamma^{*}$ form factors},''
  \href{http://dx.doi.org/10.1007/JHEP09(2021)139}{JHEP {\bfseries 21} (2020)
  139}, \href{http://arxiv.org/abs/1911.05018}{{\ttfamily arXiv:1911.05018
  [hep-ph]}}.

\bibitem{Beneke:2020fot}
M.~Beneke, C.~Bobeth, and Y.-M. Wang, ``{$B_{d,s}\to\gamma\ell\bar{\ell}$ decay
  with an energetic photon},''
  \href{http://dx.doi.org/10.1007/JHEP12(2020)148}{JHEP {\bfseries 12} (2020)
  148}, \href{http://arxiv.org/abs/2008.12494}{{\ttfamily arXiv:2008.12494
  [hep-ph]}}.

\bibitem{Chen:2020szf}
S.-L. Chen, A.~Dutta~Banik, Z.~Kang, Q.~Qin, and Y.~Shigekami, ``{Signatures of
  a flavor changing $Z'$ boson in $B_q \to \gamma Z'$},''
  \href{http://dx.doi.org/10.1016/j.nuclphysb.2020.115237}{Nucl. Phys. B
  {\bfseries 962} (2021) 115237},
  \href{http://arxiv.org/abs/2006.03383}{{\ttfamily arXiv:2006.03383
  [hep-ph]}}.

\bibitem{Carvunis:2021jga}
A.~Carvunis, F.~Dettori, S.~Gangal, D.~Guadagnoli, and C.~Normand, ``{On the
  effective lifetime of $B_{s}\to\mu\mu\gamma$},''
  \href{http://dx.doi.org/10.1007/JHEP12(2021)078}{JHEP {\bfseries 12} (2021)
  078}, \href{http://arxiv.org/abs/2102.13390}{{\ttfamily arXiv:2102.13390
  [hep-ph]}}.

\bibitem{Greljo:2022jac}
A.~Greljo, J.~Salko, A.~Smolkovi\v{c}, and P.~Stangl, ``{Rare $b$ decays meet
  high-mass Drell-Yan},'' \href{http://arxiv.org/abs/2212.10497}{{\ttfamily
  arXiv:2212.10497 [hep-ph]}}.

\bibitem{Guadagnoli:2017quo}
D.~Guadagnoli, M.~Reboud, and R.~Zwicky, ``{$B_{s}^{0} \to \ell^{+} \ell^{-}
  \gamma$ as a test of lepton flavor universality},''
  \href{http://dx.doi.org/10.1007/JHEP11(2017)184}{JHEP {\bfseries 11} (2017)
  184}, \href{http://arxiv.org/abs/1708.02649}{{\ttfamily arXiv:1708.02649
  [hep-ph]}}.

\bibitem{Korchemsky:1999qb}
G.~P. Korchemsky, D.~Pirjol, and T.-M. Yan, ``{Radiative leptonic decays of B
  mesons in QCD},'' \href{http://dx.doi.org/10.1103/PhysRevD.61.114510}{Phys.
  Rev. D {\bfseries 61} (2000) 114510},
  \href{http://arxiv.org/abs/hep-ph/9911427}{{\ttfamily arXiv:hep-ph/9911427}}.

\bibitem{Beneke:1999br}
M.~Beneke, G.~Buchalla, M.~Neubert, and C.~T. Sachrajda, ``{QCD factorization
  for $B \to \pi \pi$ decays: Strong phases and CP violation in the heavy quark
  limit},'' \href{http://dx.doi.org/10.1103/PhysRevLett.83.1914}{Phys. Rev.
  Lett. {\bfseries 83} (1999) 1914--1917},
  \href{http://arxiv.org/abs/hep-ph/9905312}{{\ttfamily arXiv:hep-ph/9905312}}.

\bibitem{Descotes-Genon:2002crx}
S.~Descotes-Genon and C.~T. Sachrajda, ``{Factorization, the light cone
  distribution amplitude of the $B$ meson and the radiative decay $B \to \gamma
  l \nu_l$},'' \href{http://dx.doi.org/10.1016/S0550-3213(02)01066-0}{Nucl.
  Phys. B {\bfseries 650} (2003) 356--390},
  \href{http://arxiv.org/abs/hep-ph/0209216}{{\ttfamily arXiv:hep-ph/0209216}}.

\bibitem{Lunghi:2002ju}
E.~Lunghi, D.~Pirjol, and D.~Wyler, ``{Factorization in leptonic radiative $B
  \to \gamma e\nu$ decays},''
  \href{http://dx.doi.org/10.1016/S0550-3213(02)01032-5}{Nucl. Phys. B
  {\bfseries 649} (2003) 349--364},
  \href{http://arxiv.org/abs/hep-ph/0210091}{{\ttfamily arXiv:hep-ph/0210091}}.

\bibitem{Braun:2012kp}
V.~M. Braun and A.~Khodjamirian, ``{Soft contribution to $B\to \gamma \ell
  \nu_\ell$ and the $B$-meson distribution amplitude},''
  \href{http://dx.doi.org/10.1016/j.physletb.2012.11.047}{Phys. Lett. B
  {\bfseries 718} (2013) 1014--1019},
  \href{http://arxiv.org/abs/1210.4453}{{\ttfamily arXiv:1210.4453 [hep-ph]}}.

\bibitem{Wang:2016qii}
Y.-M. Wang, ``{Factorization and dispersion relations for radiative leptonic
  $B$ decay},'' \href{http://dx.doi.org/10.1007/JHEP09(2016)159}{JHEP
  {\bfseries 09} (2016) 159}, \href{http://arxiv.org/abs/1606.03080}{{\ttfamily
  arXiv:1606.03080 [hep-ph]}}.

\bibitem{Beneke:2018wjp}
M.~Beneke, V.~Braun, Y.~Ji, and Y.-B. Wei, ``{Radiative leptonic decay $B\to
  \gamma \ell \nu_\ell$ with subleading power corrections},''
  \href{http://dx.doi.org/10.1007/JHEP07(2018)154}{JHEP {\bfseries 07} (2018)
  154}, \href{http://arxiv.org/abs/1804.04962}{{\ttfamily arXiv:1804.04962
  [hep-ph]}}.

\bibitem{Wang:2018wfj}
Y.-M. Wang and Y.-L. Shen, ``{Subleading-power corrections to the radiative
  leptonic $B \to \gamma \ell \nu$ decay in QCD},''
  \href{http://dx.doi.org/10.1007/JHEP05(2018)184}{JHEP {\bfseries 05} (2018)
  184}, \href{http://arxiv.org/abs/1803.06667}{{\ttfamily arXiv:1803.06667
  [hep-ph]}}.

\bibitem{Shen:2018abs}
Y.-L. Shen, Z.-T. Zou, and Y.-B. Wei, ``{Subleading power corrections to $B\to
  \gamma l\nu $ decay in PQCD approach},''
  \href{http://dx.doi.org/10.1103/PhysRevD.99.016004}{Phys. Rev. D {\bfseries
  99} no.~1, (2019) 016004}, \href{http://arxiv.org/abs/1811.08250}{{\ttfamily
  arXiv:1811.08250 [hep-ph]}}.

\bibitem{Shen:2020hsp}
Y.-L. Shen, Y.-B. Wei, X.-C. Zhao, and S.-H. Zhou, ``{Revisiting radiative
  leptonic $B$ decay},''
  \href{http://dx.doi.org/10.1088/1674-1137/abb6df}{Chin. Phys. C {\bfseries
  44} no.~12, (2020) 123106}, \href{http://arxiv.org/abs/2009.03480}{{\ttfamily
  arXiv:2009.03480 [hep-ph]}}.

\bibitem{E787:2000ehx}
{\bfseries E787} Collaboration, S.~Adler {\em et~al.}, ``{Measurement of
  structure dependent $K^+ \to \mu^+ \nu_\mu \gamma$ decay},''
  \href{http://dx.doi.org/10.1103/PhysRevLett.85.2256}{Phys. Rev. Lett.
  {\bfseries 85} (2000) 2256--2259},
  \href{http://arxiv.org/abs/hep-ex/0003019}{{\ttfamily arXiv:hep-ex/0003019}}.

\bibitem{Bychkov:2008ws}
M.~Bychkov {\em et~al.}, ``{New Precise Measurement of the Pion Weak Form
  Factors in $\pi^+ \to e^+ \nu \gamma$ Decay},''
  \href{http://dx.doi.org/10.1103/PhysRevLett.103.051802}{Phys. Rev. Lett.
  {\bfseries 103} (2009) 051802},
  \href{http://arxiv.org/abs/0804.1815}{{\ttfamily arXiv:0804.1815 [hep-ex]}}.

\bibitem{KLOE:2009urs}
{\bfseries KLOE} Collaboration, F.~Ambrosino {\em et~al.}, ``{Precise
  measurement of $\Gamma(K \to e \nu(\gamma)) / \Gamma(K \to \mu \nu(\gamma))$
  and study of $K \to e \nu \gamma$},''
  \href{http://dx.doi.org/10.1140/epjc/s10052-009-1217-6}{Eur. Phys. J. C
  {\bfseries 64} (2009) 627--636},
  \href{http://arxiv.org/abs/0907.3594}{{\ttfamily arXiv:0907.3594 [hep-ex]}}.
  [Erratum: Eur.Phys.J. 65, 703 (2010)].

\bibitem{ISTRA:2010smy}
{\bfseries ISTRA+} Collaboration, V.~A. Duk {\em et~al.}, ``{Extraction of Kaon
  Formfactors from $K^- \to \mu^- \nu_\mu \gamma$ Decay at ISTRA+ Setup},''
  \href{http://dx.doi.org/10.1016/j.physletb.2010.10.043}{Phys. Lett. B
  {\bfseries 695} (2011) 59--66},
  \href{http://arxiv.org/abs/1005.3517}{{\ttfamily arXiv:1005.3517 [hep-ex]}}.

\bibitem{OKA:2019gav}
{\bfseries OKA} Collaboration, V.~I. Kravtsov {\em et~al.}, ``{Measurement of
  the $K^+\rightarrow{\mu^+}{\nu_{\mu}}{\gamma}$ decay form factors in the OKA
  experiment},'' \href{http://dx.doi.org/10.1140/epjc/s10052-019-7145-1}{Eur.
  Phys. J. C {\bfseries 79} no.~7, (2019) 635},
  \href{http://arxiv.org/abs/1904.10078}{{\ttfamily arXiv:1904.10078
  [hep-ex]}}.

\bibitem{Kobayashi:2022hwh}
A.~Kobayashi {\em et~al.}, ``{New determination of the branching ratio of the
  structure dependent radiative $K^{+} \to e^{+} \nu_{e} \gamma$ decay},''
  \href{http://arxiv.org/abs/2212.10702}{{\ttfamily arXiv:2212.10702
  [hep-ex]}}.

\bibitem{BESIII:2017whk}
{\bfseries BESIII} Collaboration, M.~Ablikim {\em et~al.}, ``{Search for the
  radiative leptonic decay $D^{+}\to \gamma e^{+} {\nu}_{e}$},''
  \href{http://dx.doi.org/10.1103/PhysRevD.95.071102}{Phys. Rev. D {\bfseries
  95} no.~7, (2017) 071102}, \href{http://arxiv.org/abs/1702.05837}{{\ttfamily
  arXiv:1702.05837 [hep-ex]}}.

\bibitem{BESIII:2019pjk}
{\bfseries BESIII} Collaboration, M.~Ablikim {\em et~al.}, ``{Search for the
  decay $D_s^+\rightarrow \gamma e^+\nu_e$},''
  \href{http://dx.doi.org/10.1103/PhysRevD.99.072002}{Phys. Rev. D {\bfseries
  99} no.~7, (2019) 072002}, \href{http://arxiv.org/abs/1902.03351}{{\ttfamily
  arXiv:1902.03351 [hep-ex]}}.

\bibitem{Belle:2018jqd}
{\bfseries Belle} Collaboration, M.~Gelb {\em et~al.}, ``{Search for the rare
  decay of $B^+ \to \ell^{\,+} \nu_{\ell} \gamma$ with improved hadronic
  tagging},'' \href{http://dx.doi.org/10.1103/PhysRevD.98.112016}{Phys. Rev. D
  {\bfseries 98} no.~11, (2018) 112016},
  \href{http://arxiv.org/abs/1810.12976}{{\ttfamily arXiv:1810.12976
  [hep-ex]}}.

\bibitem{Belle-II:2018jsg}
{\bfseries Belle-II} Collaboration, W.~Altmannshofer {\em et~al.}, ``{The Belle
  II Physics Book},'' \href{http://dx.doi.org/10.1093/ptep/ptz106}{PTEP
  {\bfseries 2019} no.~12, (2019) 123C01},
  \href{http://arxiv.org/abs/1808.10567}{{\ttfamily arXiv:1808.10567
  [hep-ex]}}. [Erratum: PTEP 2020, 029201 (2020)].

\bibitem{BaBar:2007lky}
{\bfseries BaBar} Collaboration, B.~Aubert {\em et~al.}, ``{Search for the
  decays $B^0 \to e^{+} e^{-} \gamma$ and $B^0 \to \mu^{+} \mu^{-} \gamma$},''
  \href{http://dx.doi.org/10.1103/PhysRevD.77.011104}{Phys. Rev. D {\bfseries
  77} (2008) 011104}, \href{http://arxiv.org/abs/0706.2870}{{\ttfamily
  arXiv:0706.2870 [hep-ex]}}.

\bibitem{LHCb:2021awg}
{\bfseries LHCb} Collaboration, R.~Aaij {\em et~al.}, ``{Measurement of the
  $B^0_s\to\mu^+\mu^-$ decay properties and search for the $B^0\to\mu^+\mu^-$
  and $B^0_s\to\mu^+\mu^-\gamma$ decays},''
  \href{http://dx.doi.org/10.1103/PhysRevD.105.012010}{Phys. Rev. D {\bfseries
  105} no.~1, (2022) 012010}, \href{http://arxiv.org/abs/2108.09283}{{\ttfamily
  arXiv:2108.09283 [hep-ex]}}.

\bibitem{Bijnens:1996wm}
J.~Bijnens and P.~Talavera, ``{$\pi\to\ell\nu\gamma$ form-factors at two
  loop},'' \href{http://dx.doi.org/10.1016/S0550-3213(97)00069-2}{Nucl. Phys. B
  {\bfseries 489} (1997) 387--404},
  \href{http://arxiv.org/abs/hep-ph/9610269}{{\ttfamily arXiv:hep-ph/9610269}}.

\bibitem{Geng:2003mt}
C.~Q. Geng, I.-L. Ho, and T.~H. Wu, ``{Axial vector form-factors for $K_{l
  2\gamma}$ and $\pi_{l 2\gamma}$ at $O(p^6)$ in chiral perturbation theory},''
  \href{http://dx.doi.org/10.1016/j.nuclphysb.2003.12.039}{Nucl. Phys. B
  {\bfseries 684} (2004) 281--317},
  \href{http://arxiv.org/abs/hep-ph/0306165}{{\ttfamily arXiv:hep-ph/0306165}}.

\bibitem{Mateu:2007tr}
V.~Mateu and J.~Portoles, ``{Form-factors in radiative pion decay},''
  \href{http://dx.doi.org/10.1140/epjc/s10052-007-0393-5}{Eur. Phys. J. C
  {\bfseries 52} (2007) 325--338},
  \href{http://arxiv.org/abs/0706.1039}{{\ttfamily arXiv:0706.1039 [hep-ph]}}.

\bibitem{Unterdorfer:2008zz}
R.~Unterdorfer and H.~Pichl, ``{On the Radiative Pion Decay},''
  \href{http://dx.doi.org/10.1140/epjc/s10052-008-0584-8}{Eur. Phys. J. C
  {\bfseries 55} (2008) 273--283},
  \href{http://arxiv.org/abs/0801.2482}{{\ttfamily arXiv:0801.2482 [hep-ph]}}.

\bibitem{Cirigliano:2011ny}
V.~Cirigliano, G.~Ecker, H.~Neufeld, A.~Pich, and J.~Portoles, ``{Kaon Decays
  in the Standard Model},''
  \href{http://dx.doi.org/10.1103/RevModPhys.84.399}{Rev. Mod. Phys. {\bfseries
  84} (2012) 399}, \href{http://arxiv.org/abs/1107.6001}{{\ttfamily
  arXiv:1107.6001 [hep-ph]}}.

\bibitem{Atwood:1994za}
D.~Atwood, G.~Eilam, and A.~Soni, ``{Pure leptonic radiative decays $B^{+-},
  D_s \to l\nu\gamma$ and the annihilation graph},''
  \href{http://dx.doi.org/10.1142/S0217732396001090}{Mod. Phys. Lett. A
  {\bfseries 11} (1996) 1061--1968},
  \href{http://arxiv.org/abs/hep-ph/9411367}{{\ttfamily arXiv:hep-ph/9411367}}.

\bibitem{Colangelo:1996ct}
P.~Colangelo, F.~De~Fazio, and G.~Nardulli, ``{On the decay mode $B^- \to \mu^-
  \bar{\nu}_\mu \gamma$},''
  \href{http://dx.doi.org/10.1016/0370-2693(96)00955-0}{Phys. Lett. B
  {\bfseries 386} (1996) 328--334},
  \href{http://arxiv.org/abs/hep-ph/9606219}{{\ttfamily arXiv:hep-ph/9606219}}.

\bibitem{Chang:1997re}
C.-H. Chang, J.-P. Cheng, and C.-D. Lu, ``{Radiative leptonic decays of $B_c$
  meson},'' \href{http://dx.doi.org/10.1016/S0370-2693(98)00177-4}{Phys. Lett.
  B {\bfseries 425} (1998) 166--170},
  \href{http://arxiv.org/abs/hep-ph/9712325}{{\ttfamily arXiv:hep-ph/9712325}}.

\bibitem{Geng:2000fs}
C.~Q. Geng, C.~C. Lih, and W.-M. Zhang, ``{Study of $B_{s,d} \to l^+ l^-
  \gamma$ decays},'' \href{http://dx.doi.org/10.1103/PhysRevD.62.074017}{Phys.
  Rev. D {\bfseries 62} (2000) 074017},
  \href{http://arxiv.org/abs/hep-ph/0007252}{{\ttfamily arXiv:hep-ph/0007252}}.

\bibitem{Chelkov:2001qx}
G.~A. Chelkov, M.~I. Gostkin, and Z.~K. Silagadze, ``{Radiative leptonic B
  decays in the instantaneous Bethe-Salpeter approach},''
  \href{http://dx.doi.org/10.1103/PhysRevD.64.097503}{Phys. Rev. D {\bfseries
  64} (2001) 097503}, \href{http://arxiv.org/abs/hep-ph/0104172}{{\ttfamily
  arXiv:hep-ph/0104172}}.

\bibitem{Hwang:2005uk}
C.-W. Hwang, ``{Radiative leptonic decays of heavy mesons in heavy quark
  limit},'' \href{http://dx.doi.org/10.1140/epjc/s2006-02510-2}{Eur. Phys. J. C
  {\bfseries 46} (2006) 379--384},
  \href{http://arxiv.org/abs/hep-ph/0512006}{{\ttfamily arXiv:hep-ph/0512006}}.

\bibitem{Barik:2008zza}
N.~Barik, S.~Naimuddin, P.~C. Dash, and S.~Kar, ``{Radiative leptonic decay:
  $B^-\to \mu^-\bar{\nu}_\mu\gamma$ in a relativistic independent quark
  model},'' \href{http://dx.doi.org/10.1103/PhysRevD.77.014038}{Phys. Rev. D
  {\bfseries 77} (2008) 014038}.

\bibitem{Shen:2013oua}
Y.-L. Shen and G.~Li, ``{Radiative $D(D_{s})$ decays in the covariant light
  front approach},''
  \href{http://dx.doi.org/10.1140/epjc/s10052-013-2441-7}{Eur. Phys. J. C
  {\bfseries 73} (2013) 2441}.

\bibitem{Kozachuk:2017mdk}
A.~Kozachuk, D.~Melikhov, and N.~Nikitin, ``{Rare FCNC radiative leptonic
  $B_{s,d}\to \gamma l^+l^-$ decays in the standard model},''
  \href{http://dx.doi.org/10.1103/PhysRevD.97.053007}{Phys. Rev. D {\bfseries
  97} no.~5, (2018) 053007}, \href{http://arxiv.org/abs/1712.07926}{{\ttfamily
  arXiv:1712.07926 [hep-ph]}}.

\bibitem{Dubnicka:2018gqg}
S.~Dubni\v{c}ka, A.~Z. Dubni\v{c}kov\'a, M.~A. Ivanov, A.~Liptaj,
  P.~Santorelli, and C.~T. Tran, ``{Study of $B_s\to\ell^+\ell^-\gamma$ decays
  in covariant quark model},''
  \href{http://dx.doi.org/10.1103/PhysRevD.99.014042}{Phys. Rev. D {\bfseries
  99} no.~1, (2019) 014042}, \href{http://arxiv.org/abs/1808.06261}{{\ttfamily
  arXiv:1808.06261 [hep-ph]}}.

\bibitem{Khodjamirian:1995uc}
A.~Khodjamirian, G.~Stoll, and D.~Wyler, ``{Calculation of long distance
  effects in exclusive weak radiative decays of B meson},''
  \href{http://dx.doi.org/10.1016/0370-2693(95)00972-N}{Phys. Lett. B
  {\bfseries 358} (1995) 129--138},
  \href{http://arxiv.org/abs/hep-ph/9506242}{{\ttfamily arXiv:hep-ph/9506242}}.

\bibitem{Ali:1995uy}
A.~Ali and V.~M. Braun, ``{Estimates of the weak annihilation contributions to
  the decays $B \to \rho \gamma$ and $B \to \omega \gamma$},''
  \href{http://dx.doi.org/10.1016/0370-2693(95)01087-7}{Phys. Lett. B
  {\bfseries 359} (1995) 223--235},
  \href{http://arxiv.org/abs/hep-ph/9506248}{{\ttfamily arXiv:hep-ph/9506248}}.

\bibitem{Eilam:1995zv}
G.~Eilam, I.~E. Halperin, and R.~R. Mendel, ``{Radiative decay $B \to
  l\nu\gamma$ in the light cone QCD approach},''
  \href{http://dx.doi.org/10.1016/0370-2693(95)01088-8}{Phys. Lett. B
  {\bfseries 361} (1995) 137--145},
  \href{http://arxiv.org/abs/hep-ph/9506264}{{\ttfamily arXiv:hep-ph/9506264}}.

\bibitem{Aliev:1996ud}
T.~M. Aliev, A.~Ozpineci, and M.~Savci, ``{$B_q \to l^+ l^- \gamma$ decays in
  light cone QCD},'' \href{http://dx.doi.org/10.1103/PhysRevD.55.7059}{Phys.
  Rev. D {\bfseries 55} (1997) 7059--7066},
  \href{http://arxiv.org/abs/hep-ph/9611393}{{\ttfamily arXiv:hep-ph/9611393}}.

\bibitem{Ball:2003fq}
P.~Ball and E.~Kou, ``{$B \to \gamma e \nu$ transitions from QCD sum rules on
  the light cone},''
  \href{http://dx.doi.org/10.1088/1126-6708/2003/04/029}{JHEP {\bfseries 04}
  (2003) 029}, \href{http://arxiv.org/abs/hep-ph/0301135}{{\ttfamily
  arXiv:hep-ph/0301135}}.

\bibitem{Janowski:2021yvz}
T.~Janowski, B.~Pullin, and R.~Zwicky, ``{Charged and neutral $
  {\overline{B}}_{u,d,s} $ $\to \gamma$ form factors from light cone sum rules
  at NLO},'' \href{http://dx.doi.org/10.1007/JHEP12(2021)008}{JHEP {\bfseries
  12} (2021) 008}, \href{http://arxiv.org/abs/2106.13616}{{\ttfamily
  arXiv:2106.13616 [hep-ph]}}.

\bibitem{Burdman:1994ip}
G.~Burdman, J.~T. Goldman, and D.~Wyler, ``{Radiative leptonic decays of heavy
  mesons},'' \href{http://dx.doi.org/10.1103/PhysRevD.51.111}{Phys. Rev. D
  {\bfseries 51} (1995) 111--117},
  \href{http://arxiv.org/abs/hep-ph/9405425}{{\ttfamily arXiv:hep-ph/9405425}}.

\bibitem{SalehKhan:2004kj}
M.~Saleh~Khan, M.~Jamil~Aslam, A.~H.~S. Gilani, and Riazuddin, ``{Form-factors
  and branching ratio for the $B \to l\nu\gamma$ decay},''
  \href{http://dx.doi.org/10.1140/epjc/s10052-006-0152-z}{Eur. Phys. J. C
  {\bfseries 49} (2007) 665--674},
  \href{http://arxiv.org/abs/hep-ph/0410060}{{\ttfamily arXiv:hep-ph/0410060}}.

\bibitem{Kurten:2022zuy}
S.~K\"urten, M.~Zanke, B.~Kubis, and D.~van Dyk, ``{Dispersion relations for
  $B^- \to \ell^- \bar{\nu}_\ell \ell^{\prime-} \ell^{\prime+}$ form
  factors},'' \href{http://arxiv.org/abs/2210.09832}{{\ttfamily
  arXiv:2210.09832 [hep-ph]}}.

\bibitem{Kane:2019jtj}
C.~Kane, C.~Lehner, S.~Meinel, and A.~Soni, ``{Radiative leptonic decays on the
  lattice},'' \href{http://dx.doi.org/10.22323/1.363.0134}{PoS {\bfseries
  LATTICE2019} (2019) 134}, \href{http://arxiv.org/abs/1907.00279}{{\ttfamily
  arXiv:1907.00279 [hep-lat]}}.

\bibitem{Desiderio:2020oej}
A.~Desiderio {\em et~al.}, ``{First lattice calculation of radiative leptonic
  decay rates of pseudoscalar mesons},''
  \href{http://dx.doi.org/10.1103/PhysRevD.103.014502}{Phys. Rev. D {\bfseries
  103} no.~1, (2021) 014502}, \href{http://arxiv.org/abs/2006.05358}{{\ttfamily
  arXiv:2006.05358 [hep-lat]}}.

\bibitem{Kane:2021zee}
C.~Kane, D.~Giusti, C.~Lehner, S.~Meinel, and A.~Soni, ``{Controlling unwanted
  exponentials in lattice calculations of radiative leptonic decays},''
  \href{http://dx.doi.org/10.22323/1.396.0162}{PoS {\bfseries LATTICE2021}
  (2022) 162}, \href{http://arxiv.org/abs/2110.13196}{{\ttfamily
  arXiv:2110.13196 [hep-lat]}}.

\bibitem{Frezzotti:2020bfa}
R.~Frezzotti, M.~Garofalo, V.~Lubicz, G.~Martinelli, C.~T. Sachrajda,
  F.~Sanfilippo, S.~Simula, and N.~Tantalo, ``{Comparison of lattice QCD+QED
  predictions for radiative leptonic decays of light mesons with experimental
  data},'' \href{http://dx.doi.org/10.1103/PhysRevD.103.053005}{Phys. Rev. D
  {\bfseries 103} no.~5, (2021) 053005},
  \href{http://arxiv.org/abs/2012.02120}{{\ttfamily arXiv:2012.02120
  [hep-ph]}}.

\bibitem{RBC:2010qam}
{\bfseries RBC, UKQCD} Collaboration, Y.~Aoki {\em et~al.}, ``{Continuum Limit
  Physics from 2+1 Flavor Domain Wall QCD},''
  \href{http://dx.doi.org/10.1103/PhysRevD.83.074508}{Phys. Rev. D {\bfseries
  83} (2011) 074508}, \href{http://arxiv.org/abs/1011.0892}{{\ttfamily
  arXiv:1011.0892 [hep-lat]}}.

\bibitem{RBC:2014ntl}
{\bfseries RBC, UKQCD} Collaboration, T.~Blum {\em et~al.}, ``{Domain wall QCD
  with physical quark masses},''
  \href{http://dx.doi.org/10.1103/PhysRevD.93.074505}{Phys. Rev. D {\bfseries
  93} no.~7, (2016) 074505}, \href{http://arxiv.org/abs/1411.7017}{{\ttfamily
  arXiv:1411.7017 [hep-lat]}}.

\bibitem{Beneke:2011nf}
M.~Beneke and J.~Rohrwild, ``{$B$ meson distribution amplitude from $B \to
  \gamma l \nu$},''
  \href{http://dx.doi.org/10.1140/epjc/s10052-011-1818-8}{Eur. Phys. J. C
  {\bfseries 71} (2011) 1818}, \href{http://arxiv.org/abs/1110.3228}{{\ttfamily
  arXiv:1110.3228 [hep-ph]}}.

\bibitem{Boyle:2018knm}
{\bfseries RBC/UKQCD} Collaboration, P.~A. Boyle, L.~Del~Debbio, N.~Garron,
  A.~Juttner, A.~Soni, J.~T. Tsang, and O.~Witzel, ``{SU(3)-breaking ratios for
  $D_{(s)}$ and $B_{(s)}$ mesons},''
  \href{http://arxiv.org/abs/1812.08791}{{\ttfamily arXiv:1812.08791
  [hep-lat]}}.

\bibitem{Shintani:2014vja}
E.~Shintani, R.~Arthur, T.~Blum, T.~Izubuchi, C.~Jung, and C.~Lehner,
  ``{Covariant approximation averaging},''
  \href{http://dx.doi.org/10.1103/PhysRevD.91.114511}{Phys. Rev. D {\bfseries
  91} no.~11, (2015) 114511}, \href{http://arxiv.org/abs/1402.0244}{{\ttfamily
  arXiv:1402.0244 [hep-lat]}}.

\bibitem{Hashimoto:1999yp}
S.~Hashimoto, A.~X. El-Khadra, A.~S. Kronfeld, P.~B. Mackenzie, S.~M. Ryan, and
  J.~N. Simone, ``{Lattice QCD calculation of $\bar{B} \to D l \bar{\nu}$ decay
  form-factors at zero recoil},''
  \href{http://dx.doi.org/10.1103/PhysRevD.61.014502}{Phys. Rev. D {\bfseries
  61} (1999) 014502}, \href{http://arxiv.org/abs/hep-ph/9906376}{{\ttfamily
  arXiv:hep-ph/9906376}}.

\bibitem{El-Khadra:2001wco}
A.~X. El-Khadra, A.~S. Kronfeld, P.~B. Mackenzie, S.~M. Ryan, and J.~N. Simone,
  ``{The Semileptonic decays $B \to \pi l \nu$ and $D \to \pi l \nu$ from
  lattice QCD},'' \href{http://dx.doi.org/10.1103/PhysRevD.64.014502}{Phys.
  Rev. D {\bfseries 64} (2001) 014502},
  \href{http://arxiv.org/abs/hep-ph/0101023}{{\ttfamily arXiv:hep-ph/0101023}}.

\bibitem{RBC:2018dos}
{\bfseries RBC, UKQCD} Collaboration, T.~Blum, P.~A. Boyle, V.~G\"ulpers,
  T.~Izubuchi, L.~Jin, C.~Jung, A.~J\"uttner, C.~Lehner, A.~Portelli, and J.~T.
  Tsang, ``{Calculation of the hadronic vacuum polarization contribution to the
  muon anomalous magnetic moment},''
  \href{http://dx.doi.org/10.1103/PhysRevLett.121.022003}{Phys. Rev. Lett.
  {\bfseries 121} no.~2, (2018) 022003},
  \href{http://arxiv.org/abs/1801.07224}{{\ttfamily arXiv:1801.07224
  [hep-lat]}}.

\bibitem{Feng:2018qpx}
X.~Feng and L.~Jin, ``{QED self energies from lattice QCD without power-law
  finite-volume errors},''
  \href{http://dx.doi.org/10.1103/PhysRevD.100.094509}{Phys. Rev. D {\bfseries
  100} no.~9, (2019) 094509}, \href{http://arxiv.org/abs/1812.09817}{{\ttfamily
  arXiv:1812.09817 [hep-lat]}}.

\bibitem{Donald:2013sra}
G.~C. Donald, C.~T.~H. Davies, J.~Koponen, and G.~P. Lepage, ``{Prediction of
  the $D_s^*$ width from a calculation of its radiative decay in full lattice
  QCD},'' \href{http://dx.doi.org/10.1103/PhysRevLett.112.212002}{Phys. Rev.
  Lett. {\bfseries 112} (2014) 212002},
  \href{http://arxiv.org/abs/1312.5264}{{\ttfamily arXiv:1312.5264 [hep-lat]}}.

\bibitem{Pullin:2021ebn}
B.~Pullin and R.~Zwicky, ``{Radiative decays of heavy-light mesons and the $
  {f}_{H,{H}^{\ast },{H}_1}^{(T)} $ decay constants},''
  \href{http://dx.doi.org/10.1007/JHEP09(2021)023}{JHEP {\bfseries 09} (2021)
  023}, \href{http://arxiv.org/abs/2106.13617}{{\ttfamily arXiv:2106.13617
  [hep-ph]}}.

\bibitem{Tuo:2021ewr}
X.-Y. Tuo, X.~Feng, L.-C. Jin, and T.~Wang, ``{Lattice QCD calculation of
  $K\to\ell\nu_\ell\ell^{\prime +}\ell^{\prime-}$ decay width},''
  \href{http://dx.doi.org/10.1103/PhysRevD.105.054518}{Phys. Rev. D {\bfseries
  105} no.~5, (2022) 054518}, \href{http://arxiv.org/abs/2103.11331}{{\ttfamily
  arXiv:2103.11331 [hep-lat]}}.

\bibitem{supplemental}
See the ancillary files included in the arXiv submission.

\bibitem{Christ:2006us}
N.~H. Christ, M.~Li, and H.-W. Lin, ``{Relativistic Heavy Quark Effective
  Action},'' \href{http://dx.doi.org/10.1103/PhysRevD.76.074505}{Phys. Rev. D
  {\bfseries 76} (2007) 074505},
  \href{http://arxiv.org/abs/hep-lat/0608006}{{\ttfamily
  arXiv:hep-lat/0608006}}.

\bibitem{Giusti:2021rsf}
D.~Giusti and C.~Lehner, ``{A new framework to tune an improved relativistic
  heavy-quark action},'' \href{http://dx.doi.org/10.22323/1.396.0042}{PoS
  {\bfseries LATTICE2021} (2022) 042},
  \href{http://arxiv.org/abs/2111.15614}{{\ttfamily arXiv:2111.15614
  [hep-lat]}}.

\bibitem{Bali:2018spj}
G.~S. Bali, V.~M. Braun, B.~Gl\"a\ss{}le, M.~G\"ockeler, M.~Gruber, F.~Hutzler,
  P.~Korcyl, A.~Sch\"afer, P.~Wein, and J.-H. Zhang, ``{Pion distribution
  amplitude from Euclidean correlation functions: Exploring universality and
  higher-twist effects},''
  \href{http://dx.doi.org/10.1103/PhysRevD.98.094507}{Phys. Rev. D {\bfseries
  98} no.~9, (2018) 094507}, \href{http://arxiv.org/abs/1807.06671}{{\ttfamily
  arXiv:1807.06671 [hep-lat]}}.

\bibitem{Gerardin:2019vio}
A.~G\'erardin, H.~B. Meyer, and A.~Nyffeler, ``{Lattice calculation of the pion
  transition form factor with $N_f=2+1$ Wilson quarks},''
  \href{http://dx.doi.org/10.1103/PhysRevD.100.034520}{Phys. Rev. D {\bfseries
  100} no.~3, (2019) 034520}, \href{http://arxiv.org/abs/1903.09471}{{\ttfamily
  arXiv:1903.09471 [hep-lat]}}.

\bibitem{XSEDE}
J.~{Towns}, T.~{Cockerill}, M.~{Dahan}, I.~{Foster}, K.~{Gaither},
  A.~{Grimshaw}, V.~{Hazlewood}, S.~{Lathrop}, D.~{Lifka}, G.~D. {Peterson},
  R.~{Roskies}, J.~R. {Scott}, and N.~{Wilkins-Diehr}, ``{XSEDE: Accelerating
  Scientific Discovery},''
  \href{http://dx.doi.org/10.1109/MCSE.2014.80}{Computing in Science
  Engineering {\bfseries 16} no.~5, (2014) 62--74}.

\bibitem{Boyle:2022lsi}
P.~Boyle {\em et~al.}, ``{Isospin-breaking corrections to light-meson leptonic
  decays from lattice simulations at physical quark masses},''
  \href{http://dx.doi.org/10.1007/JHEP02(2023)242}{JHEP {\bfseries 02} (2023)
  242}, \href{http://arxiv.org/abs/2211.12865}{{\ttfamily arXiv:2211.12865
  [hep-lat]}}.

\end{thebibliography}
\end{document}